\documentclass[a4paper,12pt]{cas-sc}

\usepackage[numbers]{natbib}
\usepackage{subcaption}
\usepackage{graphicx}
\usepackage{algorithm} 
\usepackage{algpseudocode}
\usepackage{ulem,xpatch}
\usepackage{smartdiagram}
\usepackage{caption}
\usepackage{setspace}

\usepackage{algorithm}

\def\tsc#1{\csdef{#1}{\textsc{\lowercase{#1}}\xspace}}
\tsc{WGM}
\tsc{QE}
\tsc{EP}
\tsc{PMS}
\tsc{BEC}
\tsc{DE}

\begin{document}
\def\floatpagepagefraction{1}
\def\textpagefraction{.001}

\shorttitle{An encoder-decoder deep surrogate for reverse time migration}
\shortauthors{R. S. M. Freitas et~al.}

\title [mode = title]{An encoder-decoder deep surrogate for reverse time migration in seismic imaging under uncertainty}
\author[1]{Rodolfo S. M. Freitas}[orcid=0000-0001-6036-8534]
\author[1]{Carlos H. S. Barbosa}[orcid=0000-0002-1420-9118]
\author[2]{Gabriel M. Guerra}[orcid=0000-0002-5430-3475]
\author[1]{Alvaro L. G. A. Coutinho}[orcid= 0000-0002-4764-1142]
\author[1]{Fernando A. Rochinha}[orcid=0000-0001-8035-9651]
\cormark[1]

\address[1]{COPPE, Federal University of Rio de Janeiro, Rio de Janeiro, Rio de Janeiro, 21941-598, Brazil}
\address[2]{Department of Mechanical Engineering, Federal Fluminense University, Niterói, Brazil}

\cortext[cor1]{Corresponding author. Email: rodolfosmfreitas@gmail.com (Rodolfo S. M. Freitas) faro@mecanica.coppe.ufrj.br (Fernando A. Rochinha)}

\begin{abstract}
Seismic imaging faces challenges due to the presence of several uncertainty sources. Uncertainties exist in data measurements, source positioning, and subsurface geophysical properties. Reverse time migration (RTM) is a high-resolution depth migration approach useful for extracting information such as reservoir localization and boundaries. RTM, however, is time-consuming and data-intensive as it requires computing twice the wave equation to generate and store an imaging condition. RTM, when embedded in an uncertainty quantification algorithm (like the Monte Carlo method), shows a many-fold increase in its computational complexity due to the high input-output dimensionality. In this work, we propose an encoder-decoder deep learning surrogate model for RTM under uncertainty. Inputs are an ensemble of velocity fields, expressing the uncertainty, and outputs the seismic images. We show by numerical experimentation that the surrogate model can reproduce the seismic images accurately, and, more importantly, the uncertainty propagation from the input velocity fields to the image ensemble. 
\end{abstract}


\maketitle

\section{Introduction}
\label{intro}
Seismic imaging is employed to delineate the salient geological features of the Earth subsurface. Imaging methods are popular in the Oil \& Gas industry as they are designed to be focused on the more essential characteristics: the horizons bounding the regions of interest. They can also be used in conjunction with inverse methods such as Full Waveform Inversion \cite{doi:10.1063/1.5034122}. Imaging methods are designed and built departing from the integration of specialized optical (illuminating) principles and physics-based models describing the wave propagation through heterogeneous media. A critical aspect arising from such arrangement is the potential computational cost required, as a large domain is to be illuminated, which implies solving partial differential equations (PDEs) associated with the wave models in that area. The situation tends to be more complicated as the excitation signals bear high-frequency content, which demands very fine grids in space and time. Such time-consuming tasks often hamper the use of high-fidelity codes constructed upon physics-based models. That becomes a more critical issue whenever one faces many-query applications like sensitivity analysis, design, optimization, or uncertainty quantification. 

In this work, we develop a machine-learning model to alleviate computational costs to provide seismic images with quantified uncertainty \cite{barbosa2020workflow}. In this context, we propose a Monte Carlo method (MC) to sweep a large ensemble of plausible velocity fields obtained by approximate methods, and, therefore, prone to uncertainties, to compute an ensemble of images aiming at characterizing the propagated uncertainties along with the seismic image processing. Moreover, we embed the MC sampling as an outer loop of a larger computational workflow proposed in \cite{barbosa2020workflow} and detailed in Algorithm \ref{algworkflow}. This algorithm is structured in three sequential stages, enabling a probabilistic framework for seismic imaging.

The first stage aims at generating plausible subsurface velocity fields honoring seismic data. Probabilistic inversion, such as Bayesian tomography \cite{barbosa2020workflow, Botteroetal2016, Belhadjetal2018, Brantut2018}, and stochastic FWI \cite{martin2012stochastic, zhu2016bayesian, biswas20172d, gebraad2019bayesian, zhao2019gradient} can provide a velocity field ensemble used as input to the second stage. Hence, in {\textbf{Stage 2}}, an imaging technique migrates the seismogram information using each velocity field sample. This strategy wraps a seismic migration tool into an MC algorithm aiming to build a set of migrated seismic images.  We have chosen the Reverse Time Migration (RTM) as the seismic migration technique to localize the seismic reflectors in the correct depth location in the subsurface \cite{Zhouetal2018}. RTM is a depth migration approach based on the two-way wave equation, frequently used in industry, that provides reliable subsurface high-resolution seismic images useful for seismic interpretation and reservoir characterization \cite{Zhouetal2018}. The last stage of the workflow post-processes the RTM seismic images ensemble,  calculating uncertainty maps and extracting features, such as horizons and faults, that characterizes uncertainty in the resulting images.

\begin{algorithm}
\caption{Workflow for seismic imaging with quantified uncertainty}\label{algworkflow}
{\bf{Input}}: source signals, seismograms, and spatial domain (raw data).\\
{\bf{Output}}: ensemble of seismic images.
\begin{algorithmic}
\State {\bf{Stage 1}}: Generate an ensemble of velocity fields:
\begin{itemize}
\item Bayesian inversion with simplified physics-based models
\end{itemize}
\State {\bf{Stage 2}}: Propagate uncertainties -- migrate the seismograms for the velocity field ensemble using RTM, producing a corresponding ensemble of seismic images
\begin{itemize}
    \item Monte Carlo loop over samples produced in Stage 1
\end{itemize}
\State {\bf{Stage 3}}: Post-process the RTM seismic images 
\begin{itemize}
    \item Uncertainty maps;
    \item Automatic features (horizons) detection;
    \item Probabilistic characterization of such features;
\end{itemize}
\end{algorithmic}
\end{algorithm}

Due to its flexibility by design, it is possible to generate different workflow versions, by, for instance, replacing components within the stages (e.g., different strategies for the uncertain velocity fields estimation in {\bf{Stage 1}}) targeting to accommodate different demands or efficiency requirements. Nonetheless, we would still be facing a time-consuming computational task in the many-query UQ analysis of {\bf{Stage 2}}. That is what motivates us to follow a consolidated trend, replacing the original physics-based model by a cheap-to-evaluate surrogate. Recently, Machine Learning techniques, like Gaussian Processes \cite{rasmussen,MGP,MGP2,karniadakis1,karniadakis2,comgeosciences} and Deep Neural Networks (DNNs), \cite{Zabaras1,Zabaras2,Zabaras4,Zabaras5,perdikaris,Durlofsky,KAHANA2020109458,Sun2020} have deployed  efficient surrogates for UQ analysis. Gaussian Processes have achieved considerable success with computer models with inputs and outputs of moderate dimensionality, including the ability to blend data of different sources, leading to multi-fidelity approaches \cite{karniadakis1,karniadakis2}. On the other hand, Deep Neural Networks have gained traction due to their unique profile of being flexible and scalable nonlinear function approximators.  Another aspect worth highlighting is the substantial amount of computer libraries and tools available to enable their use.

Here, we apply the deep learning surrogate architecture proposed in \cite{Zabaras1} for systems governed by PDEs cast as an image-to-image problem. The performance of such architecture was tested in uncertainty quantification of flows in heterogeneous media \cite{Zabaras2}, extended to semi-supervised learning in \cite{Zabaras4}, and inverse problems in \cite{Zabaras5}, with excellent results. This architecture is composed of convolutional layers and dense blocks,  following an encoder-decoder neural network arrangement to handle the potential high-dimensionality of inputs and outputs. More specifically, we employ the deep learning architecture for constructing efficient proxies for RTM imaging by avoiding the high costs of solving twice a wave propagation equation in a heterogeneous medium. Such surrogates are nonlinear mappings linking the uncertain velocity field to the seismic images. It is worth to highlight that differently from usual surrogates, we do not replace only a forward solver associated to a PDE, but the whole more expensive imaging process. The surrogate can handle the high-dimensional inputs (velocity fields) and outputs (seismic images), leading to cost savings in processing and memory storage. We demonstrate through two examples that such an approach enables producing seismic images with quantified uncertainty. Indeed, it can accurately reproduce the ensemble of images resulting from the MC uncertain propagation with much less computational effort. Moreover, it uses a limited training data as expected, which was confirmed by our results and efficiency estimation.

The remainder of this paper is organized as follows. The next section details the RTM mathematical problem along with its computational implementation within {\bf{Stage 2}} MC uncertainty propagation. Section 3 presents our deep learning, surrogate architecture and training strategy. In Section 4, we present numerical experiments where we investigate the accuracy, convergence, and cost-effectiveness of our surrogate model to replace the high-fidelity RTM under uncertainty. The paper ends with a summary of our main findings.

\section{Reverse Time Migration under Uncertainty}\label{sec:rtm}

RTM is a high-resolution depth migration technique providing useful subsurface images for extracting information such as reservoir localization and boundaries \cite{Zhouetal2018}. The raw data for RTM consists of recorded seismic signals induced by a seismic source (a shot). The group of seismic signals represents a seismogram that captures information related to reflections coming from the subsurface. RTM relies on the two-way wave propagation equation, resulting in an imaging condition (IC) \cite{schuster2017} computed over the space-time domain to be imaged. More specifically, the wave equation is solved twice, the first time to compute the forward-propagated wave due to seismic sources, followed by the computation of the backward-propagated wave induced by the recorded seismograms. Both solutions are needed to compute the IC. We calculate the forward wave isotropic acoustic case by solving,

\begin{eqnarray} \label{eq.secondOrder}
\nabla^{2} p(\textbf{r},t) - \frac{1}{v^{2}(\textbf{r})}\frac{\partial^{2} p(\textbf{r},t)}{\partial t^{2}} &=& f(\mathbf{r}_{s},t),  \nonumber \\
p(\textbf{r},t) = 0 \quad \rm \text{on } \partial\Omega_{D} \quad \text{and} \quad \mathcal{B}p(\textbf{r},t) &=& 0 \quad \rm \text{on } \partial\Omega_{inf},  \\
p(\textbf{r},0) = 0 \quad \text{and} \quad
\frac{\partial p(\textbf{r},0)}{\partial t} &=& 0, \quad \textbf{r} \in \Omega, \nonumber
\end{eqnarray}

\noindent where $\Omega \subset \mathbb{R}^{3}$ denotes the domain to be imaged, $\partial\Omega = \partial\Omega_{D} \cup \, \partial\Omega_{inf} \subset \mathbb{R}^{2}$ is the domain boundary and  $\partial\Omega_{D}$ and $\partial\Omega_{inf}$ are non-overlapping boundary partitions. $\partial\Omega_{D}$ is the portion of the boundary where Dirichlet boundary conditions are applied, representing, for instance, the free-surface. The operator $\mathcal{B}$ represents the non-reflecting boundary condition \cite{cerjan1985nonreflecting} applied on $\partial\Omega_{inf}$. The pressure (the forward-propagated source wavefield) $p(\textbf{r},t)$ is defined  at the position $\textbf{r} = (r_{x}, r_{y}, r_{z}) \in \Omega$ and time $t \in [0,T]$. Moreover, $v(\mathbf{r})$ is the compressional wave velocity spatial field, and $f(\mathbf{r}_{s},t)$ is the seismic source. The vector $\mathbf{r}_{s}$ represents the seismic source position. The backward-propagated wavefield is calculated solving,

\begin{eqnarray} \label{eq.secondOrderBackward}
\nabla^{2} \Bar{p}(\textbf{r},\tau) - \frac{1}{v^{2}(\textbf{r})}\frac{\partial^{2} \Bar{p}(\textbf{r},\tau)}{\partial \tau^{2}} &=& s(\mathbf{r}_{r},\tau),  \nonumber \\
\Bar{p}(\textbf{r},\tau) = 0 \quad \rm \text{on } \partial\Omega_D \quad \text{and} \quad \mathcal{B}\Bar{p}(\textbf{r},\tau) &=& 0 \quad \rm \text{on } \partial\Omega_{inf} \\
\Bar{p}(\textbf{r},0) = p(\textbf{r},T) \quad \text{and} \quad
\frac{\partial \Bar{p}(\textbf{r},0)}{\partial \tau} &=& \frac{\partial p(\textbf{r},T)}{\partial \tau}, \quad \textbf{r} \in \Omega. \nonumber
\end{eqnarray}

\noindent which is an equation similar to \eqref{eq.secondOrder}, but with a different source $s(\mathbf{r}_{r},\tau)$, that is, the recorded signals at the receivers positioned in $\mathbf{r}_{r}$.  Besides, the evolution in Eq. \eqref{eq.secondOrderBackward} is over the reverse time  $\tau = T - t$. Thus, the backward-propagated wavefield $\Bar{p}(\textbf{r},\tau)$ is defined in $\Omega$ and $\tau \in [0,T]$. 


The IC dictates the quality and fidelity of the final RTM image. There are several possibilities, for instance, excitation ICs \cite{chang1986reverse, chang1987elastic, nguyen2013excitation}, extend ICs \cite{sava2006time, wang2009subsalt, sava2011extended}, wavefield decomposition ICs \cite{liu2011effective}, and the zero-lag cross-correlation ICs \citep{Zhouetal2018, chattopadhyay2008imaging}. We have chosen the zero-lag cross-correlation between the forward and backward propagated waves at each point in $\Omega$,

\begin{eqnarray} \label{eq.imageCondition}
I(\textbf{r}) = \int_{0}^{T} p(\textbf{r},t) \, \Bar{p}(\textbf{r},\tau)\ dt.
\end{eqnarray}

The IC amplitudes in equation \eqref{eq.imageCondition} do not provide an explicit physical relationship with the reflection coefficients. In \citep{chattopadhyay2008imaging}, we find a detailed explanation of the relation between the imaging condition and the reflection coefficient. Nevertheless, the resulting image provides the correct amplitude contrast locations of the geological interfaces of rocks with different physical properties \cite{Zhouetal2018}. The amplitude contrast patterns are the main feature of the migrated seismic images explored in the present work.

We apply an explicit $2^{nd}$-order in time and $4^{th}$-order in space finite difference numerical scheme \cite{stability} to approximate equations \eqref{eq.secondOrder} and \eqref{eq.secondOrderBackward}, leading to the vector $\mathbf{v}$, the discrete version of the velocity field, and similarly the vectors $\mathbf{p}$, $\mathbf{\Bar{p}}$, $\mathbf{s}$, $\mathbf{f}$ at each time step.  Note that the vectors $\mathbf{p}$, $\mathbf{\Bar{p}}$, and $\mathbf{v}$ have the same dimension, that is $N = N_{x}*N_{y}$ (or $N = N_{x}*N_{y}*N_{z}$ in 3D), where $N_{x}$, $N_{y}$ (and $N_{z}$) are the number of grid points in each Cartesian direction. Each discrete seismogram is a vector of size $N_{rec} * (N_{t}+1)$, where $N_{rec}$ is the number of receivers, and $N_{t} = T / \Delta t$, with $\Delta t$ is the time step. The seismic source $\mathbf{f}$ has dimension $N_{t}$. RTM is not only computationally intensive but also data-intensive due to the high dimensional inputs, the amount of data to manage, for instance, storing and retrieving  $\mathbf{p}$, and the computational costs associated with imposing the stability and dispersion conditions in the discrete two-way wave equation \citep{Zhouetal2018}. The dispersion relation takes into account the number of grid points per wavelength and the medium properties, which in our isotropic acoustic case is the P-wave velocity. Thus, complexity in estimating the migrated image increases with high heterogeneous media and seismic source cutoff frequency. Least-squares RTM (LSRTM) extends the basic method to an iterative method that minimizes a data misfit term \citep{schuster2017, zand2020}. However, in the present work, we restrict ourselves to the standard RTM technique.

As we wrap RTM  into a sampling method in Algorithm \ref{algworkflow}, for taking into consideration the input uncertainties, the overall computational cost of {\bf{Stage 2}} rises proportionally to $N_{MC}$, the cardinality of the ensemble of possible velocity fields. Typically, seismic raw data recording sets are split into multiple steps ($N_{shots}$) to cope with the potentially high spatial dimensions to be covered and to enhance the signal to noise ratio (SNR) in processing stages. Each step covering fully or partially the domain corresponds to a different arrangement of sources and receivers.
It is essential to mention that RTM iterates over the number of shots producing partial migrated images characterized by equation \eqref{eq.imageCondition}. When this loop ends, a process called staking sums the partially migrated seismic images into a single one \cite{kearey2013introduction, yilmaz2001seismic}, gathering into this single image all information related to the seismogram set. Algorithm \ref{alg.rtmuq} details the generation of the RTM images ensemble, where a set of seismograms, $\{ \mathbf{s}_{1}, \cdot \cdot \cdot, \mathbf{s}_{N_{shots}}\}$, a set of velocity fields, $\{ \mathbf{v}_{1}, \cdot \cdot \cdot,\mathbf{v}_{N_{MC}}\}$, and a seismic source ($\mathbf{f}$) are given as inputs. The indexes $N_{shots}$ and $N_{MC}$ represent the number of shots and the number of samples for the MC method.
For each MC iteration, we solve the wave equation twice, one for the seismic source and other for the seismograms associated with it. The computation of the imaging condition uses both solutions (source wavefield, and receiver wavefield), retrieving from persistent storage the source wavefield to build the migrated seismic section and stacking the partial results over time ($\mathbf{I}_{\sum t}$), and over the number of seismograms ($\mathbf{I}_{\sum shot\_id}$). At the end of Algorithm \ref{alg.rtmuq}, we have the discrete imaging condition set $\{\mathbf{I}_1, \mathbf{I}_2, \cdots, \mathbf{I}_{N_{MC}}\}$ where each $\mathbf{I}_i$ is a vector in $\mathbb{R}^N$. It is usual to filter each image to sharpen its features. Nevertheless, we do not apply any filter to the ensemble of seismic images. Summarizing, migrations of seismograms for the set of velocity fields produce the corresponding set of migrated seismic images, where each one has a direct relation with one velocity sample.

\begin{algorithm}
    \caption{Reverse Time Migration under Uncertainty.}\label{alg.rtmuq}
        \begin{algorithmic}[1]
            \Require $\{ \mathbf{v}_{1}, \cdot \cdot \cdot,\mathbf{v}_{N_{samples}}\}$, $\{ \mathbf{s}_{1}, \cdot \cdot \cdot, \mathbf{s}_{N_{shots}}\}$, and $\mathbf{f}$
            \Function{rtm\_uq}{ vectors $\{ \mathbf{v}_{1}, \cdot \cdot \cdot,\mathbf{v}_{N_{samples}}\}$, vectors $\{ \mathbf{s}_{1}, \cdot \cdot \cdot, \mathbf{s}_{N_{shots}}\}$, vector $\mathbf{f}$ }
            \For{$sample\_id = 1$ to $N_{MC}$}
                \State read $\mathbf{v}_{sample\_id}$, and $\mathbf{f}$
                \State initialize image condition $\mathbf{I}_{\sum shot\_id} = 0$
                \For{$shot\_id = 1$ to $N_{shots}$}
                    \State initialize $n_{t} = 0$
                    \State apply initial conditions for $i_{t} = 0$
                    \For{$i_{t} = 1$ to $N_{t}$}
                        \State $n_{t} = n_{t} + i_{t} * \Delta t$
                        \State solve equation \eqref{eq.secondOrder} \Comment{source wavefield}
                        \State store $\mathbf{p}_{n_{t}}$ 
                    \EndFor
                    \State initialize $n_{\tau} = 0$, and $\mathbf{I}_{\sum \tau} = 0$
                    \State apply initial conditions for $i_{\tau} = 0$
                    \For{$i_{\tau} = 1$ to $N_{t}$}
                        \State $n_{\tau} = N_{t} - (n_{\tau} + i_{\tau} * \Delta \tau)$ \Comment{reverse time}
                        \State read $\mathbf{p}_{n_{\tau}}$, and $\mathbf{s}_{shot\_id}$
                        \State solve equation \eqref{eq.secondOrderBackward} \Comment{receiver wavefield}
                        \State calculate $\mathbf{I}_{\sum n_{\tau}} = \mathbf{I}_{\sum n_{\tau}} + \mathbf{p}_{n_{\tau}} \Bar{\mathbf{p}}_{n_{\tau}}$ \Comment{imaging condition}
                    \EndFor
                    \State stack $\mathbf{I}_{\sum shot\_id} = \mathbf{I}_{\sum shot\_id} + \mathbf{I}_{\sum n_{\tau}}$ \Comment{stacking}
                \EndFor
                \State store $\mathbf{I}_{\sum shot\_id} \enskip \forall \enskip  sample\_id \enskip \in \enskip 1 \leq sample\_id \leq N_{MC}$
            \EndFor
            \EndFunction
        \end{algorithmic}
\end{algorithm}

Different strategies can be pursued in order to make feasible Algorithm \ref{algworkflow} by reducing the inherent computational costs of processing and storage. They could rely, for instance, on data compression \cite{bai2020high, kukreja2018combining, lindstrom2016reducing, witte2019compressive} or more effective stochastic sampling \cite{ballesio2019multilevel}, but here, as mentioned before, our option is for using deep learning surrogates for the RTM imaging, what we describe in the following section.

\section{Deep Learning Surrogate}

The main goal of surrogate models is to replicate the multivariate input-output mapping provided by physical models governed by PDEs to save computational costs. Performing uncertainty quantification in such conditions is often hampered whenever one faces high-dimensionality,  the so-called curse of dimensionality. As pointed out in the literature, DNNs have proved successful in such situations by exploiting low-dimensional latent spaces and sophisticated training methods \cite{Goodfellow-et-al-2016}. We aim to construct and evaluate the performance of DNNs acting as a surrogate model for the RTM imaging under uncertainty, using as baseline the encoder-decoder architecture proposed by \cite{Zabaras1} and designed for problems cast as image-to-image regressions. We briefly review the building blocks of the network in this section.

\begin{figure}
    \centering
    \includegraphics[scale=0.75]{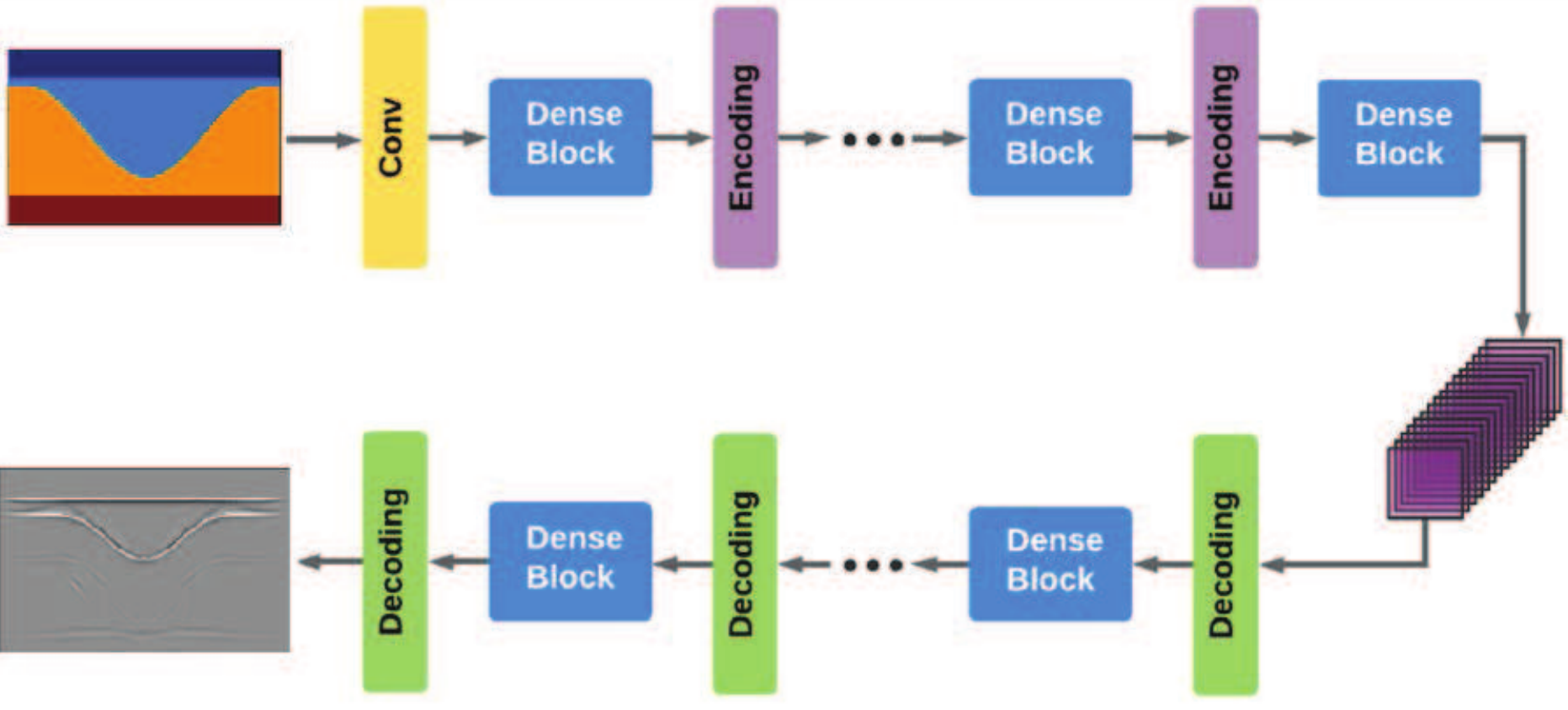}
    \caption{RTM deep convolutional encoder-decoder network architecture. Inputs: ensemble of velocity fields $\mathbf{v} \in \mathbb{R}^N$. Outputs: corresponding ensemble of image conditions $\mathbf{I} \in \mathbb{R}^N$.}
    \label{fig:encoder_decoder}
\end{figure}

Figure \ref{fig:encoder_decoder}, inspired in \cite{Zabaras1}, provides the big picture of our end-to-end solution, depicting the main components of the encoder-decoder network. In our particular application, the input for the deep learning surrogate is the ensemble of heterogeneous velocity fields, and the outputs are the corresponding imaging conditions given by Eq. \eqref{eq.imageCondition} for each sample of the velocity field ensemble. Inputs and outputs are high-dimensional spatial fields, and the surrogate modeling cast as a field-to-field regression. This approach is image-inspired. Then, it relies on connecting each pixel of the input field to an output pixel, where pixels correspond to grid points in the input computational mesh and output fields. The trained network maps the velocity $\mathbf{v} \in \mathbb{R}^N$ into the IC field $\mathbf{I} \in \mathbb{R}^N$.

The encoder-decoder architecture displayed in Fig. \ref{fig:encoder_decoder} consists of two sequential main phases. The first is the encoder, wherein dimension reduction occurs, followed by the decoder that allows expressing the network output with its original dimension. Alternating dense blocks and transition layers constitute both. Indeed, this architecture merges key characteristics of fully connected with convolutional networks, suited for the present application.  On one side, convolutional networks are quite effective in dimension reduction \cite{Goodfellow-et-al-2016} and are capable of capturing spatial correlations present in the input velocity fields. Fully connected layers enhance the information transmitted across the network, improving the overall efficiency reflected in reasonable sizes of the needed training data set \cite{DenseNet}.

Dense-blocks connect all layers directly to each other,  helping the training process with the improvement of information flow and gradients across the network \citep{DenseNet}.  Inputs of the $l$-th layer are the concatenated outputs from the previous layers, that is, $ \textbf{z}^l=H^l([\textbf{z}^0, \textbf{z}^1,\dots,\textbf{z}^{l-1}])$,
with $\textbf{z}^l$ the output of $l$-th layer, and $[\textbf{z}^0, \textbf{z}^1,\dots,\textbf{z}^{l-1}]$ refers to their concatenation, and $[0,\dots,l-1]$. $H^l$ is a non-linear transformation. Here, $H^l$ results from applying three consecutive operations, batch normalization \cite{Batch} followed by a ReLU \cite{glorot-relu} and, convolution. The dense-block has two design parameters, the number of layers $L$, and the growth rate, $K$, the number of output features of every single layer. The transition layers here,  in the encoder (decoder), are convolutional (deconvolutional) and, therefore, handle the dimension inputs or outputs of dense blocks. As shown in Figure \ref{fig:encoder_decoder}, during the encoder phase, the high dimensional velocity fields are immersed in an alternating series of layers of dense blocks and encoders. The last layer of the encoder phase produces low-dimensional feature maps that characterize the high dimensional field, as shown in the purple maps. Such maps are immersed in the decoder phase, which is composed of an alternating series of layers of dense-blocks and decoders, returning ICs to its (high) dimension.


The surrogate  $\mathbf{g}$ is expressed formally in a compact notation as,

\begin{equation} 
    \mathbf{\hat{y}} = \mathbf{g}(\mathbf{x}; \mathbf{w}),
\end{equation}

\noindent where $\mathbf{\hat{y}}$ is the surrogate prediction (imaging condition, $\mathbf{I}$) for an input $\mathbf{x}$ (that is, a velocity field $\mathbf{v}$), and vector $\mathbf{w}$ contains the parameters of the neural network. Training  the neural network means learning   parameters $\mathbf{w}$ using data from the set $\mathcal{D}=\{(\mathbf{x}_i,\mathbf{y}_i)\}, i=1 \cdots N_{train}$ obtained from simulations with the RTM algorithm, where $N_{train}$ is the number of samples in the training set. The stochastic gradient descent algorithm computes the unknown network elements for a given loss function \cite{Chollet}. We consider for training our surrogate, the following $\mathcal{L}_2$ regularized mean squared error (MSE) function,

\begin{equation}
    L_{MSE} = \frac{1}{N_{train}}\sum_{i=1}^{N_{train}}\left\lVert\mathbf{\hat y}_i  - \mathbf{y}_i\right\rVert_{2}^{2} + \alpha \ \Omega(\mathbf{w})
    \label{eq:l_MSE}
\end{equation}

\noindent where $\mathbf{\hat y}_i=\mathbf{g}_i (\mathbf{x}_i,\mathbf{w})$. Here the penalty function is given by $\Omega(\mathbf{w})=\frac{1}{2}\mathbf{w}^T\mathbf{w}$ for the $\mathcal{L}_2$ regularization. Moreover, the root mean squared error ($RMSE$) is used for monitoring the convergence of the training error. The $RMSE$ is given by, 

\begin{equation}
    RMSE = \sqrt{\frac{1}{N_{train}}\sum_{i=1}^{N_{train}}\left\lVert\mathbf{y}_i-\hat{\mathbf{y}}_i\right\rVert_{2}^{2}}.
\end{equation}

\section{Numerical experiments}

Here, we present two examples to demonstrate the ability of the encoder-decoder surrogate in replacing the original two-way wave equation RTM algorithm efficiently. In these numerical experiments, we mimic {\bf{Stage 1}} outputs of Algorithm \ref{algworkflow}. That is, we need to generate an ensemble of velocities. Hence, we assign to the different geological layers random velocity magnitudes for producing synthetic data to train the neural network and perform the uncertainty analysis.  The first example deals with a medium containing two flat geological layers of constant velocity. We increase the difficulty for the surrogate in the second example, by employing a more complex medium, in which the five geological layers are no longer flat, implying in horizontal heterogeneity.    The seismic source term considered in the present work is a Ricker-type wavelet \cite{Ricker}.

The encoder-decoder networks are constructed using the open platform Tensorflow \cite{Tensorflow}. The Adam optimizer algorithm \cite{kingma2014adam} is employed for parameter learning,  considering a weight decay of $1\times10^{-5}$, and an initial learning rate of $1 \times 10^{-3}$, with a learning rate scheduler, that is used dropping two times on plateau of the rooted mean squared error. We compute a total of 1300 samples by considering the velocity magnitude constant in the interior of each geological layer. Therefore, each velocity field in the ensemble has the form,

\begin{equation} \label{eq.velocitySamples}
    \mathbf{v}= \sum_{l=1}^{n_l}\overline{v}_l\ (1 + \sigma_l \xi_l) \ \mathbf{P}_l   
\end{equation}

\noindent where $n_l$ is the number of geological layers, $\overline{v}_l$ is the mean velocity within each geological layer, $\sigma_l$ is the standard deviation, here assumed as $5\% $, $\xi_l \sim \mathbb{U}[-1,1]$ is a random variable following a uniform distribution, and $\mathbf{P}_l$ is a $N$-dimensional vector containing $1$ in the components corresponding to the $l$-th geological layer grid points and $0$ otherwise. After that, we apply a moving harmonic average to $\mathbf{v}$ with a window length of 20 grid points to mimic the velocity fields computed by parameter model building techniques like tomography or full-waveform inversion. To analyze the surrogate training convergence, out of  1300 samples computed with the original RTM model, we create four training datasets with $200, 400, 600$, and $800$ samples each. We used the remaining $N_{test}= 500$ samples to test the trained network.

Accuracy is measured using the distance between predictions with the surrogate model and those computed with the RTM original model. To evaluate the surrogate model quality, we consider the coefficient of determination ($R^2$-score) metric  \cite{Sanfordbook}. We define the coefficient of determination as,

\begin{equation}
    R^2= 1 - \frac{\sum_{i=1}^{N_{test}}\left\lVert\mathbf{y}_i-\hat{\mathbf{y}}_i\right\rVert_{2}^{2}}{\sum_{i=1}^{N_{test}}\left\lVert\mathbf{y}_i-\overline{\mathbf{y}}\right\rVert_{2}^{2}}
\end{equation}

\noindent where $\overline{\mathbf{y}}=\frac{1}{N_{test}}\sum_{i=1}^{N_{test}}\mathbf{y}_i$ is the mean of test samples. The $R^2$-score metric represents the normalized error, allowing the comparison between surrogate models trained by different datasets, with values close to 1 corresponding to the surrogate models best accuracy. Here, we consider  0.95  a good target. Also, we intend that the surrogate model returns not only good predictions of seismic images but also accurate estimations of quantities that characterize the uncertainties in such images. To measure the degree of uncertainty in the seismic images, we follow the approach proposed in \citep{LiSun2016}. In their approach, the degree of uncertainty is expressed by a confidence index that represents the pointwise normalized standard deviation, where low values represent high variabilities and high values the opposite. The confidence index is,

\begin{equation} 
    \label{eq.confidence}
    c(\textbf{r}) = \frac{\sigma_{max} - \sigma(\textbf{r})}{\sigma_{max} - \sigma_{min}},
\end{equation}

\noindent where $c(\textbf{r})$ is the confidence index at position $\textbf{r}$, and $\sigma_{min}$ and $\sigma_{max}$ are the minimum and maximum field standard deviations, respectively. Another  form of measuring the degree of uncertainty is the coefficient of variation, defined as the pointwise ratio between the standard deviation and the mean,

\begin{equation}
    \text{c}_v(\textbf{r}) = \frac{\sigma(\textbf{r})}{\mu(\textbf{r})}
\end{equation}

\noindent where $\mu(\textbf{r})$ is the expected value at position $\textbf{r}$.

\subsection{A simple geologic scenario:  efficiency and convergence analysis}
\label{sec:efficiency_analysis}

\begin{figure}
    \centering
    \includegraphics[scale=.75]{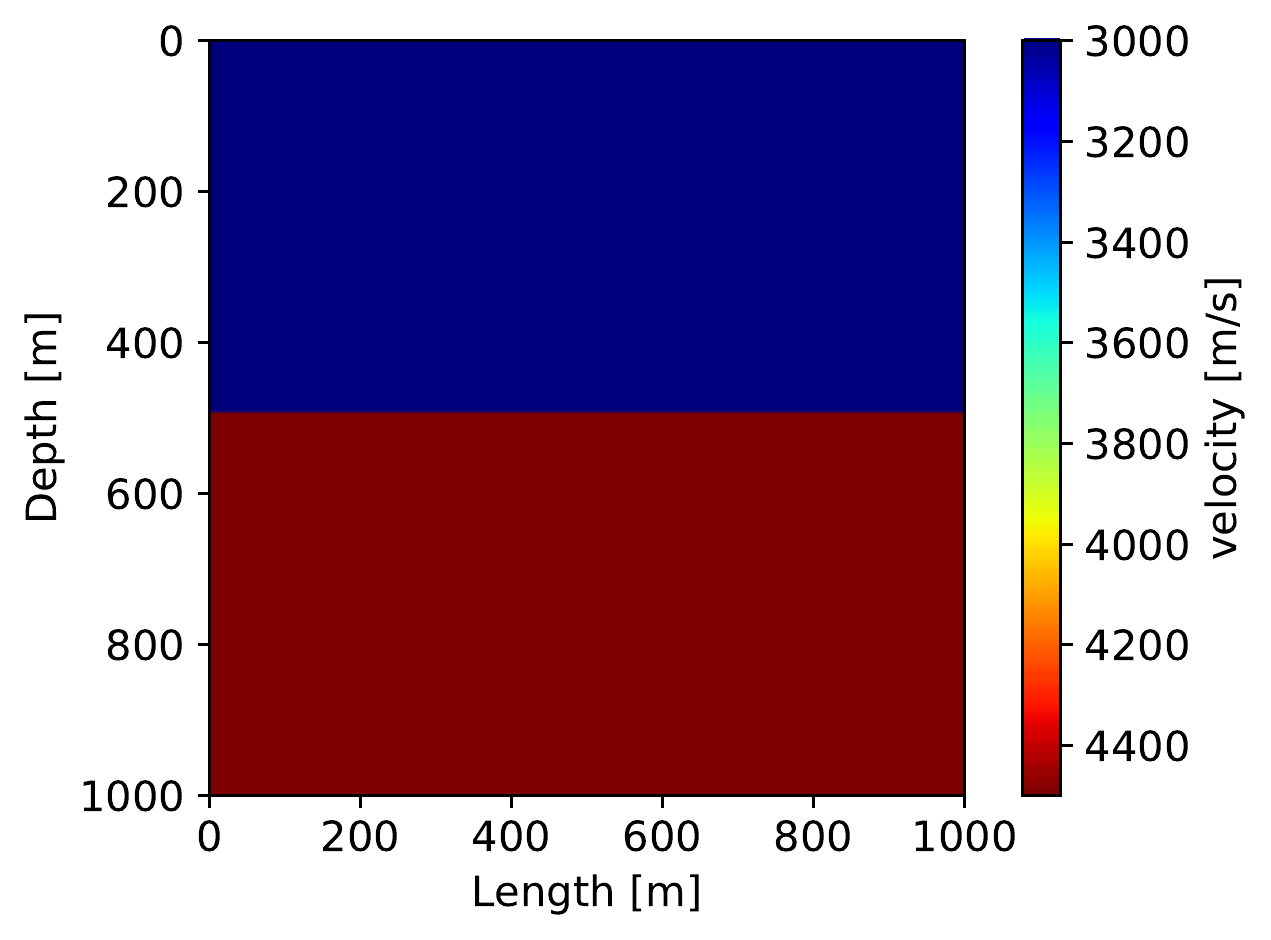}
    \caption{Simple geologic setup: two horizontal geological layers.}
    \label{fig:velocity_field_2layers}
\end{figure}

In this first example, designed to evaluate the efficacy and efficiency of the proposed surrogate, we assume a simple geologic scenario in which two horizontal homogeneous geological layers separated by a flat horizon parallel to the surface composes the subsurface, as shown in Fig \ref{fig:velocity_field_2layers}. This domain has 1000 m of depth and 1000 m of lateral extension, where the velocity in the first geological layer is 3000 m/s, and the velocity in the deeper geological layer is 4500 m/s.

We produce synthetic data using the wave propagation forward solver with the reference velocity field of Fig \ref{fig:velocity_field_2layers}, with a  seismic source with  cutoff frequency of 30 Hz.In such modeling, we simulate a fixed split-spread acquisition \cite{kearey2013introduction} comprising nine shots, where receivers are positioned near the surface for each shot and equally spaced of 20 meters. The seismic source is also placed near the surface and moved 100 meters for each shot, covering the entire domain with nine shots. Table \ref{tab:par_run1} shows the parameters used in the numerical modeling of the wave acoustic equations and the positioning of the seismic source and receivers given by the index ranges [$i_{srcx},i_{srcy}$] for the sources, and [$i_{sisx},i_{sisy}$] for the receivers. The grid size and time step respect the numerical dispersion and stability criteria \cite{stability}.

\begin{table}
\caption{RTM numerical parameters.}
\centering
\begin{tabular}{cccc}
\toprule
Parameter & Value & Description (Unit) \\ 
\midrule
$h$	& $20$ & Spatial discretization (m) \\
$\Delta t$ & $2.22\times10^{-3}$  & Temporal discretization (s) \\
$t_a$ & 0.5 & Total acquisition time (s) \\
$N_x \times N_y$ & $50\times50$ & Number of grid points \\
$i_{srcx},i_{srcy}$ & ([5:5:45], 5) & Source positions \\
$i_{sisx},i_{sisy}$ & ( [1:1:50], 5) & Receiver positions \\
\bottomrule
\end{tabular}
\label{tab:par_run1}
\end{table}

Table \ref{tab:nn_architecture1}  details the neural network architecture. The first layer is convolutional, with kernel size equals to 4 and stride 2. This first layer captures spatial relations from the velocity input. The number of features maps after the first convolutional layer is 48. The neural network has 3 dense-blocks with $L$ = 4 and $K$ = 16. Dense-blocks have a kernel size equals 3, and a stride of 1. Encoder-decoder layers have a kernel size of 3 and a stride of 2. In the decoding layer, we introduce a transposed convolution, allowing the expression of the output with its original dimensionality, equal to the computational grid. A final ReLU layer \cite{glorot-relu} imposes that the outputs are positive numbers, naturally constraining the network to output $IC > 0$ at each grid point. Thus, the neural network has $218,425$  parameters to be estimated.

\begin{table}
\caption{Neural Network Architecture. "Outputs" represents the number of features maps and "Dimension" is the dimension of the features maps.}
\centering
\begin{tabular}{ccc}
\toprule
Layers & Output & Dimension \\
\midrule
Input & 1 &  $50 \times 50$ \\
Convolution & 48 &  $24 \times 24$  \\
Dense-block 1 & 112 &  $24 \times 24$ \\
Encoding & 56 &  $12 \times 12$  \\
Dense-block 2 & 120 &  $12 \times 12$  \\
Decoding 1& 60 &  $24 \times 24$ \\
Dense-block 3 & 124 &  $24 \times 24$ \\
Decoding 2 & 1 &  $50 \times 50$  \\
ReLU & 1 &  $50 \times 50$  \\
\bottomrule
\end{tabular}
\label{tab:nn_architecture1}
\end{table}

\begin{figure}
    \centering
    \includegraphics[scale=.5]{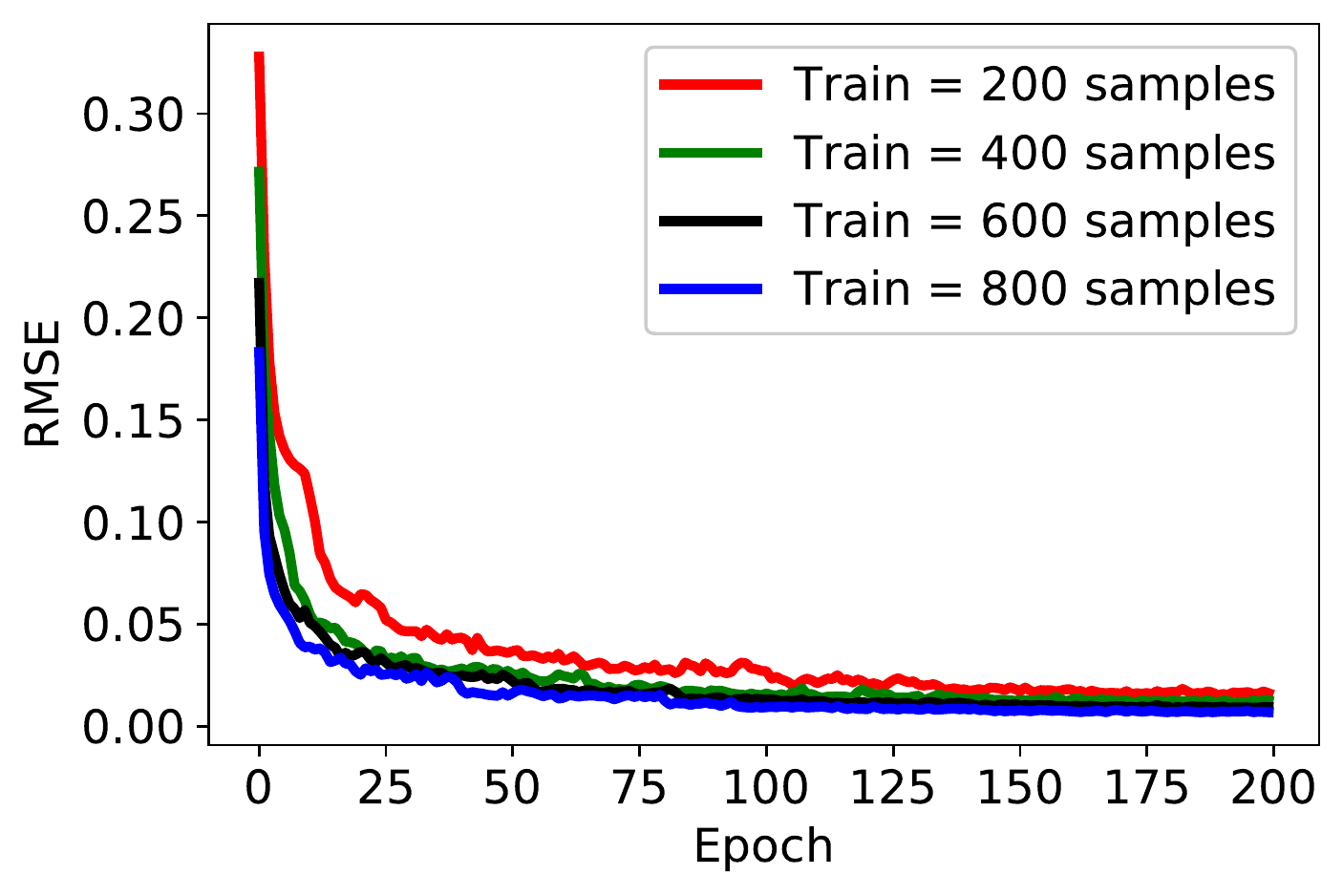}
    \caption{RMSE decay with number of epochs.}
    \label{fig:rmse_f30_2layers}
\end{figure}

Figure \ref{fig:rmse_f30_2layers} shows the decay of the $RMSE$ as a function of the number of epochs during the training process, for training data sets ranging from 200 to 800 samples. Note that the $RMSE$ stabilizes after 150 epochs for all cases and that for smaller data sets, we see higher error values. Key characteristics one is seeking when replacing the original expensive computational model by a  surrogate are efficiency and accuracy. We estimate efficiency as the ratio between $N_S$, the number of samples in the training set,  and  $N_{MC}$, the number of samples of the MC method to achieve a prescribed level of accuracy ($R^2 \geq 0.95$).  Thus, we introduce the following index to evaluate efficiency,

\begin{equation}
    \text{Efficiency}=\left(1 - \eta \frac{N_S}{N_{MC}}\right)\times 100
    \label{eq:eff}
\end{equation}

\begin{figure}
    \centering
    \includegraphics[scale=.5]{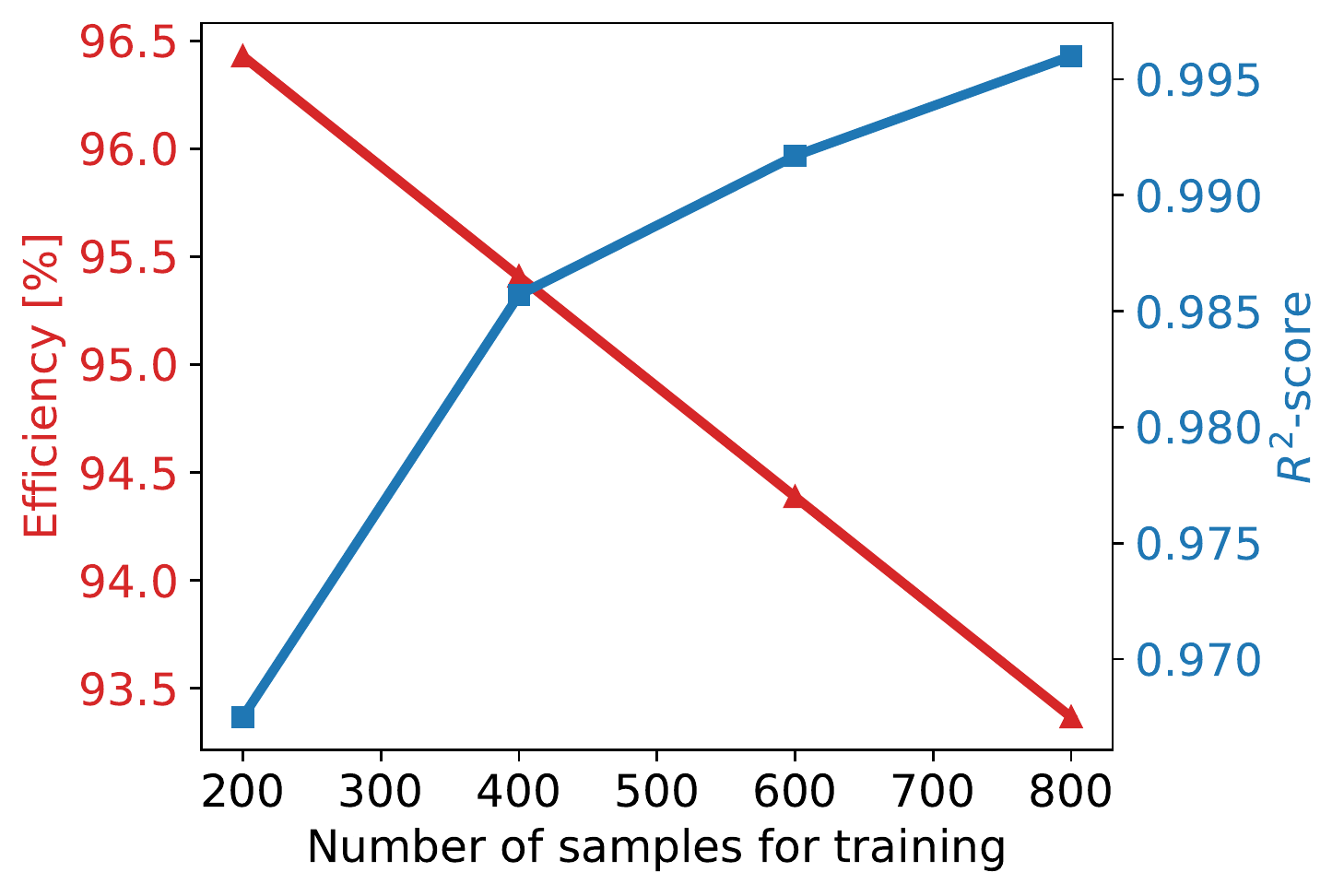}
    \caption{Test $R^2$-score and efficiency in function of the number of training samples.}
    \label{fig:efficiency}
\end{figure}

Then, the index in equation \eqref{eq:eff} represents the percentage of the saved computational costs, and $\eta$ is an adjustment factor accounting for the time spent in the construction, training, and making predictions with the surrogate model. Without loss of generality, we assume for now $\eta=1.0$. For less optimistic conditions, we recognize that $\eta>1.0$. We calculate the coefficient of determination $R^2$ to assess the accuracy of the neural network with the remaining 500 samples. We observe that the surrogate model accuracy is good even in small training data scenarios, reaching $R^2 \geq 0.95$, as shown in Fig. \ref{fig:efficiency}. Furthermore, Fig. \ref{fig:efficiency} also depicts the surrogate efficiency. Here it is worth highlighting that 600 RTM runs are necessary to compute the variance with a relative error of $1\times 10^{-3}$. Thus, we see that the efficiency is higher than 90\%, even for the larger data set with 800 samples.

\begin{figure}
    \centering
    \begin{subfigure}[b]{0.45\textwidth}
        \includegraphics[width=\textwidth]{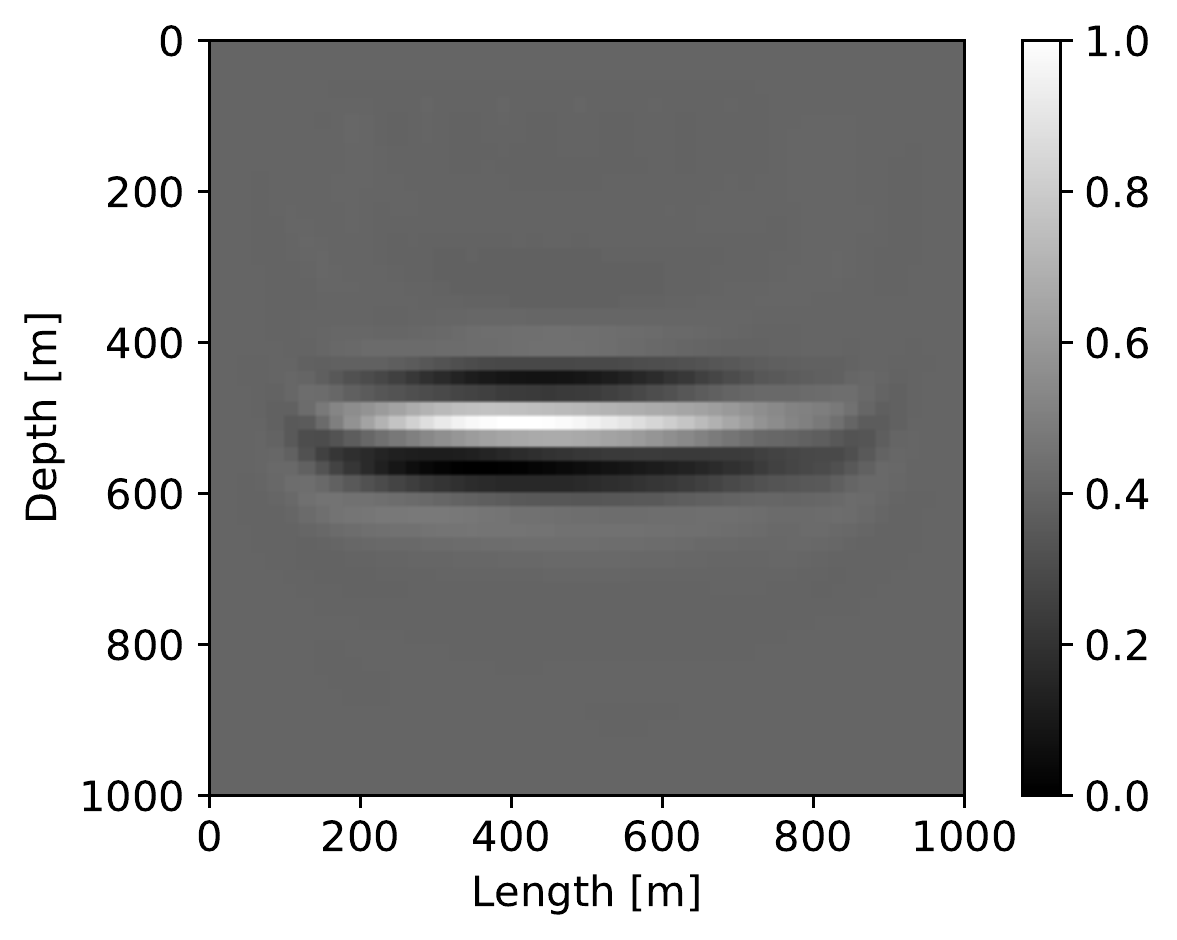}
    \end{subfigure}
    \begin{subfigure}[b]{0.45\textwidth}
        \includegraphics[width=\textwidth]{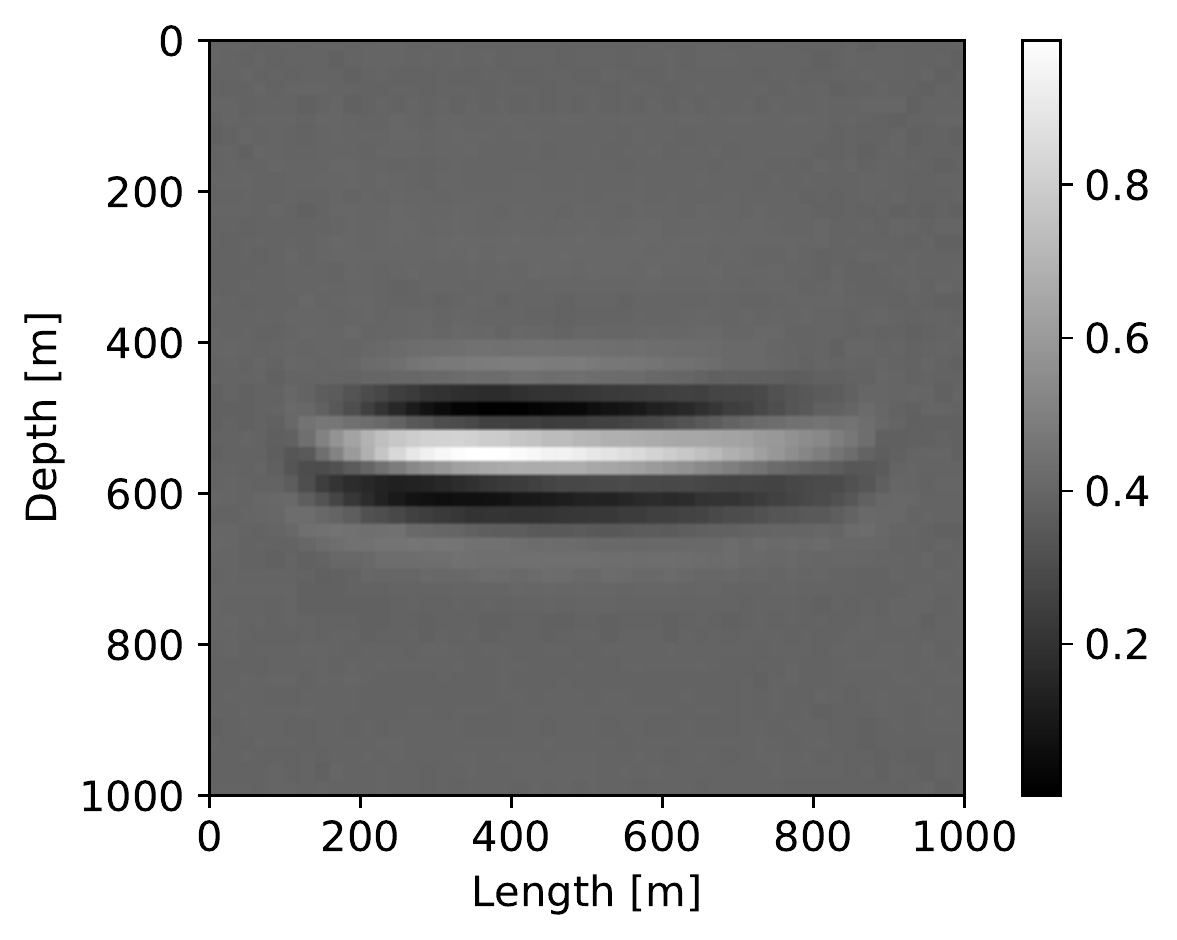}
    \end{subfigure}
    \begin{subfigure}[b]{0.45\textwidth}
        \includegraphics[width=\textwidth]{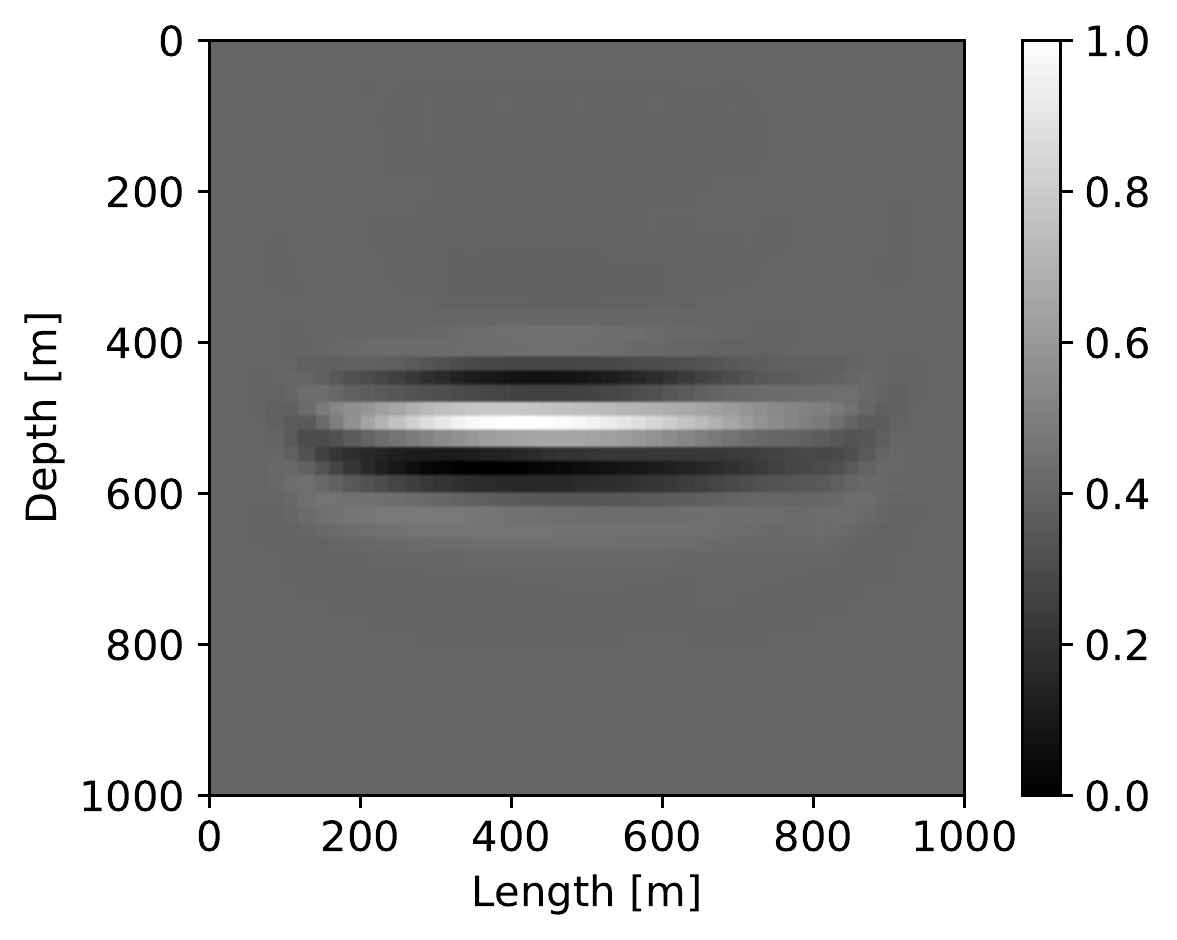}
        \caption{RTM model}
    \end{subfigure}
    \begin{subfigure}[b]{0.45\textwidth}
        \includegraphics[width=\textwidth]{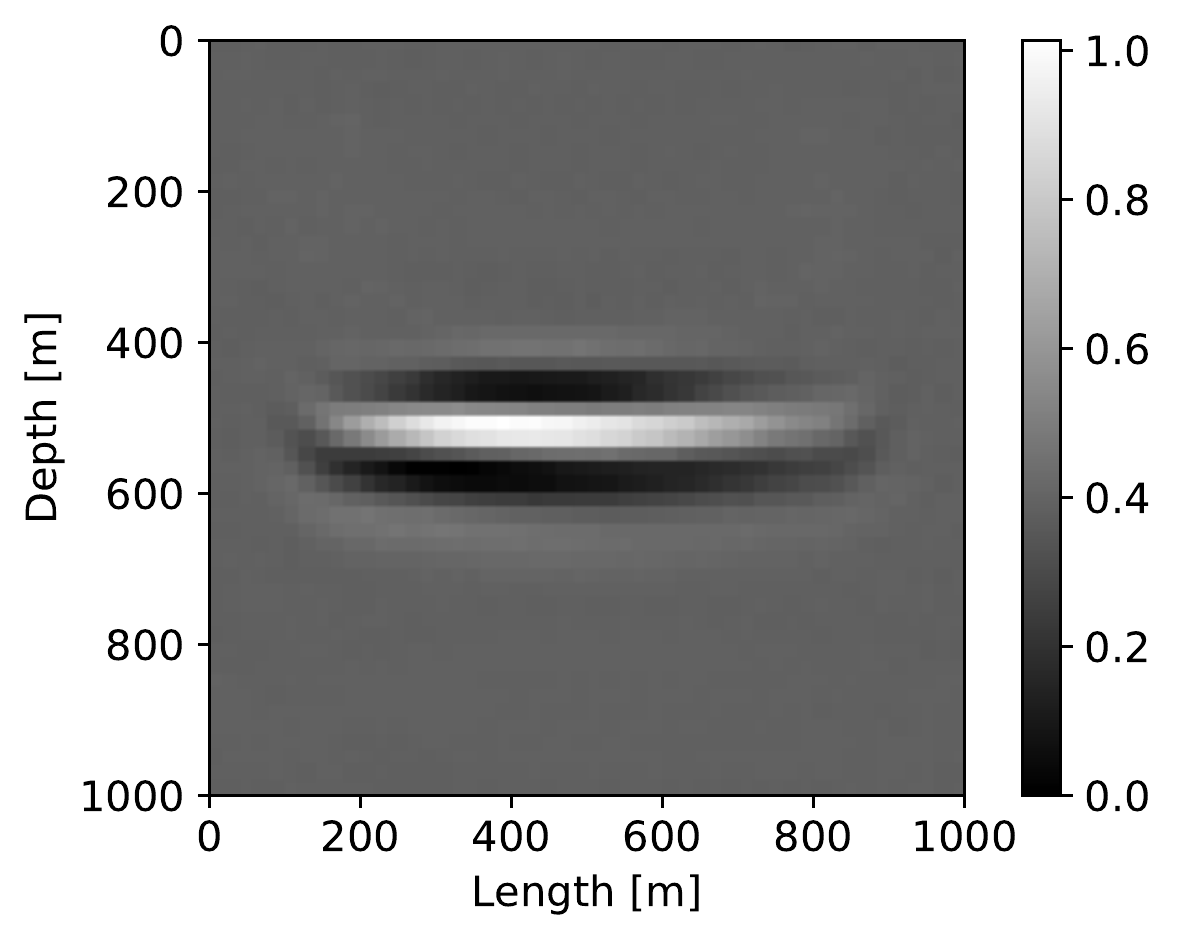}
        \caption{Surrogate model}
    \end{subfigure}
    \caption{Randomly selected  samples of seismic images in the test data set computed by the RTM (a) and the surrogate model (b). The relative  errors in the image condition, $e_I$, are lower than 2\%.}
    \label{fig:prediction_f30_2layers}
\end{figure}

\begin{figure}
    \centering
    \begin{subfigure}[b]{0.40\textwidth}
        \includegraphics[width=\textwidth]{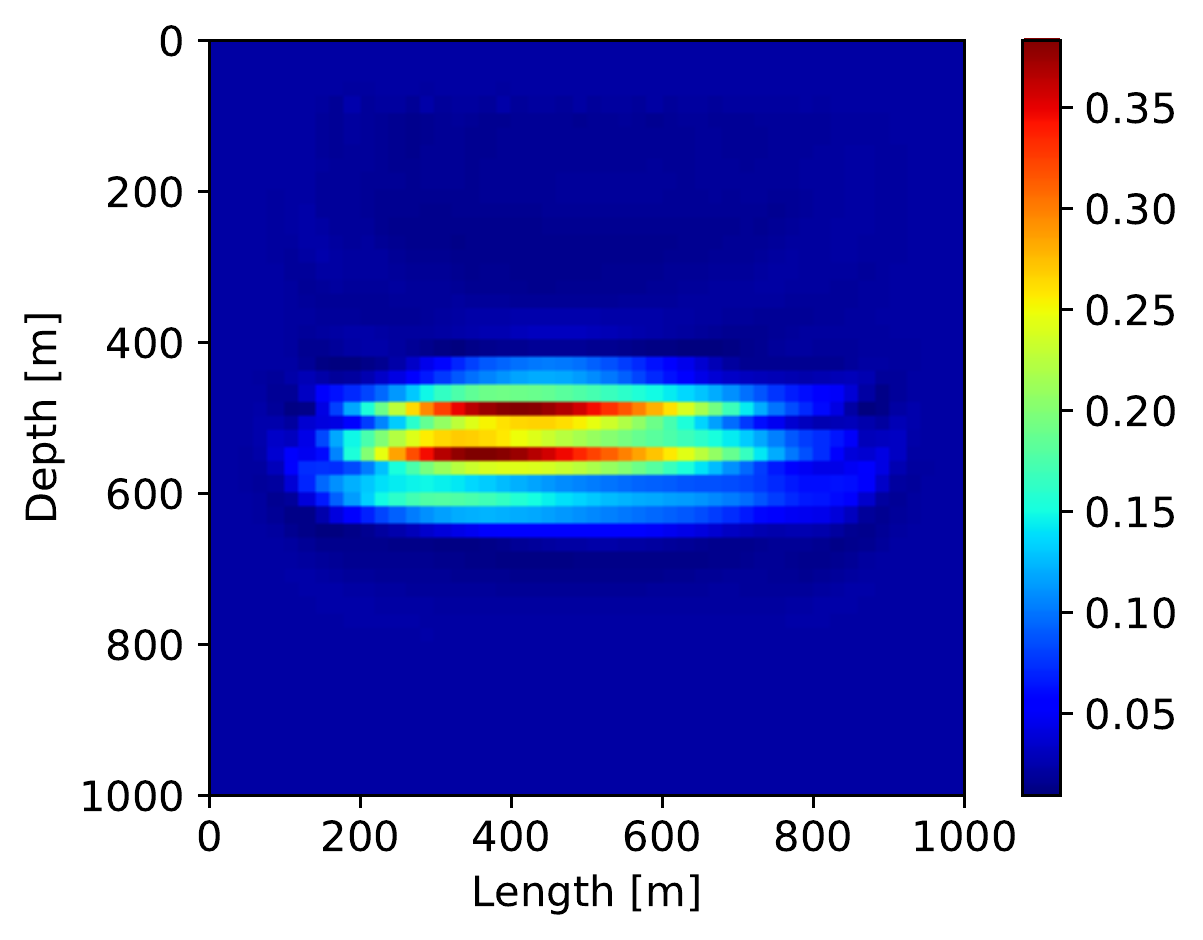}
        \caption*{Standard deviation - RTM model}
    \end{subfigure}
    \begin{subfigure}[b]{0.40\textwidth}
        \includegraphics[width=\textwidth]{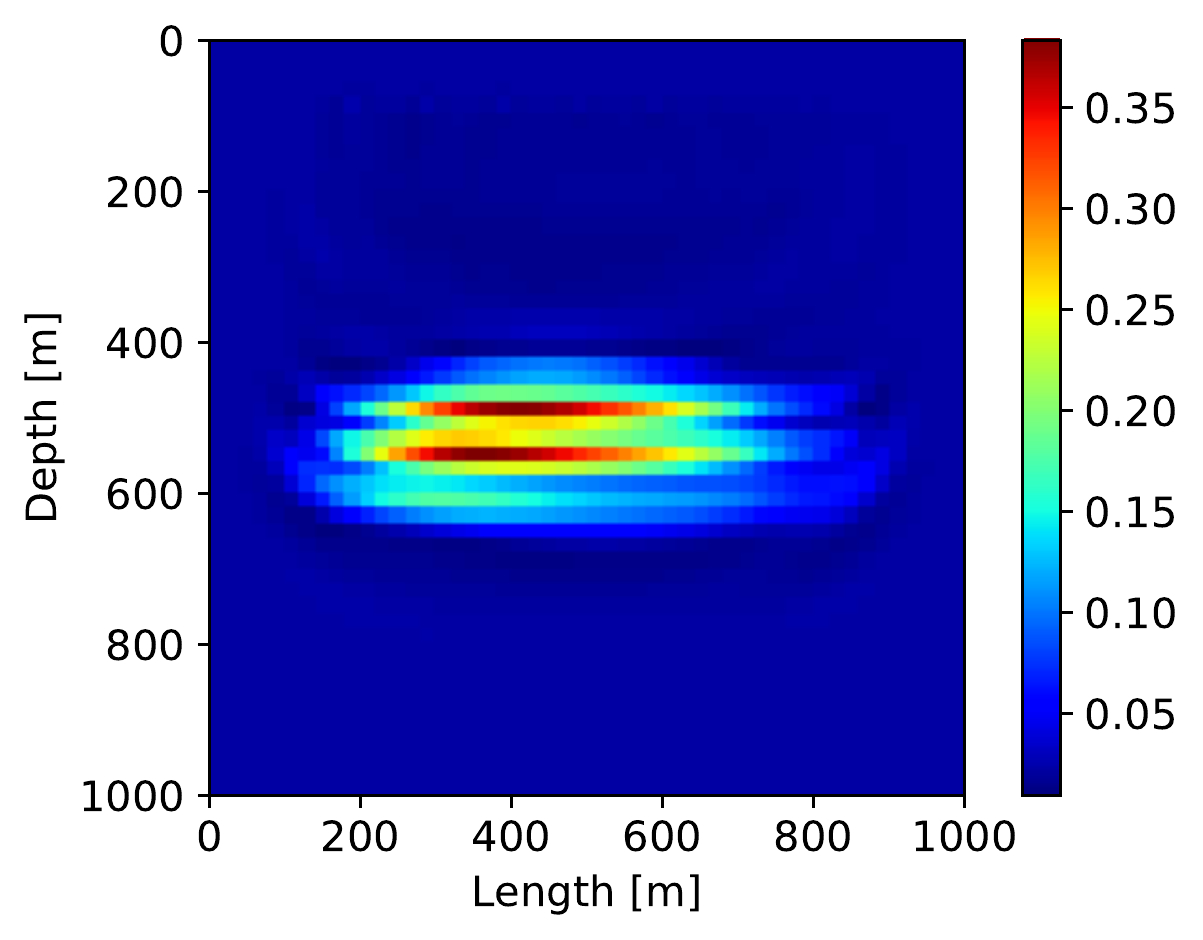}
        \caption*{Standard deviation - Surrogate model}
    \end{subfigure}
    \begin{subfigure}[b]{0.40\textwidth}
        \includegraphics[width=\textwidth]{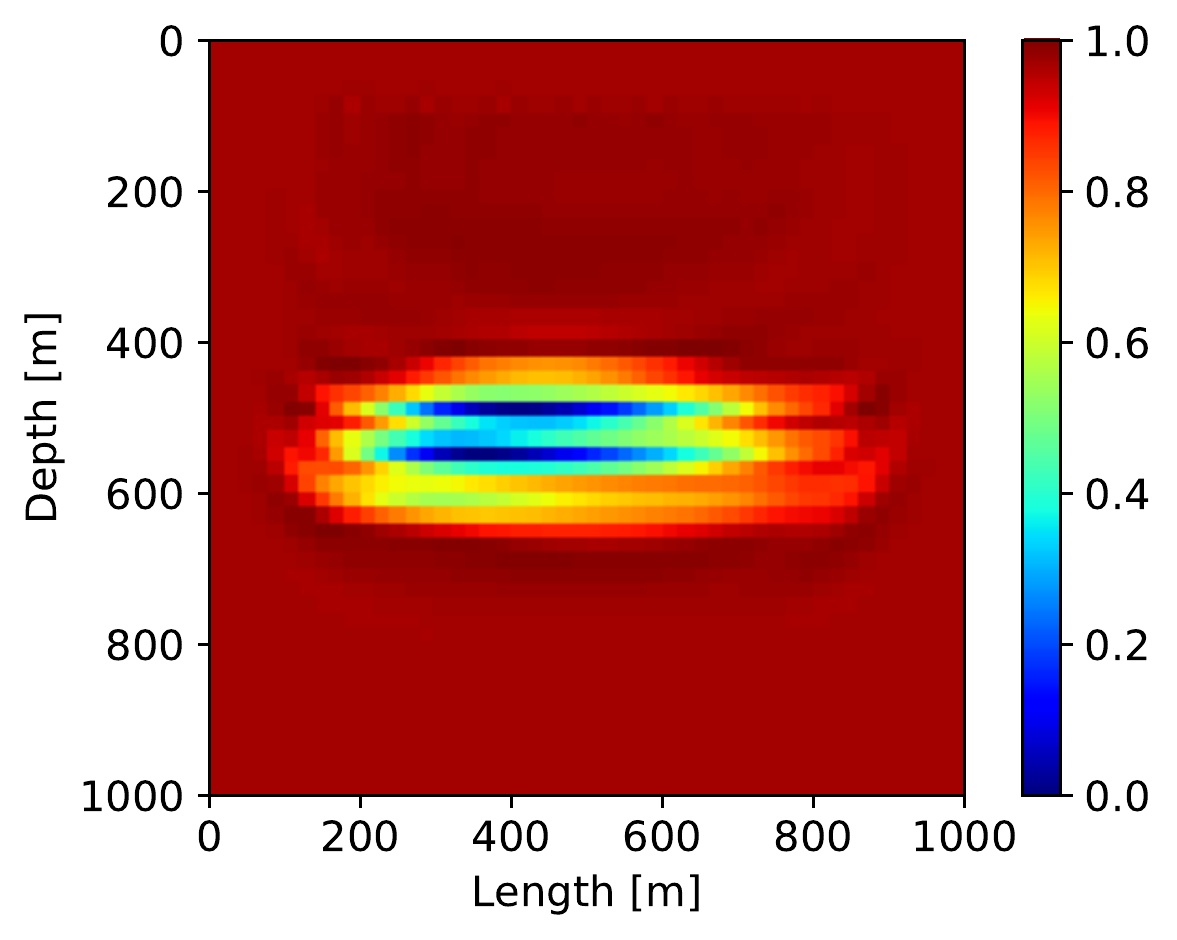}
        \caption*{Confidence index - RTM model}
    \end{subfigure}
    \begin{subfigure}[b]{0.40\textwidth}
        \includegraphics[width=\textwidth]{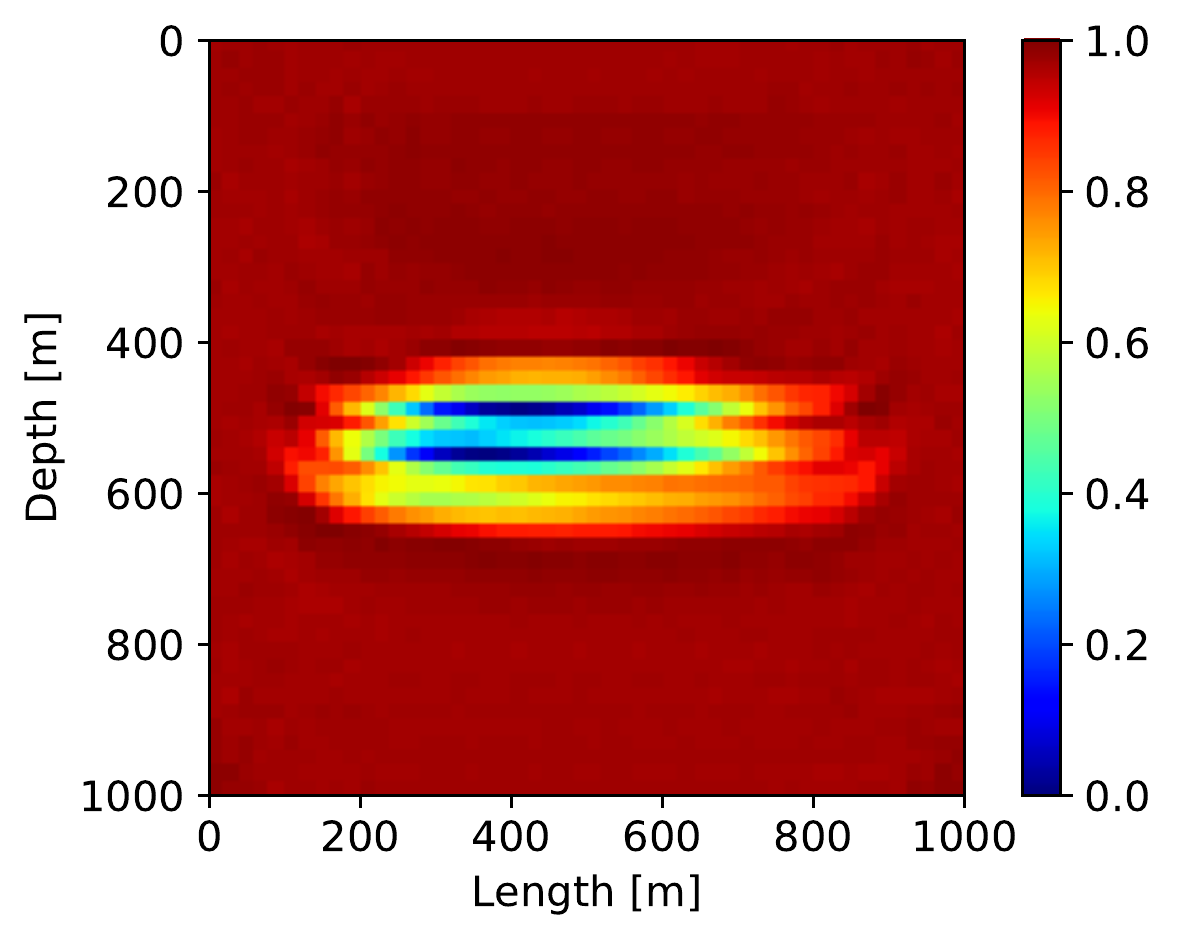}
        \caption*{Confidence index - Surrogate model}
    \end{subfigure}
    \begin{subfigure}[b]{0.40\textwidth}
        \includegraphics[width=\textwidth]{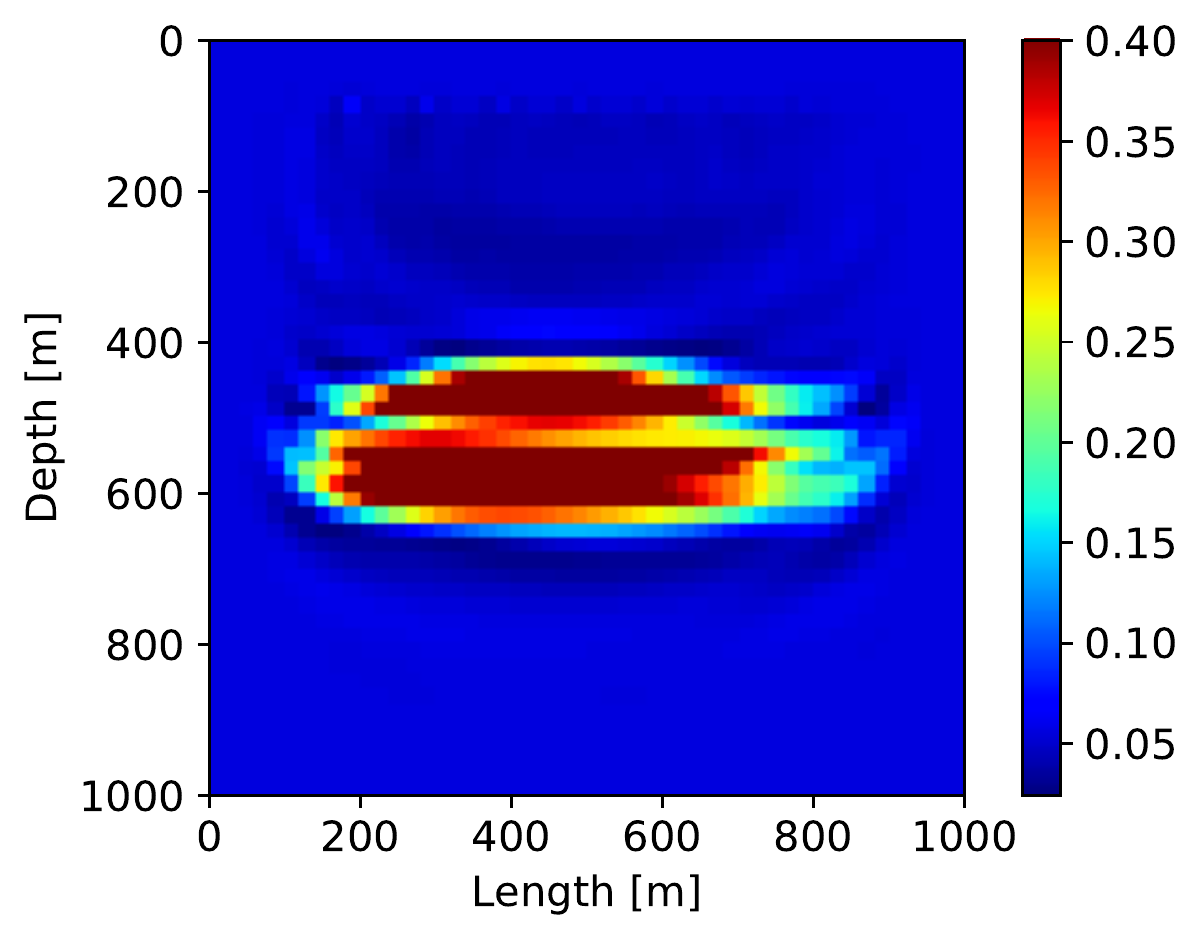}
        \caption*{Coefficient of variation - RTM model}
    \end{subfigure}
    \begin{subfigure}[b]{0.40\textwidth}
        \includegraphics[width=\textwidth]{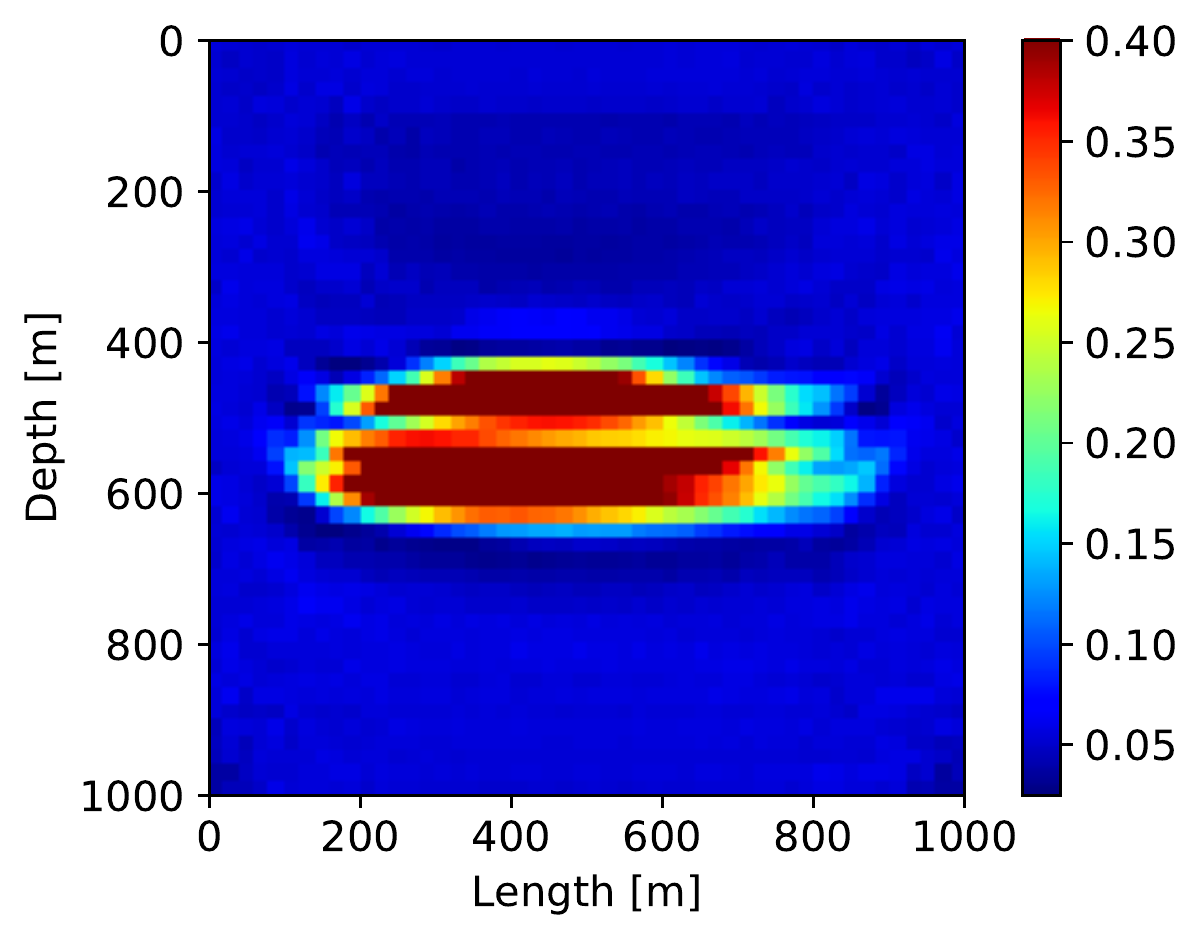}
        \caption*{Coefficient of variation - Surrogate model}
    \end{subfigure}
    \caption{UQ indexes - standard deviation, $\sigma(\mathbf{r})$, confidence index, $c(\mathbf{r})$, and coefficient of variation, $c_v(\mathbf{r})$ - predicted by the RTM (left) and surrogate models (right). The relative errors to the RTM model are lower than 1\%.}
    \label{fig:uq_2layers}
\end{figure}

To further illustrate how the surrogate model approximates the predictions of the original model with good accuracy, Fig. \ref{fig:prediction_f30_2layers} shows comparisons for two realizations chosen randomly from the test set. We observe that the surrogate presents good results, returning good predictions of the image condition. We also depict a comparison between the standard deviation, $\sigma(\mathbf{r})$, the confidence index, $c(\mathbf{r})$, and the coefficient of variation, $c_v(\mathbf{r})$, predicted by the original and surrogate models with $N_{testing} =500$ testing samples, see Fig. \ref{fig:uq_2layers}. Besides, we introduce a discrete version of a $L2$  relative error between two spatial fields $f$, one computed with the RTM and the other by the surrogate as,

 \begin{equation}
    {e_g}^2= \frac{1}{N}\sum_{i=1}^{N} \left( \frac{g^i_{RTM}-g^i_s }{g^i_{RTM}}\right)^2
    \label{relativeerror}
\end{equation} \label{eq:relativeerror}

\noindent where the subscripts refer to how we compute the field $g$. This index is an average of the pointwise relative error for all $N$ grid points. Figure \ref{fig:prediction_f30_2layers} compares randomly selected seismic images produced by RTM  with the corresponding ones obtained with the deep learning surrogate. The visual resemblance is quantified using equation \eqref{relativeerror}, leading to relative errors that stay below  2\%. We extend further our assessment of the surrogate effectiveness by comparing the uncertainty indexes computed with the two techniques displayed in Fig. (\ref{fig:uq_2layers}). For all indexes, the relative errors computed with equation \eqref{relativeerror} are less than 1\%.

\subsection{A non-flat medium with five geological layers}

\begin{figure}
    \centering
    \includegraphics[scale=.75]{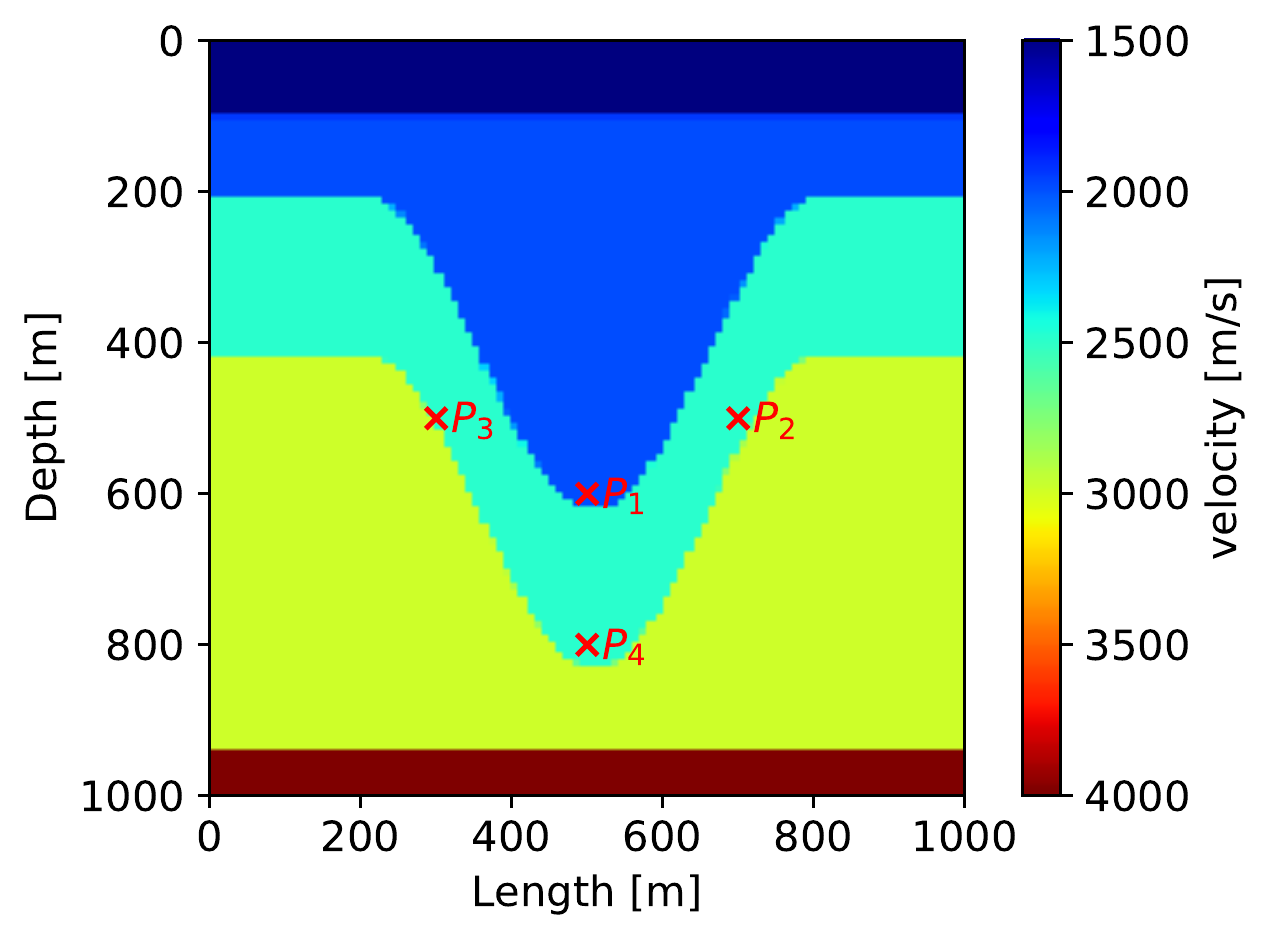}
    \caption{Geologic model with 5 layers adapted from \cite{huang_velocityfield}. $P_1$, $P_2$, $P_3$, and $P_4$ are control points in the geological model to compute statistics.}
    \label{fig:velocity_field}
\end{figure}

To challenge the encoder-decoder surrogate, we use a synthetic geologic model with five homogeneous layers similar to the one proposed in \citep*{huang_velocityfield}. The 2D velocity model consists of a water layer with velocity 1500 m/s, and four mini sedimentary basins with velocities of 2000 m/s, 2500 m/s, 3000 m/s, and 4000 m/s, respectively. Figure \ref{fig:velocity_field} display a schematic view of the velocity field with 1000 m of depth and 1000 m of lateral extension. 

Two synthetic seismograms are generated for the velocity fields shown in Fig \ref{fig:velocity_field} considering now the seismic sources with cutoff frequencies of 30 and 45Hz.Table \ref{tab:par_run} shows the RTM parameters and the positioning of sources and receivers. For the cutoff frequency of 45Hz, due to the imposition of the stability and dispersion conditions in the discrete two-way wave equation, there is a significant increase in the input dimensionality, that bears the potential to hamper the neural network training. The next sub-sections present results for the scenarios involving the two frequencies of excitation.

\begin{table}
\caption{RTM numerical parameters.}
\centering
\begin{tabular}{cccc}
\toprule
Parameter & $f_{cutoff}=30Hz$ & $f_{cutoff}=45Hz$ & Description (Unit) \\ 
\midrule
$h$	& $10$ & $6.666$ & Spatial discretization (m) \\
$\Delta t$ & $1.25\times10^{-3}$ & $8.333\times10^{-4}$ & Temporal discretization (s) \\
$t_a$ & 2.0 & 2.0 & Total acquisition time (s) \\
$N_x \times N_y$ & $100\times100$ & $150\times150$ & Number of grid points \\
$i_{srcx},i_{srcy}$ & ([5:10:95], 5) & ([7:15:142], 7) & Source positions \\
$i_{sisx},i_{sisy}$ & ( [1:1:100], 5) & ([1:1:150], 7) & Receiver positions\\
\bottomrule
\end{tabular}
\label{tab:par_run}
\end{table}

\subsubsection{Cutoff frequency - 30Hz}
\label{subsec:5_layers_30hz}

\begin{figure}
    \centering
    \includegraphics[scale=.5]{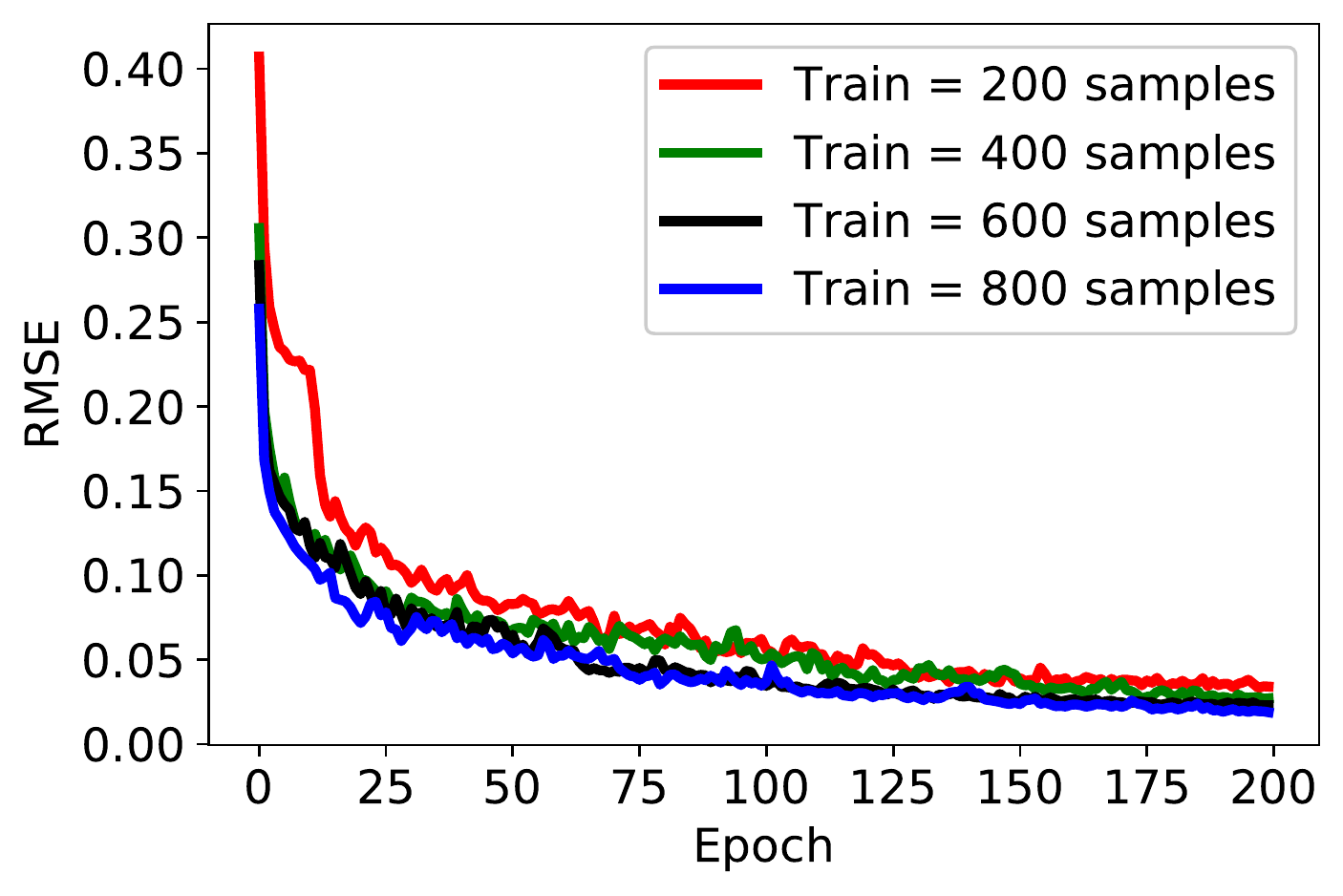}
    \caption{RMSE decay with number of epochs.}
    \label{fig:rmse_f30}
\end{figure}

\begin{figure}
    \centering
    \begin{subfigure}[b]{0.45\textwidth}
        \includegraphics[width=\textwidth]{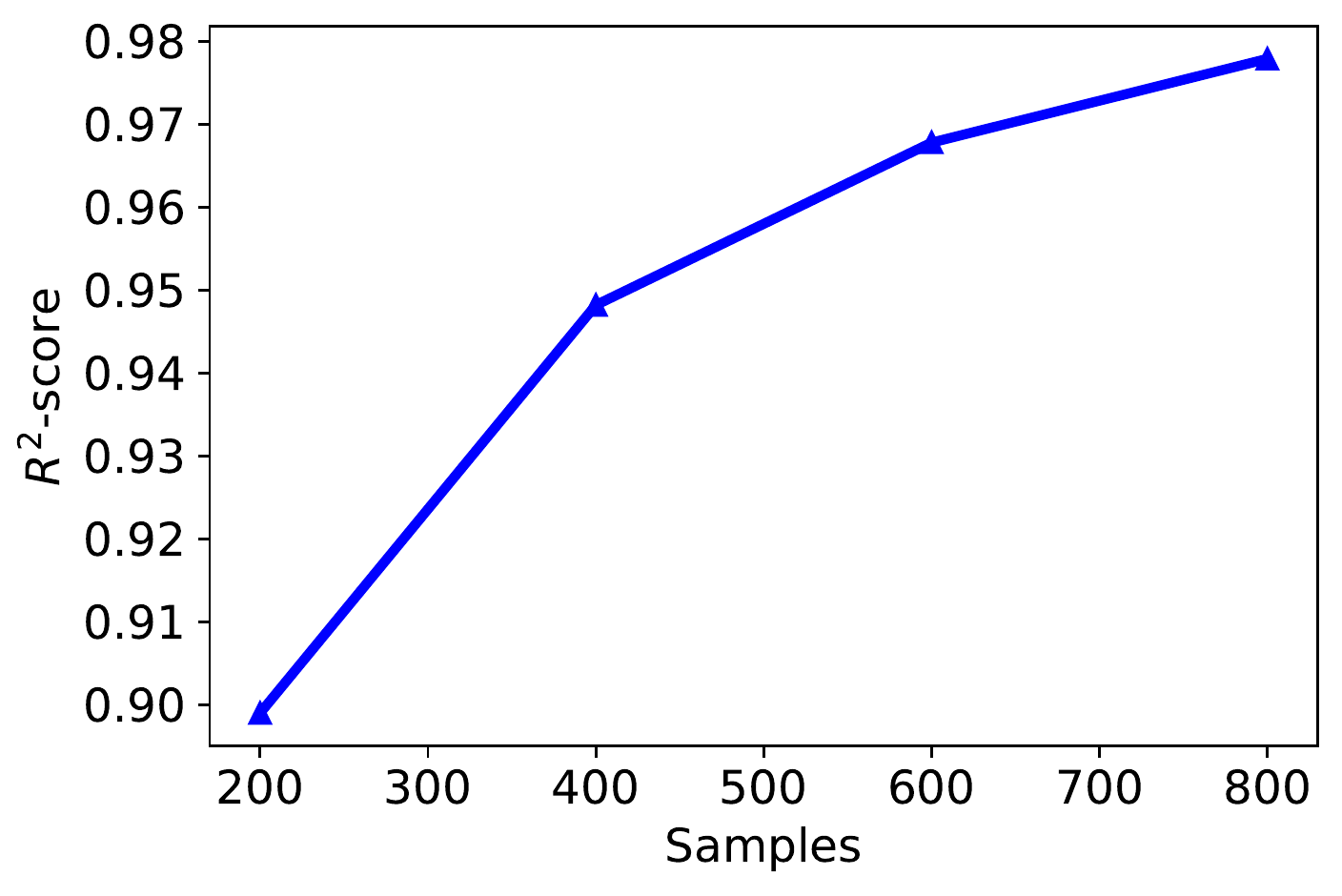}
        \caption{ $R^2$-score for the trained networks}
    \end{subfigure}
    \begin{subfigure}[b]{0.45\textwidth}
        \includegraphics[width=\textwidth]{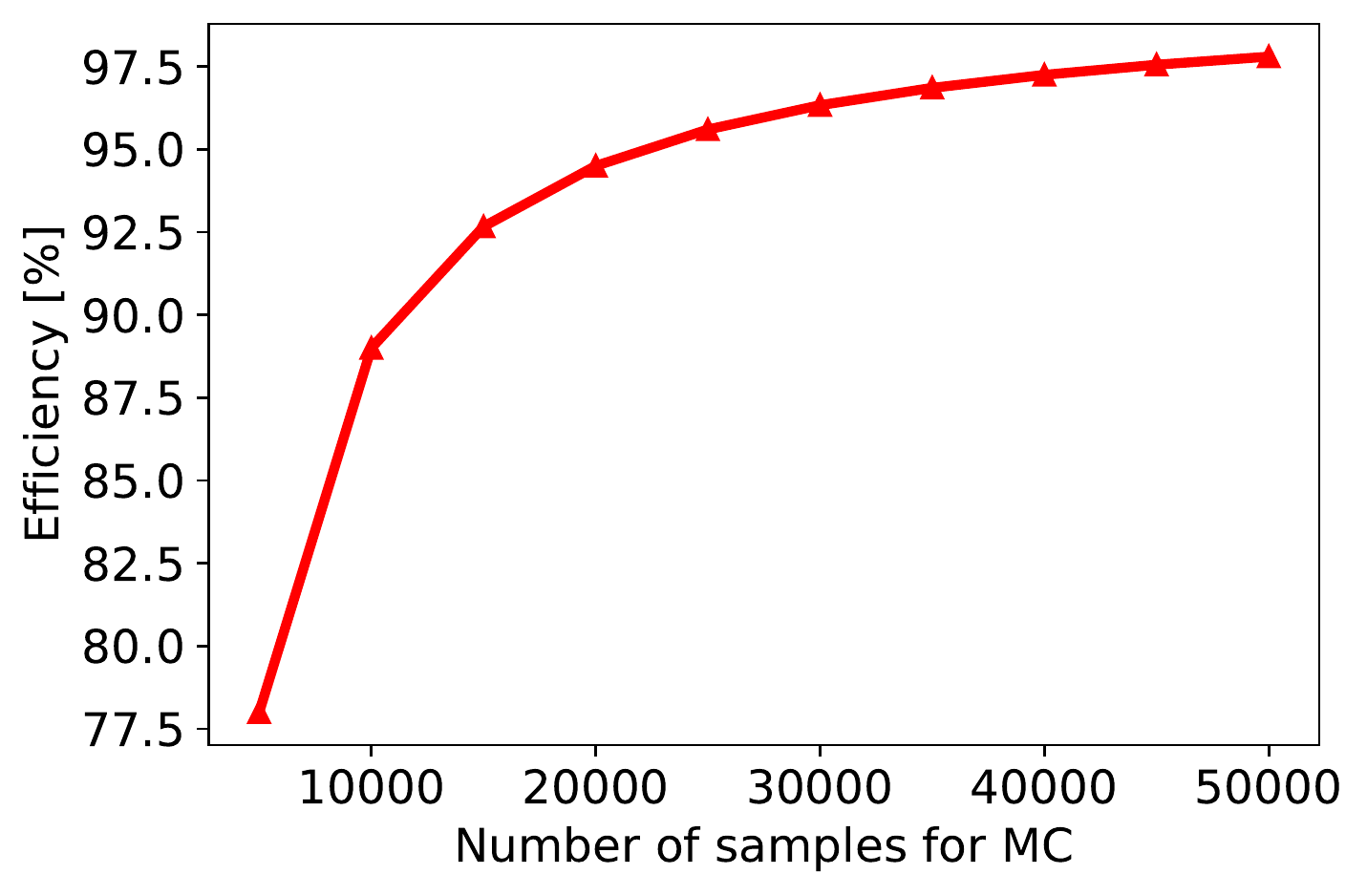}
        \caption{Efficiency}
    \end{subfigure}
    \caption{Test $R^2$-score and efficiency in function of the number of MC samples.}
    \label{fig:test_r2_eff_f30}
\end{figure}

We detail the architecture of the neural network for this scenario in Table \ref{tab:nn_architecture2}. It  contains five dense blocks, leading to $412,210$  parameters to be trained. We can see in Figure \ref{fig:rmse_f30} the $RMSE$ decay as the number of epochs increase for all training sets. We verify the accuracy of the surrogate by computing the $R^2$ score for the 500 testing samples. We find that, as expected, for networks trained with larger data sets, $R^2$ values are closer to $1.0$, as shown in Figure \ref{fig:test_r2_eff_f30}(a). Due to limitations imposed by the high cost of generating samples using the full RTM model for this example, we develop a different efficiency analysis extrapolating from the basic conditions used for the network training. We assume conservatively that $N_{MC}$ is of the same order of the case in section \ref{sec:efficiency_analysis}. Thus, we start from a scenario where only 5,000 samples are needed to characterize uncertainties in the seismic images and extrapolate to more expensive potential scenarios requiring hypothetically till 50,000 samples. Here, $N_S$ is equal to 1100, 600 samples to train the neural network with an accuracy of $R^2 \geq 0.95$, and 500 to test the surrogate model. Figure \ref{fig:test_r2_eff_f30}(b) depicts the efficiency analysis in function of $N_{MC}$. We note,  for the worst scenario, an efficiency of around 78\%, and for the scenarios where $N_{MC}$ is higher than 10,000 samples, the efficiencies reach values greater than 90\%. For the most expensive scenario, we see an efficiency close to 98\%. 

\begin{table}
\caption{Neural Network Architecture. "Outputs" represents the number of features maps and "Dimension" is the spatial dimension of the features maps.}
\centering
\begin{tabular}{cccc}
\toprule
Layers & Output & Dimension  \\
\midrule
Input & 1 &  $100 \times 100$ \\
Convolution & 48 &  $48 \times 48$  \\
Dense-block 1 & 112 &  $48 \times 48$  \\
Encoding & 56 &  $24 \times 24$  \\
Dense-block 2 & 120 &  $24 \times 24$  \\
Encoding & 60 &  $12 \times 12$  \\
Dense-block 3 & 124 &  $12 \times 12$  \\
Decoding & 62 &  $24 \times 24$  \\
Dense-block 4 & 126 &  $24 \times 24$  \\
Decoding & 63 &  $48 \times 48$  \\
Dense-block 5 & 127 &  $48 \times 48$  \\
Decoding & 1 &  $100 \times 100$  \\
ReLU & 1 &  $100 \times 100$  \\
\bottomrule
\end{tabular}
\label{tab:nn_architecture2}
\end{table}

\begin{figure}
    \centering
    \begin{subfigure}[b]{0.40\textwidth}
        \includegraphics[width=\textwidth]{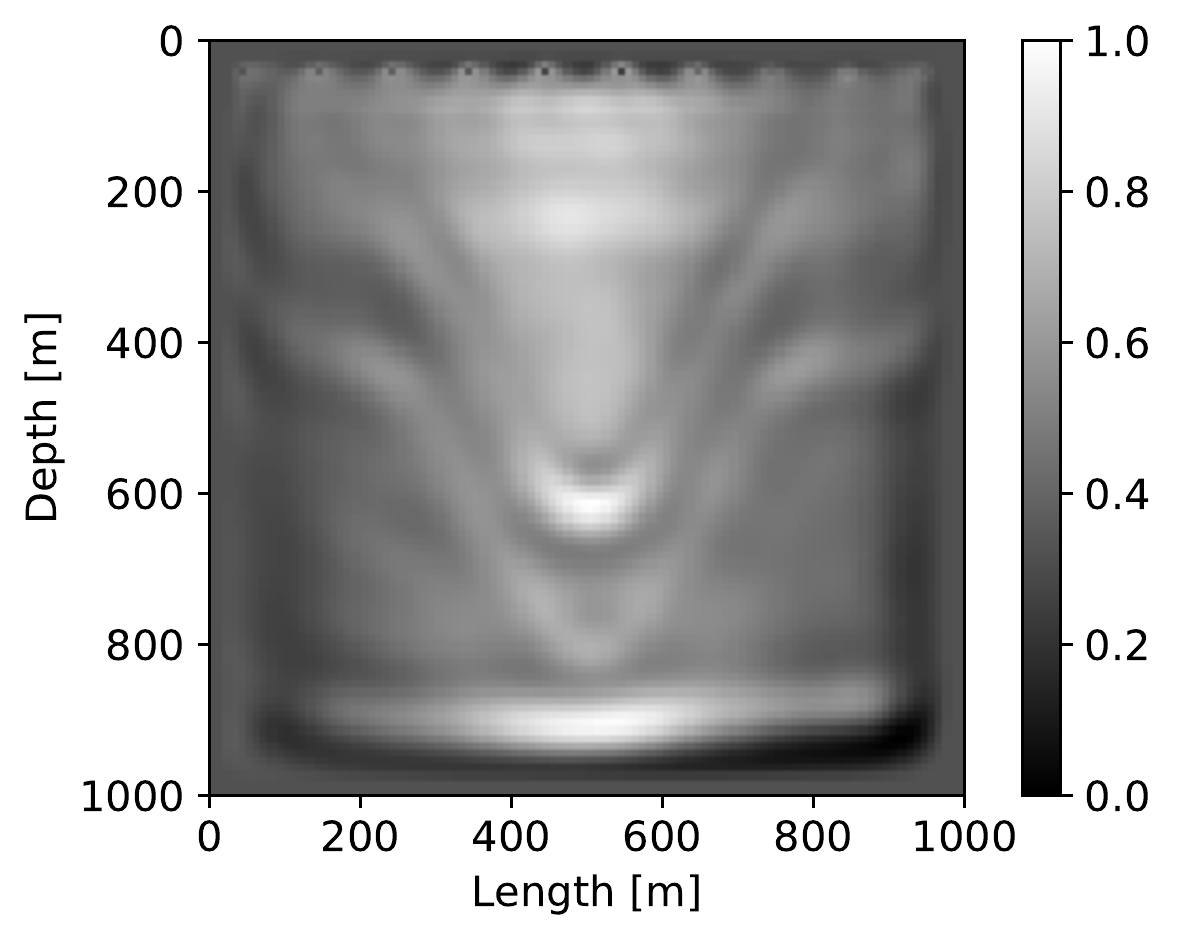}
    \end{subfigure}
    \begin{subfigure}[b]{0.40\textwidth}
        \includegraphics[width=\textwidth]{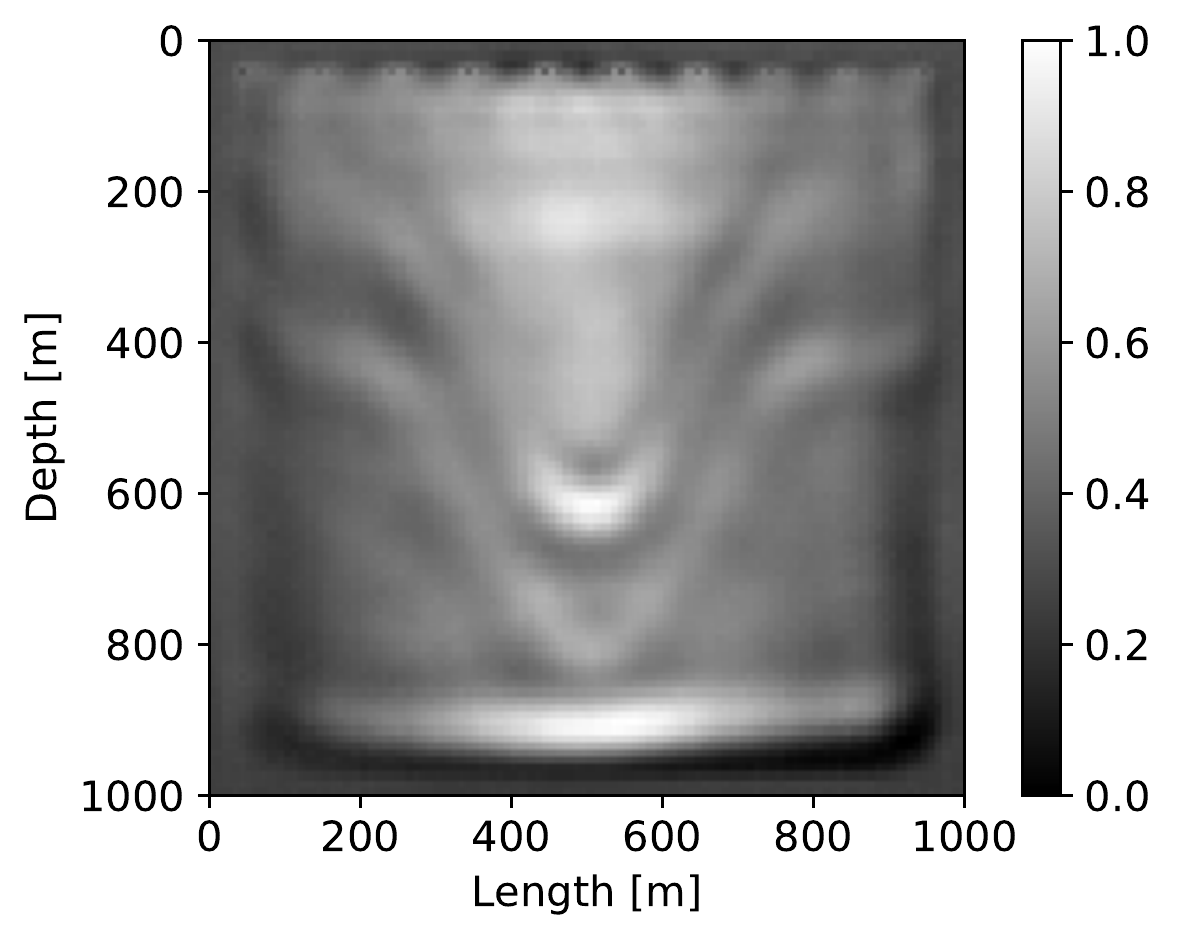}
    \end{subfigure}
    \begin{subfigure}[b]{0.40\textwidth}
        \includegraphics[width=\textwidth]{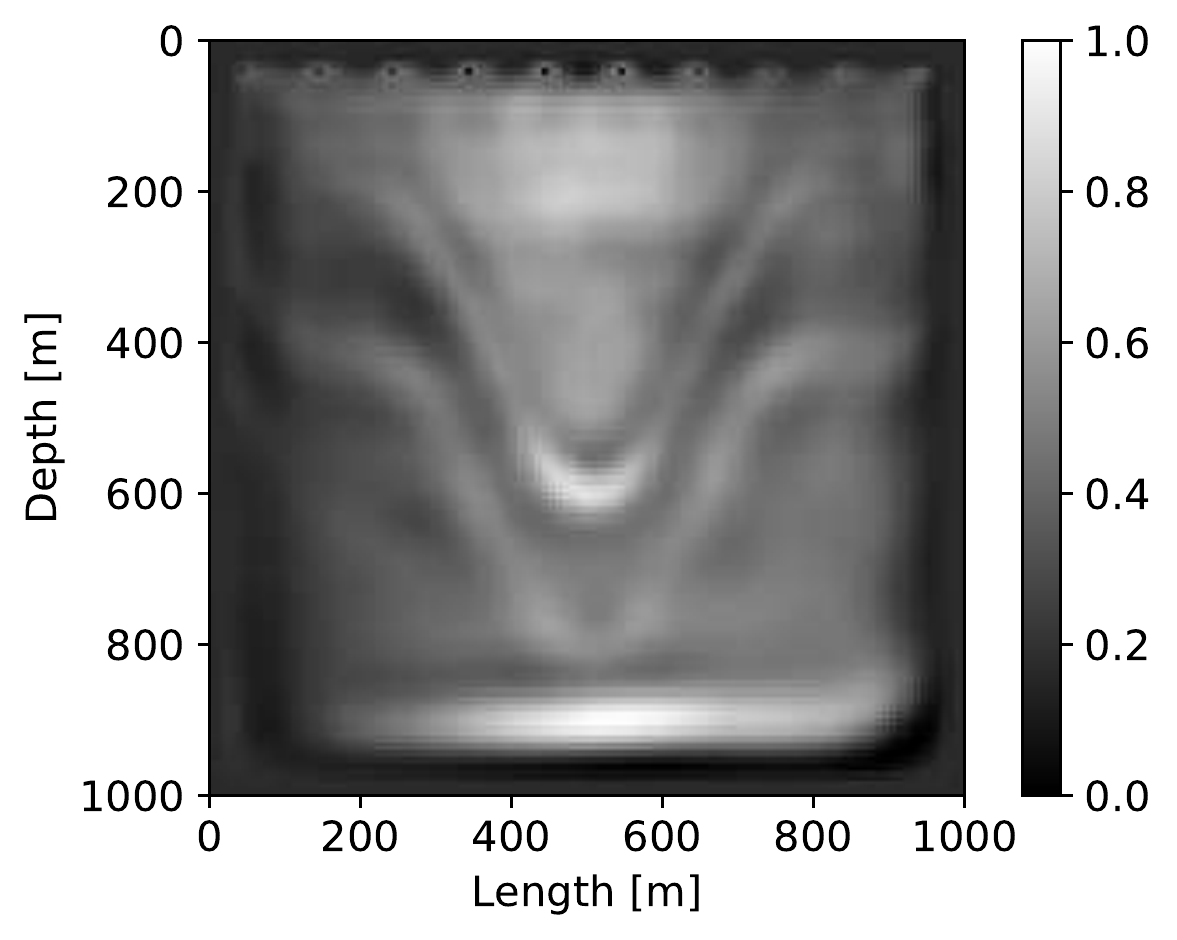}
    \end{subfigure}
    \begin{subfigure}[b]{0.40\textwidth}
        \includegraphics[width=\textwidth]{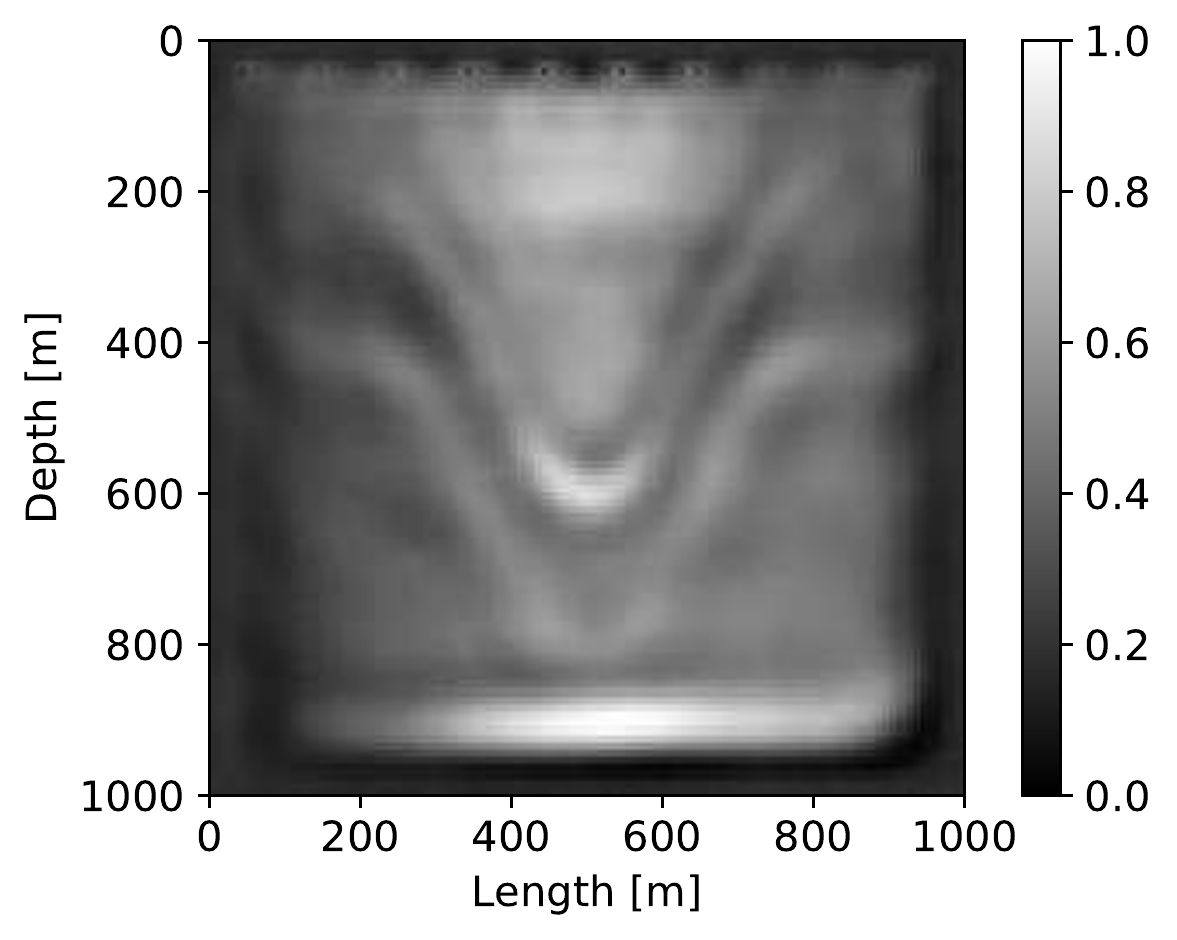}
    \end{subfigure}
    \begin{subfigure}[b]{0.40\textwidth}
        \includegraphics[width=\textwidth]{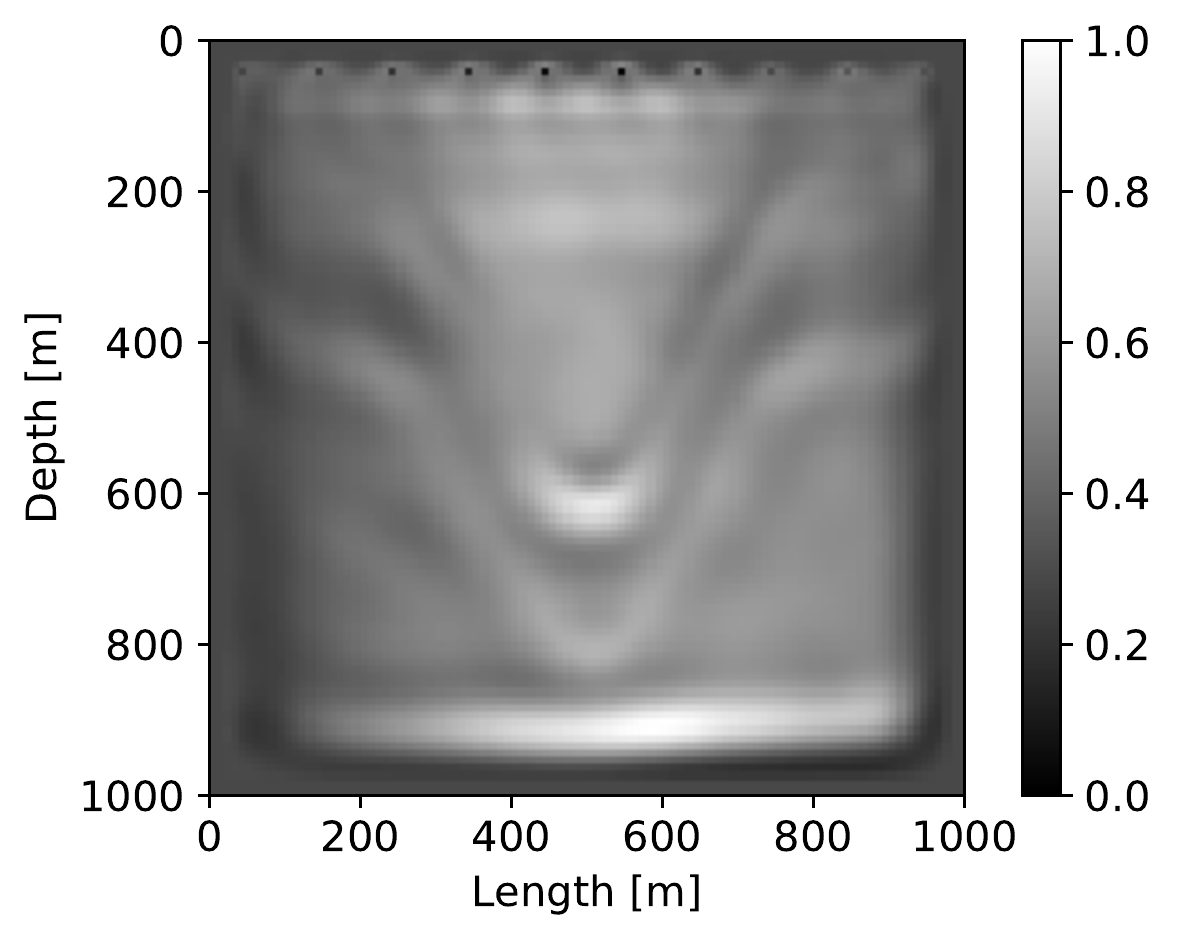}
        \caption{RTM model}
    \end{subfigure}
    \begin{subfigure}[b]{0.40\textwidth}
        \includegraphics[width=\textwidth]{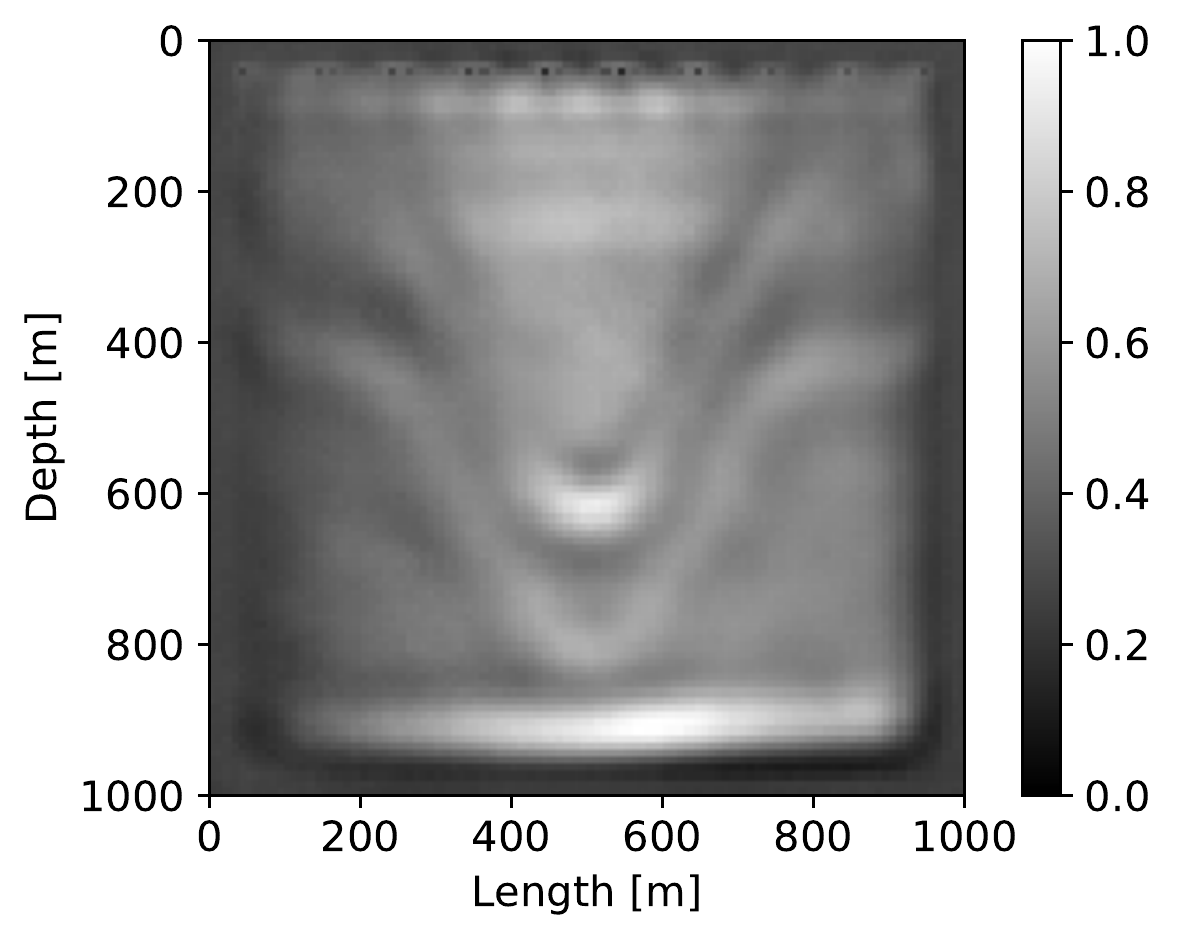}
        \caption{Surrogate model}
    \end{subfigure}
    \caption{Randomly selected images from the test data set computed by the RTM model (a) and the surrogate model (b) trained with 600 samples. The relative  errors in the image condition, $e_I$, are lower than 6\%.}
    \label{fig:prediction_f30}
\end{figure}

\begin{figure}
    \centering
    \begin{subfigure}[b]{0.39\textwidth}
        \includegraphics[width=\textwidth]{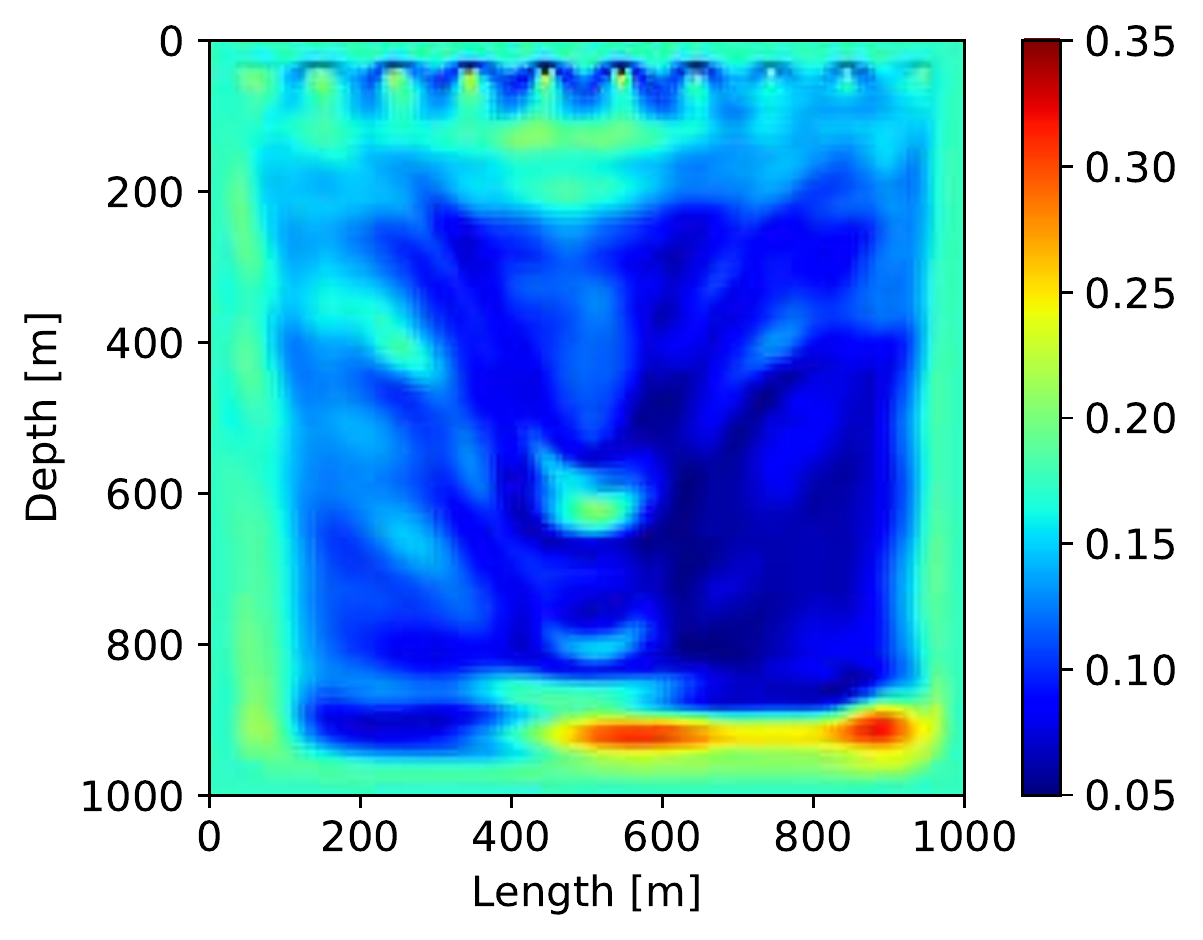}
        \caption*{Standard deviation - RTM model}
    \end{subfigure}
    \begin{subfigure}[b]{0.39\textwidth}
        \includegraphics[width=\textwidth]{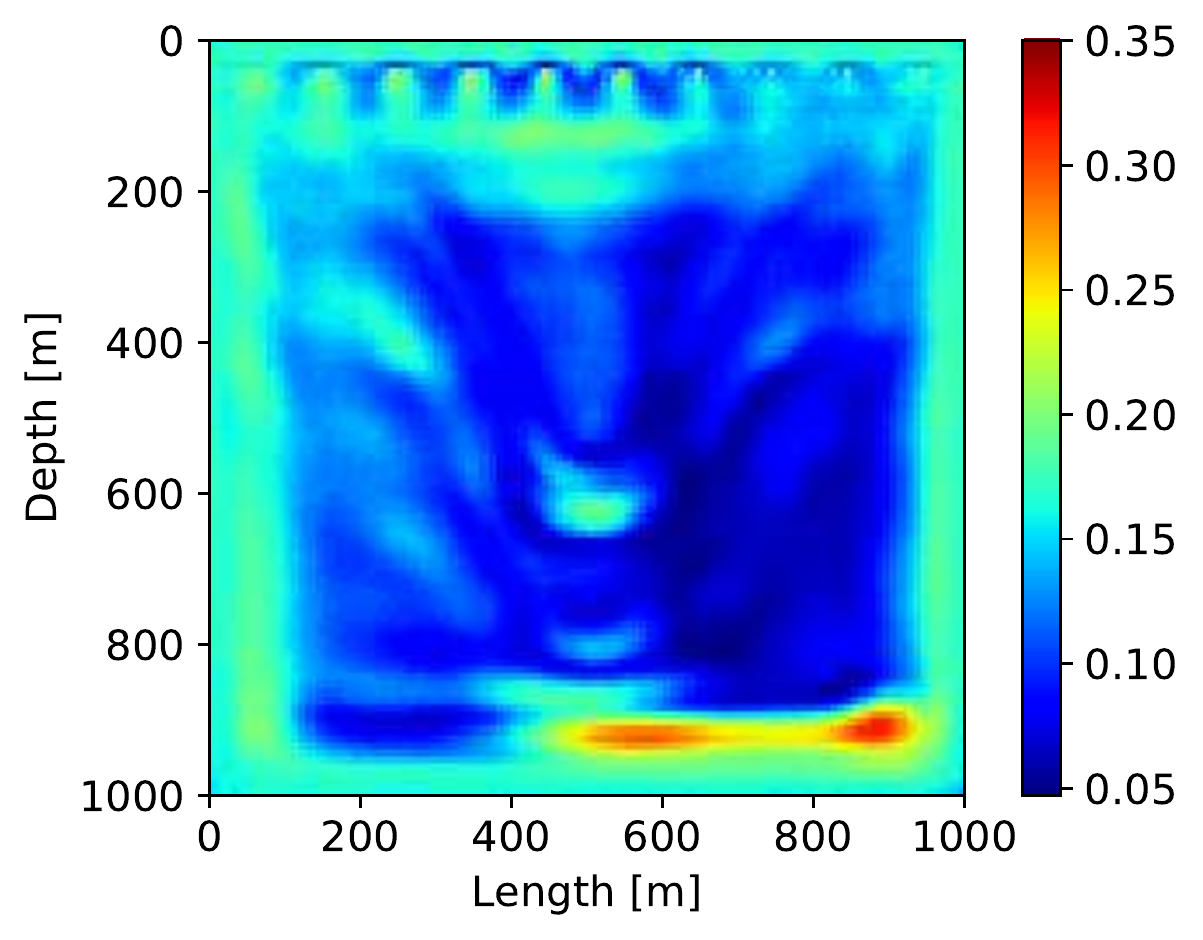}
        \caption*{Standard deviation - Surrogate model}
    \end{subfigure}
    \begin{subfigure}[b]{0.39\textwidth}
        \includegraphics[width=\textwidth]{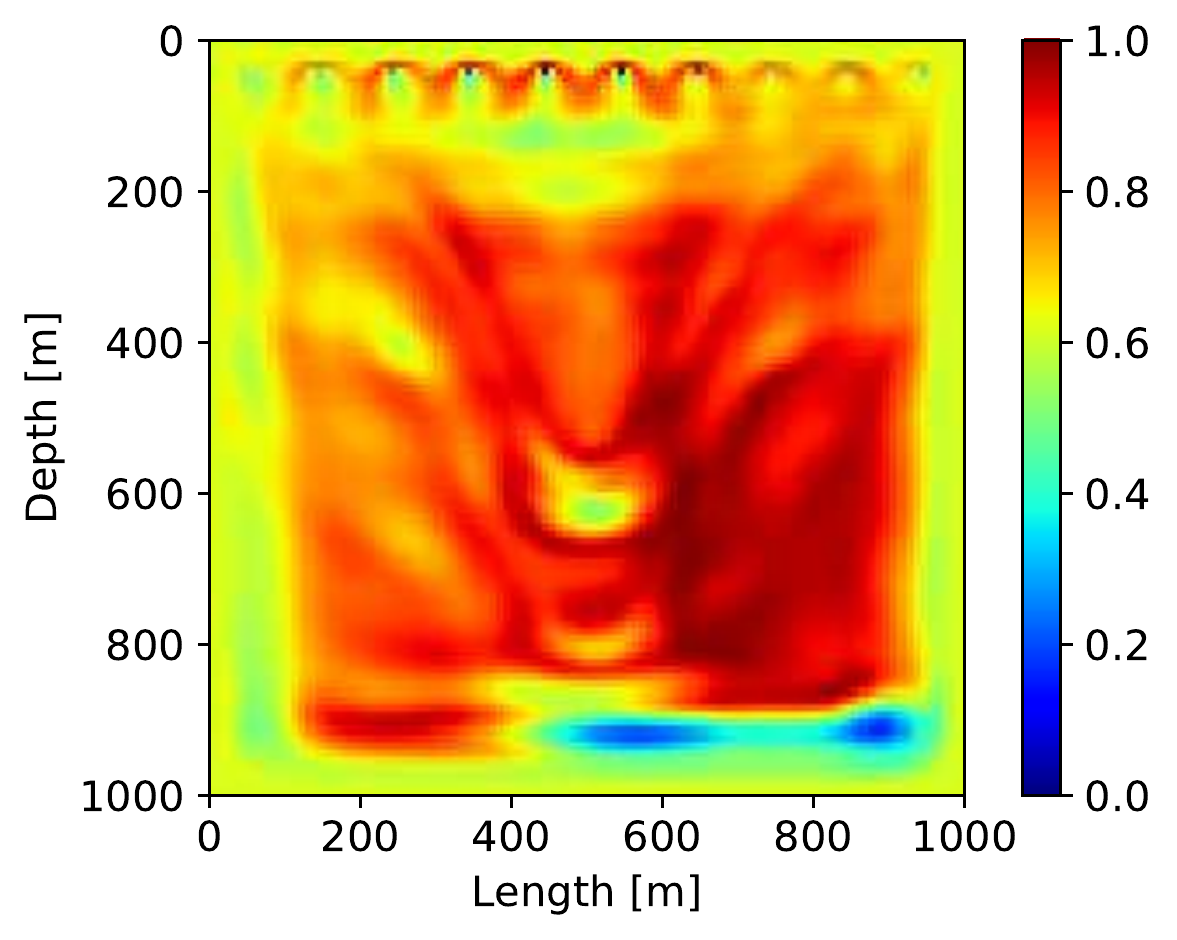}
        \caption*{Confidence index - RTM model}
    \end{subfigure}
    \begin{subfigure}[b]{0.39\textwidth}
        \includegraphics[width=\textwidth]{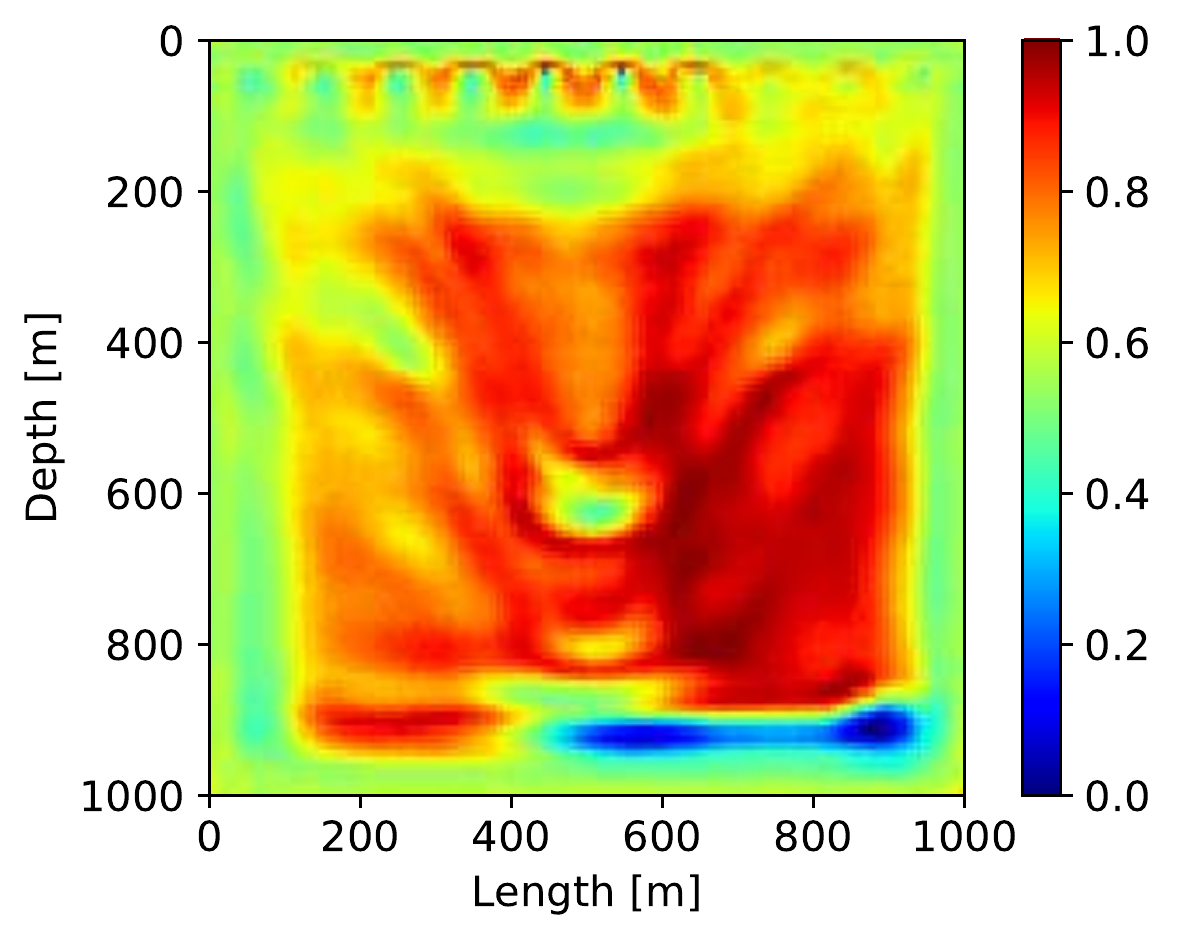}
        \caption*{Confidence index - Surrogate model}
    \end{subfigure}
    \begin{subfigure}[b]{0.39\textwidth}
        \includegraphics[width=\textwidth]{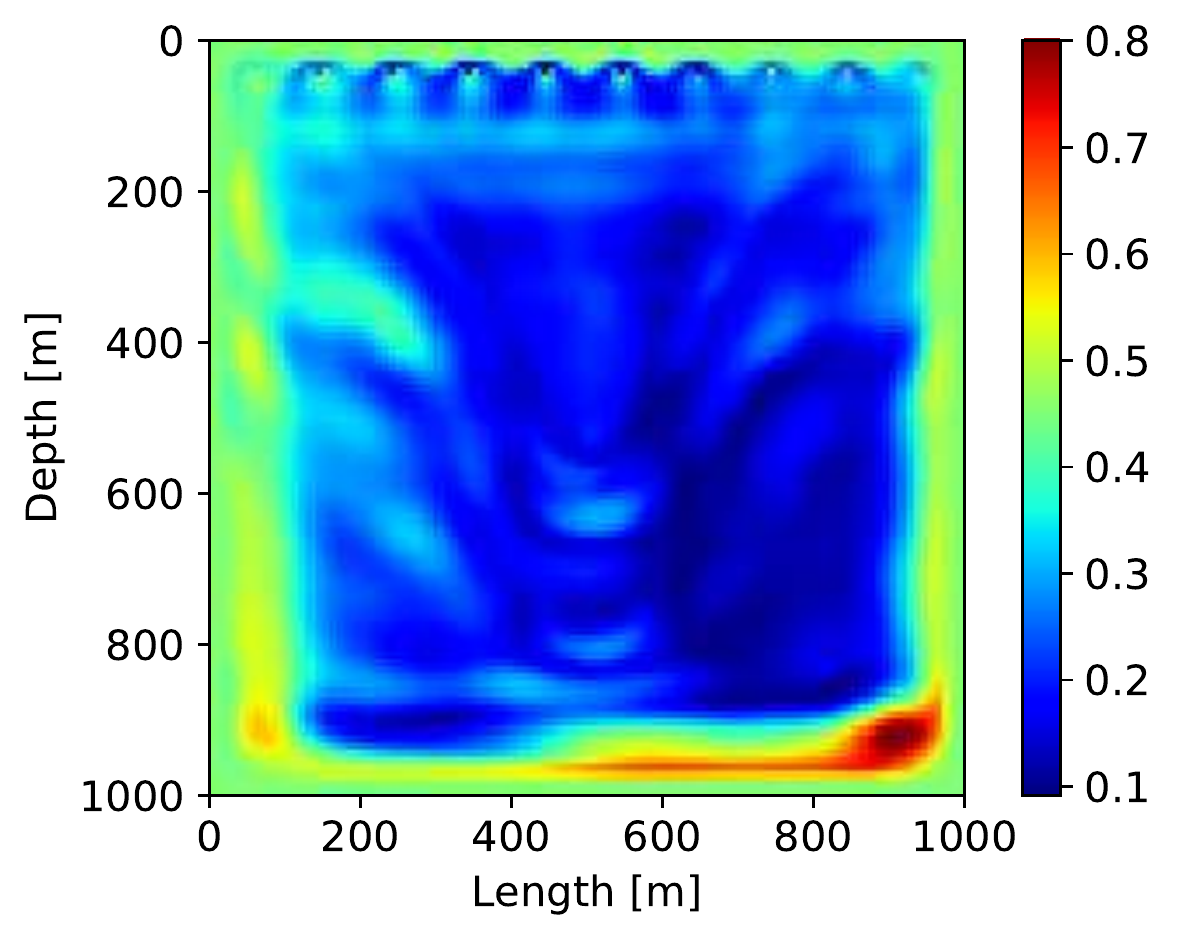}
        \caption*{Coefficient of Variation - RTM model}
    \end{subfigure}
    \begin{subfigure}[b]{0.39\textwidth}
        \includegraphics[width=\textwidth]{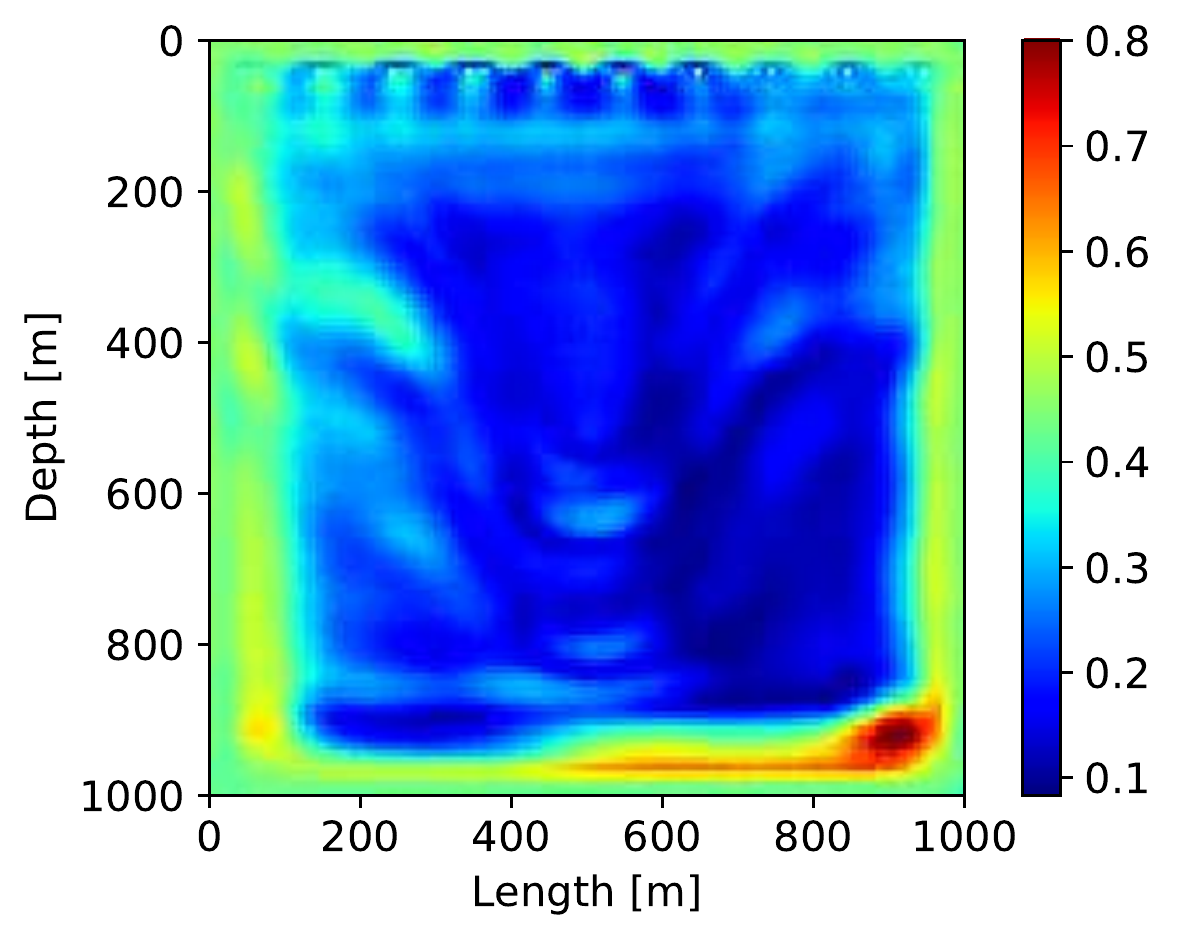}
        \caption*{Coefficient of variation - Surrogate model}
    \end{subfigure}
    \caption{UQ indexes - standard deviation, $\sigma(\mathbf{r})$, confidence index, $c(\mathbf{r})$, and coefficient of variation, $c_v(\mathbf{r})$ - predicted by the RTM model (left) and the surrogate model (right). The relative errors between the surrogate predictions to the RTM model for the UQ indexes are less than 6\%.}
    \label{fig:uq_maps_5layers}
\end{figure}

Fig. \ref{fig:prediction_f30} shows comparisons between images randomly selected from the test set and the corresponding images produced with the full RTM model. We observe that the surrogate model returns excellent predictions of the imaging condition. We can also note that the surrogate model predicts the UQ figures - standard deviation, $\sigma(\mathbf{r})$, confidence index, $c(\mathbf{r})$, and coefficient of variation, $c_v(\mathbf{r})$ - with good accuracy, as seen in Fig \ref{fig:uq_maps_5layers}, where the relative error between the surrogate predictions and the RTM model are lower than 6\%. Results in Figures \ref{fig:prediction_f30} and \ref{fig:uq_maps_5layers} shows that the encoder-decoder surrogate extrapolates the replication of the IC training targets, delivering remarkable results to assist in UQ analysis.

\begin{figure}
    \centering
    \begin{subfigure}[b]{0.45\textwidth}
        \includegraphics[width=\textwidth]{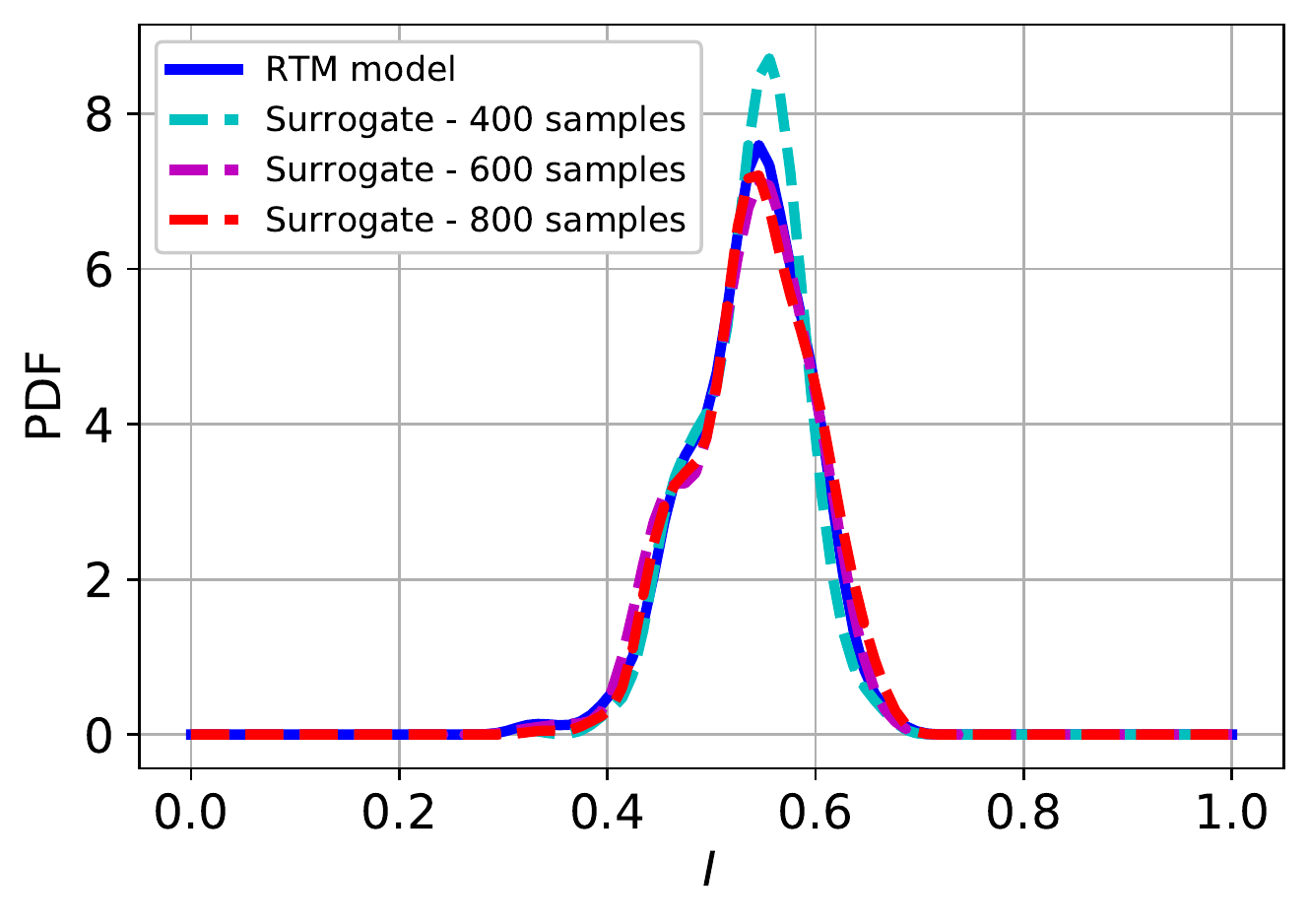}
        \caption*{$P_1$}
    \end{subfigure}
    \begin{subfigure}[b]{0.45\textwidth}
        \includegraphics[width=\textwidth]{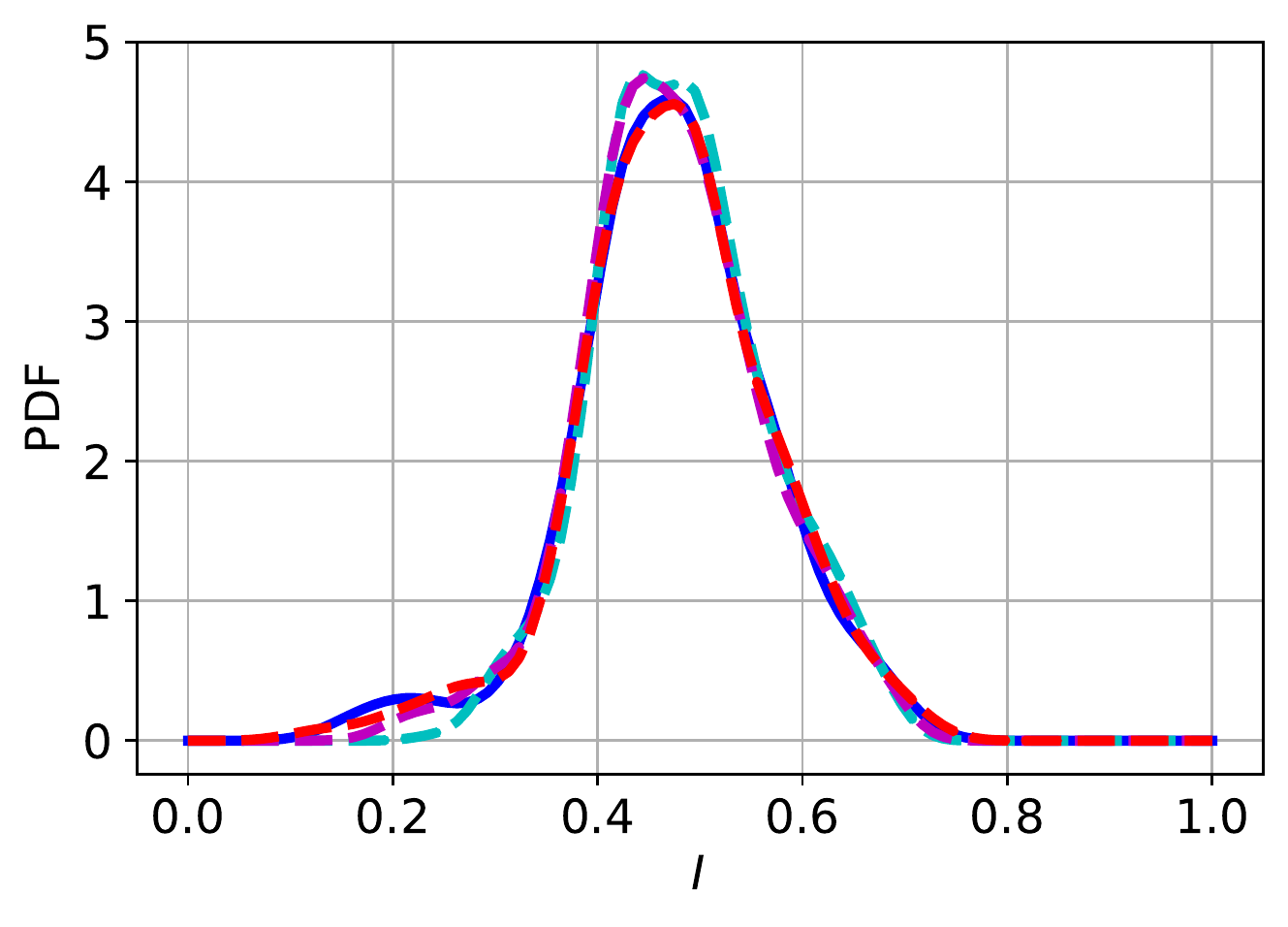}
        \caption*{$P_2$}
    \end{subfigure}
    \begin{subfigure}[b]{0.45\textwidth}
        \includegraphics[width=\textwidth]{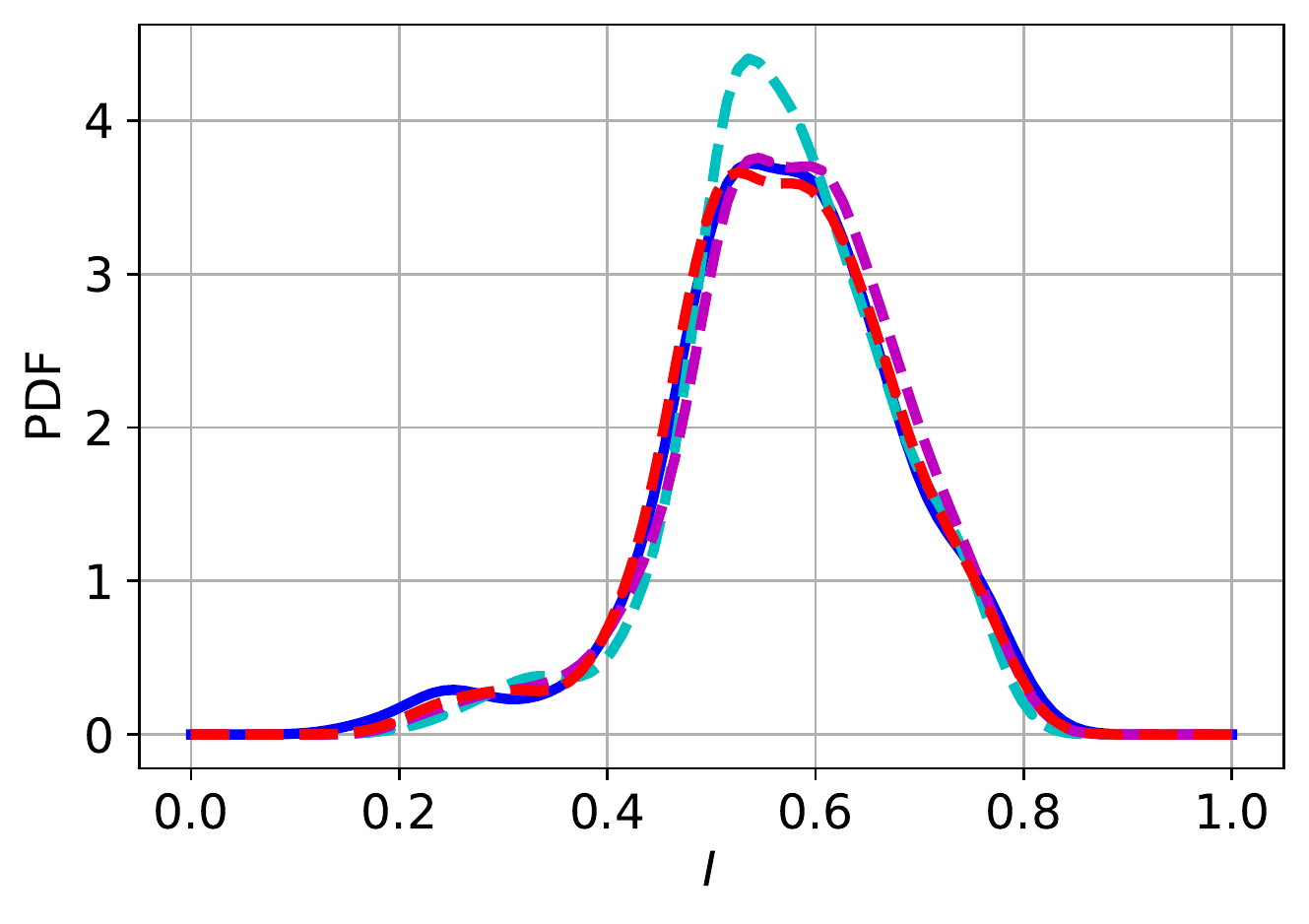}
        \caption*{$P_3$}
    \end{subfigure}
    \begin{subfigure}[b]{0.45\textwidth}
        \includegraphics[width=\textwidth]{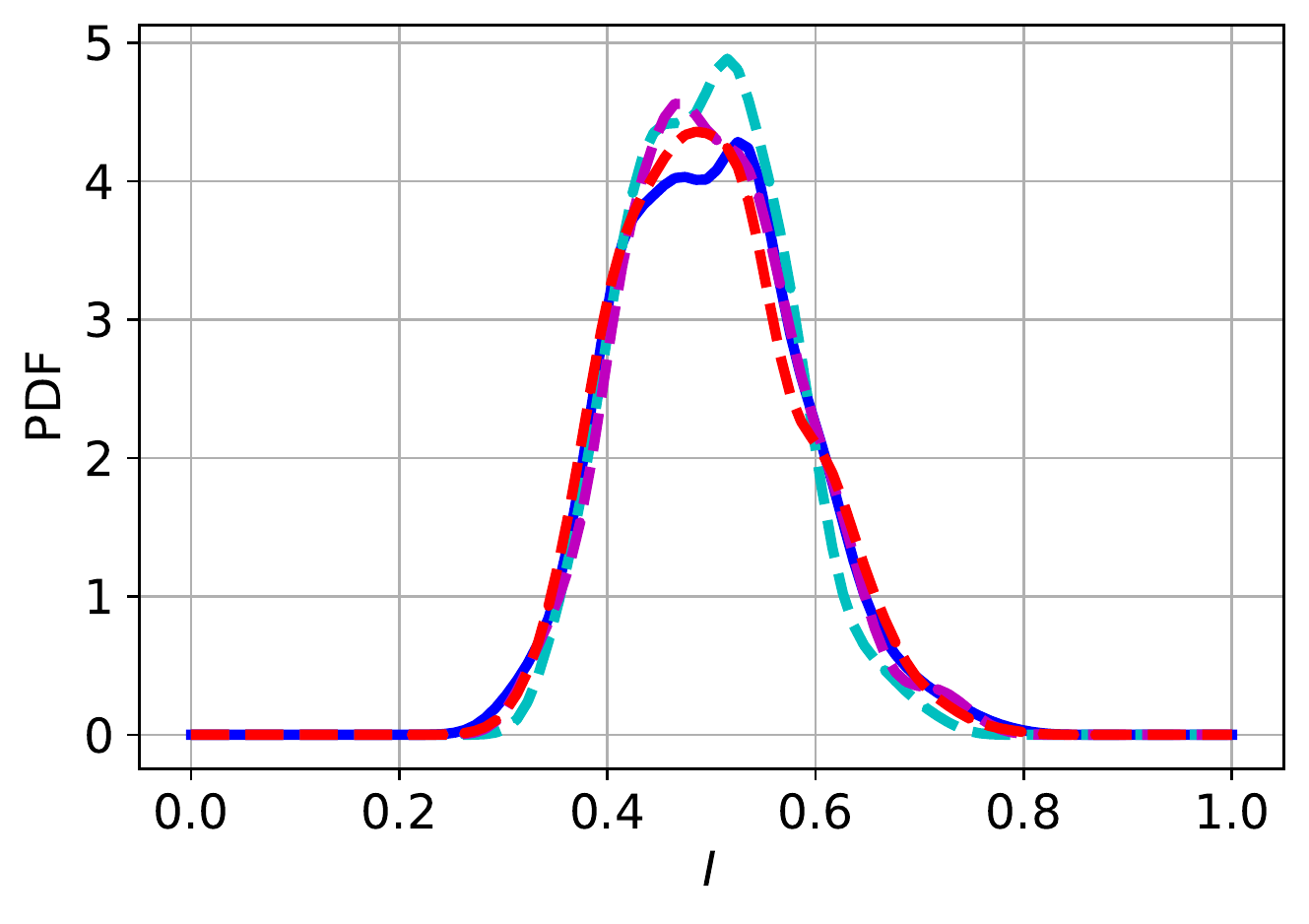}
        \caption*{$P_4$}
    \end{subfigure}
    \caption{Comparison between PDFs predicted by the RTM model and the surrogate model. }
    \label{fig:pdf_points}
\end{figure}

Next, we deepen our investigation of the surrogate's ability to reproduce the probabilistic characterization of the IC fields by plotting the probability density functions (PDFs) of the imaging condition at control points in the domain, as displayed in Fig. \ref{fig:velocity_field}. We place the control points in regions of low and high uncertainties.  As the reference solution, we use PDFs obtained by the RTM model with the 500 test samples to compare the accuracy of the surrogate model trained with different datasets to estimate the PDFs at the control points. Figure \ref{fig:pdf_points} shows the imaging condition PDFs estimated with the surrogate models trained with 400, 600, 800 samples, together with the reference PDFs. We observe that the PDFs obtained with the surrogate model are in good agreement with the reference ones for all control points, particularly at the deeper point ($P4$).

\begin{figure}
    \centering
    \begin{subfigure}[b]{0.56\textwidth}
        \includegraphics[width=\textwidth]{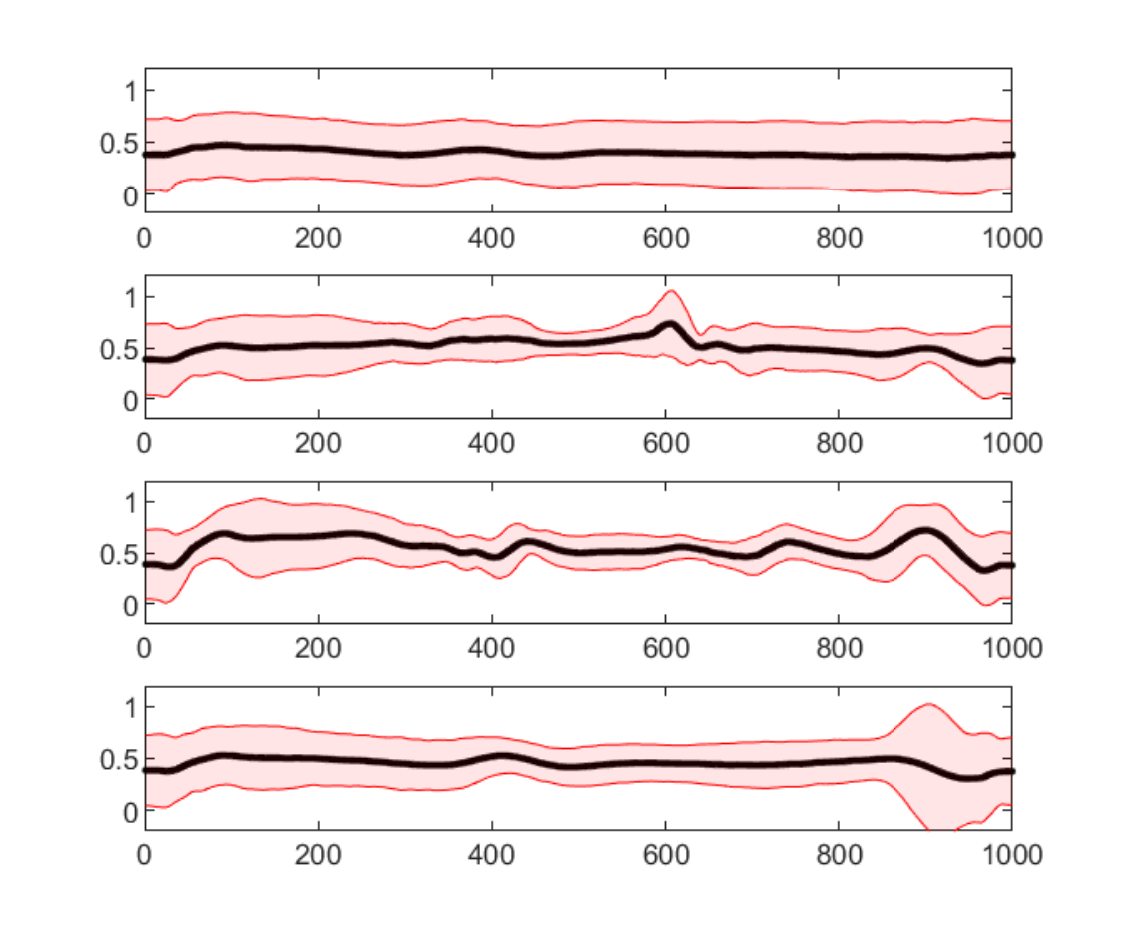}
        \caption*{Confidence bands on the interfaces for $f=30$Hz}
    \end{subfigure}
    \begin{subfigure}[b]{0.42\textwidth}
        \includegraphics[width=\textwidth]{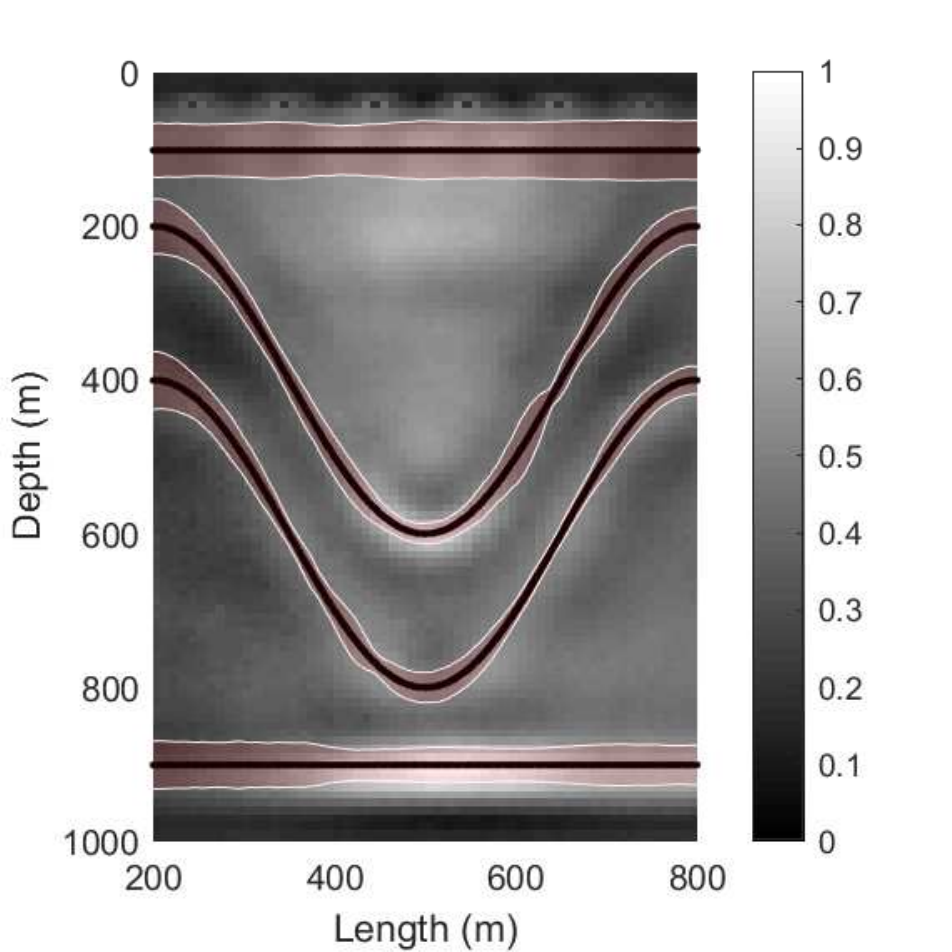}
        \caption*{Confidence bands over image condition, $f=30$Hz}
    \end{subfigure}
    
    \caption{IC confidence bands over the interface between geological layers (left), and IC confidence bands superimposed over a randomly selected image (right) for $f=30$~Hz. }
    \label{fig:targets_f30}
 \end{figure}

Now we exemplify a possible surrogate use in the feature extraction and interpretation of seismic images, {\bf{Stage 3}} of Algorithm \ref{algworkflow}. We provide, using the surrogate,  a view of the uncertainties associated with specific seismic targets, the interfaces of geological layers. This view can reveal how the propagated uncertainties can directly impact the images posterior interpretation. Figure \ref{fig:targets_f30} provides the IC mean value and associate confidence bands for the four interfaces. In the right part of the figure, we give an idea of the uncertainties spatial distribution, having as background a randomly selected image from the ensemble. To promote visual perception, we plotted amplified IC confidence bands associated with each interface. Those bands reflect the IC value dispersion amid the image ensemble, and, therefore, might lead to a lack of confidence in the reflector placement.

\subsubsection{Cutoff frequency - 45Hz}

\begin{figure}
    \centering
    \includegraphics[scale=.5]{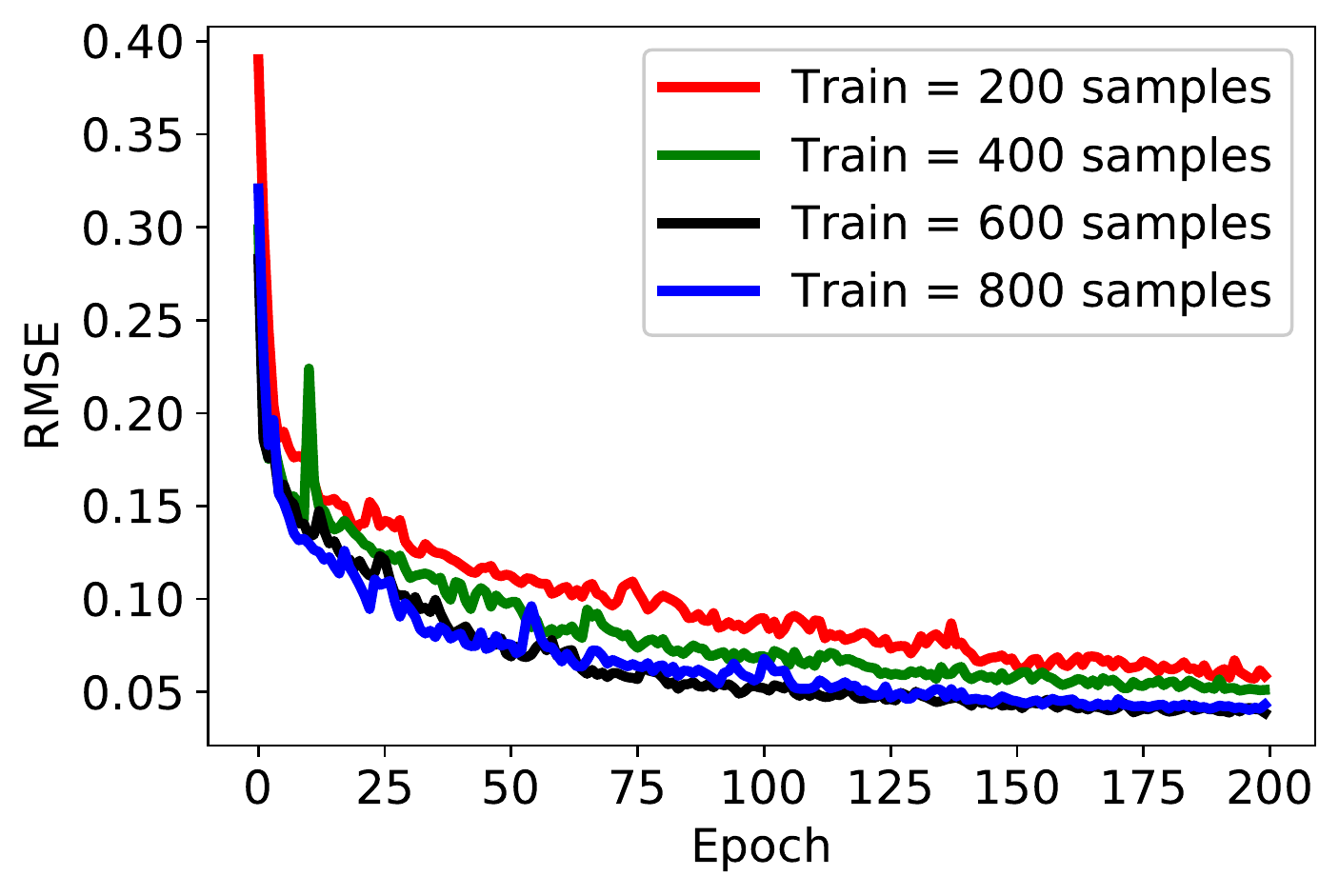}
    \caption{RMSE decay with number of epochs.}
    \label{fig:rmse_f45}
\end{figure}

We now stress the proposed deep learning surrogate by testing its performance in a scenario not considered for the training, but still focusing on the subsurface geology of Fig.\ref{fig:velocity_field}.  The  RTM image comprises a certain number of shots, each designed for illuminating the same subsurface using different seismic sources conditions.  Here we analyze a situation involving a higher excitation frequency, $f = 45 Hz$, implying, due to numerical requirements, in the necessity of a finer grid. In such a case, input and output images have different dimensions compared to the previous scenario. Still, the intrinsic dimensionality is the same for the input as we are imaging the same velocity field as before. Instead of seeking for a new architecture, we slightly changed the previous one by replacing the first and last network layers and adapting the initial convolutional layer to ensure that an integer number defines the kernel. We do not expect to obtain the surrogate's optimal performance by employing such a strategy, but that can be quite useful in practical terms if it works. Table \ref{tab:nn_architecture3} shows the neural network architecture for the 45~Hz scenario. The network architecture is the same as in section  \ref{subsec:5_layers_30hz} with small changes. The first convolutional layer has a kernel size equal to 7 and a stride of 2. The total number of parameters in the network is $416,390$. Figure \ref{fig:rmse_f45} shows $RMSE$ decay as a function of the number of epochs in the training process.

\begin{table}
\caption{Neural Network Architecture. "Outputs" represents the number of features maps and "Dimension" is the dimension of the features maps.}
\centering
\begin{tabular}{cccc}
\toprule
Layers & Output & Dimension  \\
\midrule
Input & 1 &  $150 \times 150$ \\
Convolution & 48 &  $72 \times 72$  \\
Dense-block 1 & 112 &  $72 \times 72$  \\
Encoding & 56 &  $36 \times 36$  \\
Dense-block 2 & 120 &  $36 \times 36$  \\
Encoding & 60 &  $18 \times 18$  \\
Dense-block 3 & 124 &  $18 \times 18$  \\
Decoding & 62 &  $36 \times 36$  \\
Dense-block 4 & 126 &  $36 \times 36$  \\
Decoding & 63 &  $72 \times 72$  \\
Dense-block 5 & 127 &  $72 \times 72$  \\
Decoding & 1 &  $150 \times 150$  \\
ReLU & 1 &  $150 \times 150$  \\
\bottomrule
\end{tabular}
\label{tab:nn_architecture3}
\end{table}

\begin{figure}
    \centering
    \begin{subfigure}[b]{0.45\textwidth}
        \includegraphics[width=\textwidth]{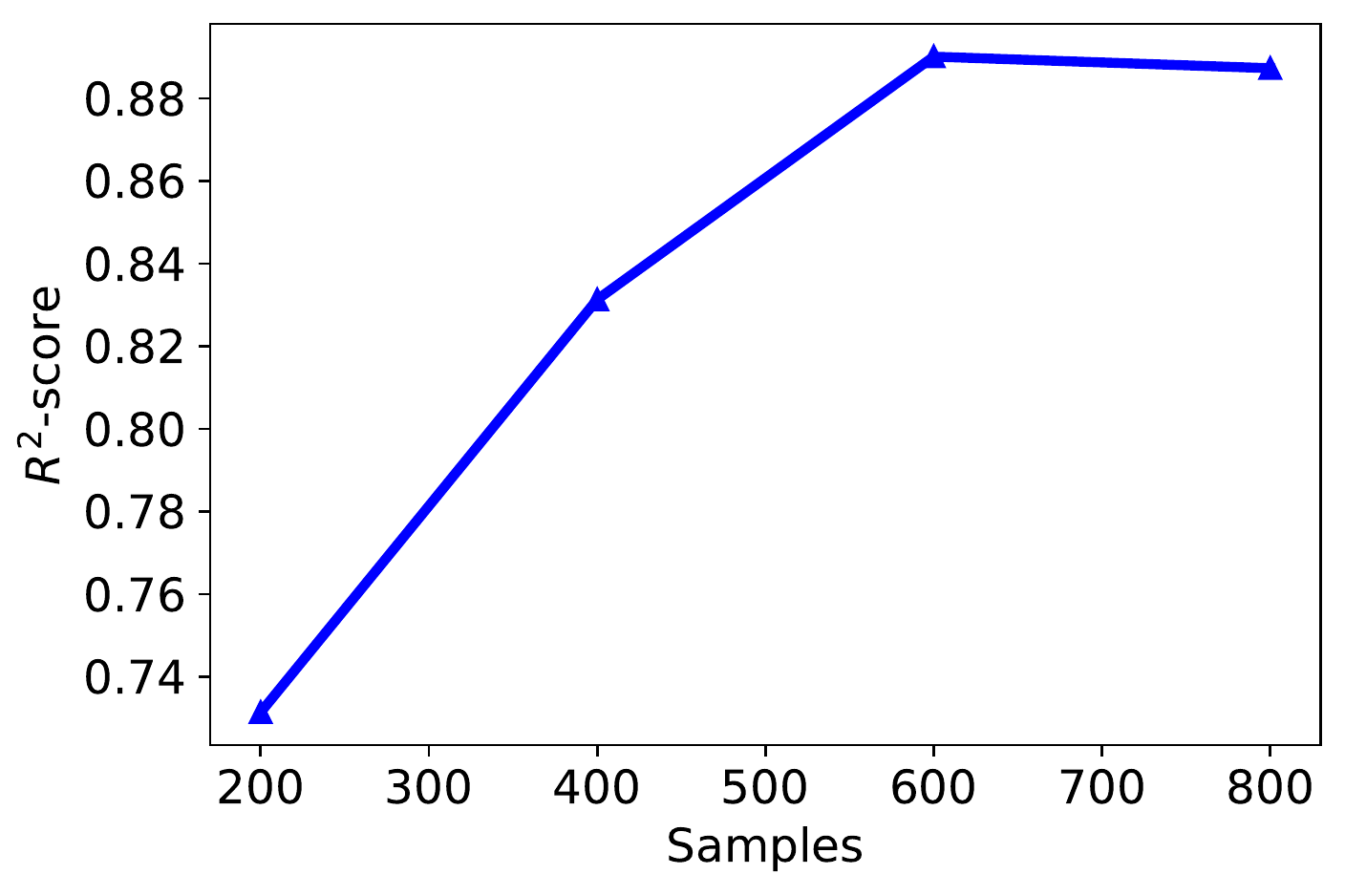}
        \caption{$R^2$-score for the trained networks}
    \end{subfigure}
    \begin{subfigure}[b]{0.45\textwidth}
        \includegraphics[width=\textwidth]{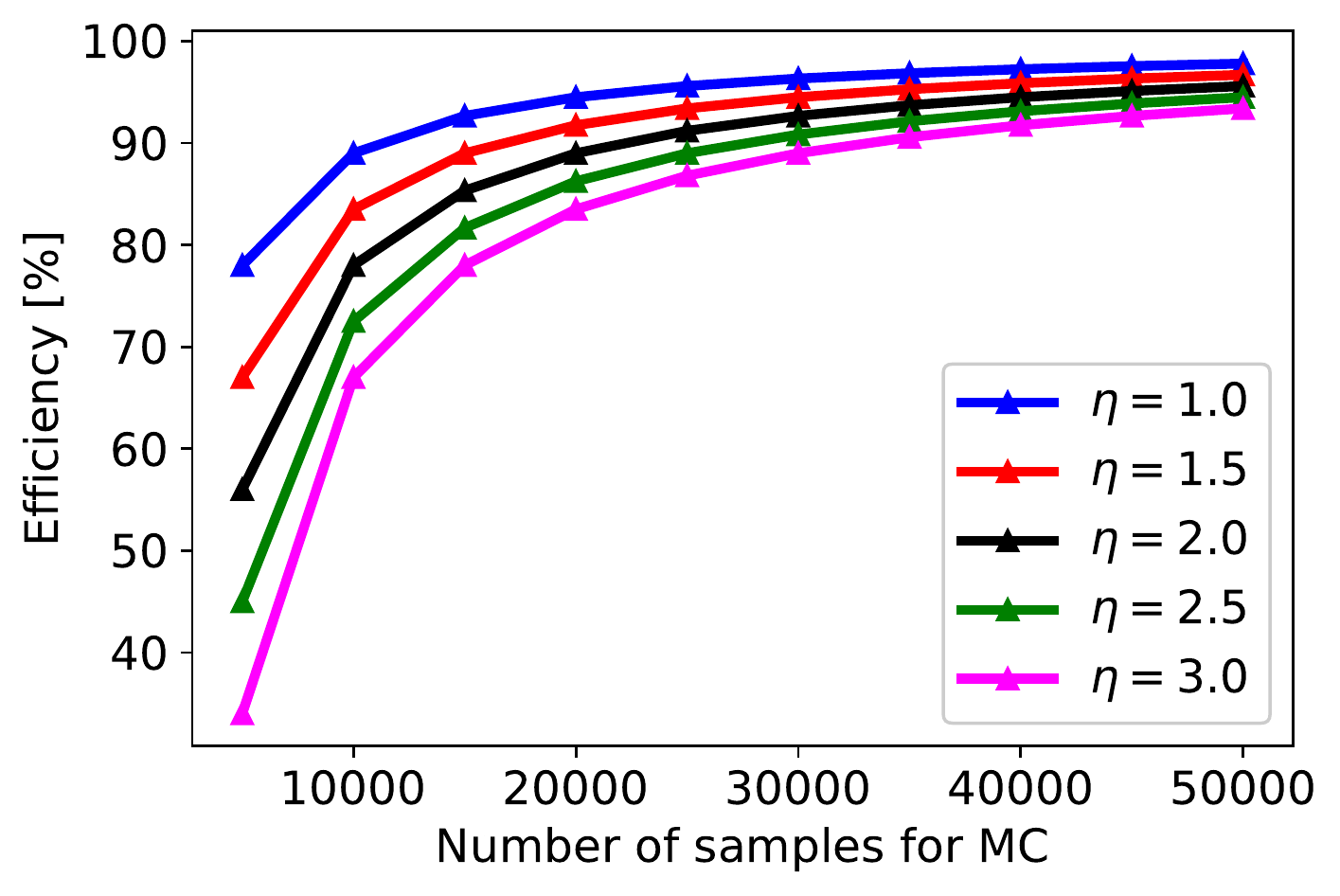}
        \caption{Efficiency}
    \end{subfigure}
    \caption{ $R^2$-score and efficiency for the trained networks.}
    \label{fig:test_r2_eff_f45}
\end{figure}

We can see in Fig. \ref{fig:test_r2_eff_f45}a the $R^2$ score for different training sets showing for this more difficult scenario a slight decrease in the neural network quality. Confirming our initial expectations of a non-optimal but acceptable performance, the coefficients of determination $R^2$ for all training datasets are lower than 0.90. Moreover, we estimate the efficiency of the surrogate model in the same manner as in the previous case. However, here we consider values for the adjustment factor $\eta > 1.0$. More precisely, the adjustment factor tries to estimate the time spent in search of the neural network hyperparameters to optimize the surrogate model accuracy and to generate larger training sets. Without loss of generality, we assume that the number of samples $N_S$ to train the neural network is equal to 1100, 600 to train, and 500 to test the surrogate model. Figure \ref{fig:test_r2_eff_f45}(b) shows the efficiency in function of $N_{MC}$, for several adjustment factors. Note that for scenarios with $N_{MC}\leq~10,000$, the efficiency drops significantly for higher adjustment factors. However, for scenarios where $N_{MC} \geq 20,000$ samples the efficiency reaches values close to 80-90\%. For scenarios where $N_{MC} \geq 40,000$ we observe an efficiency close to 90\% even for the higher adjustment factor.

\begin{figure}
    \centering
    \begin{subfigure}[b]{0.4\textwidth}
        \includegraphics[width=\textwidth]{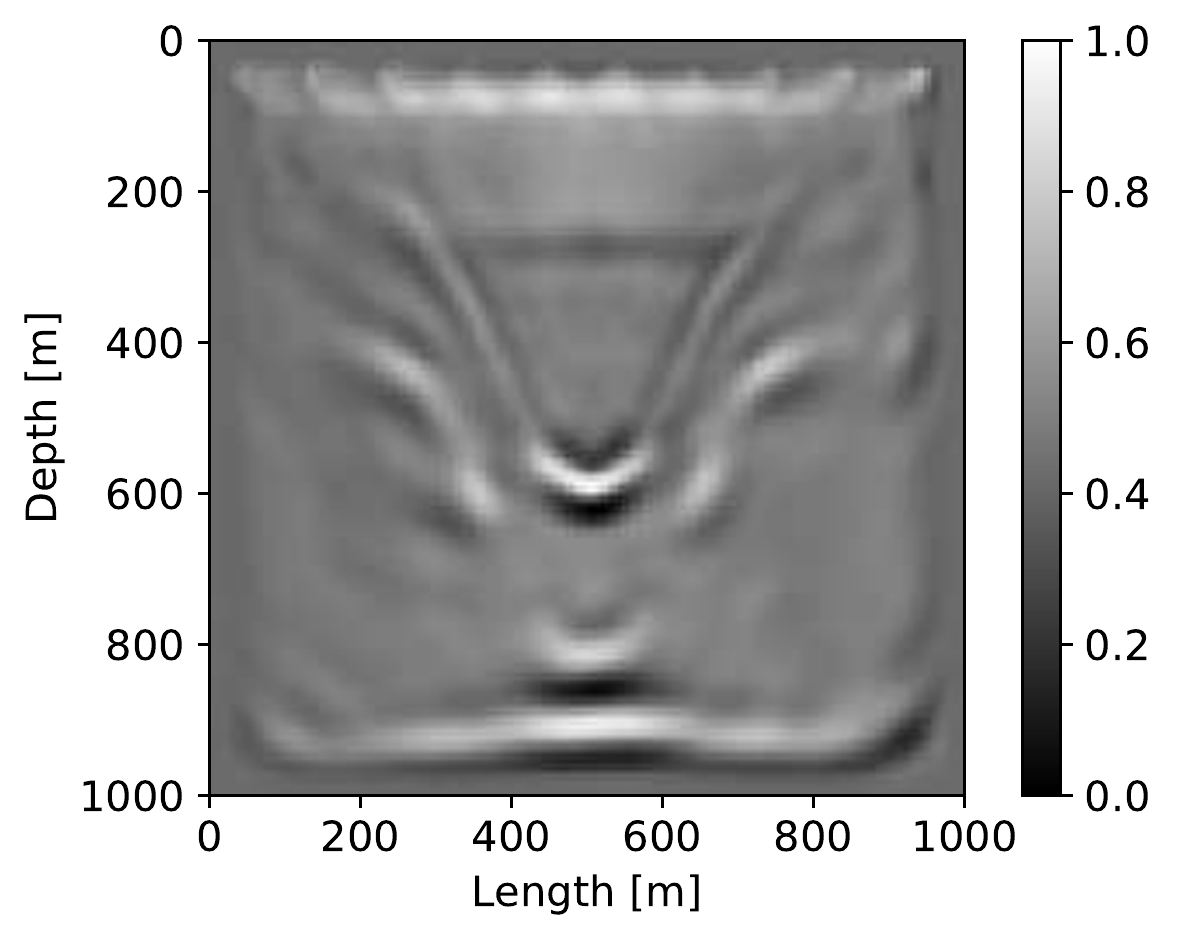}
    \end{subfigure}
    \begin{subfigure}[b]{0.4\textwidth}
        \includegraphics[width=\textwidth]{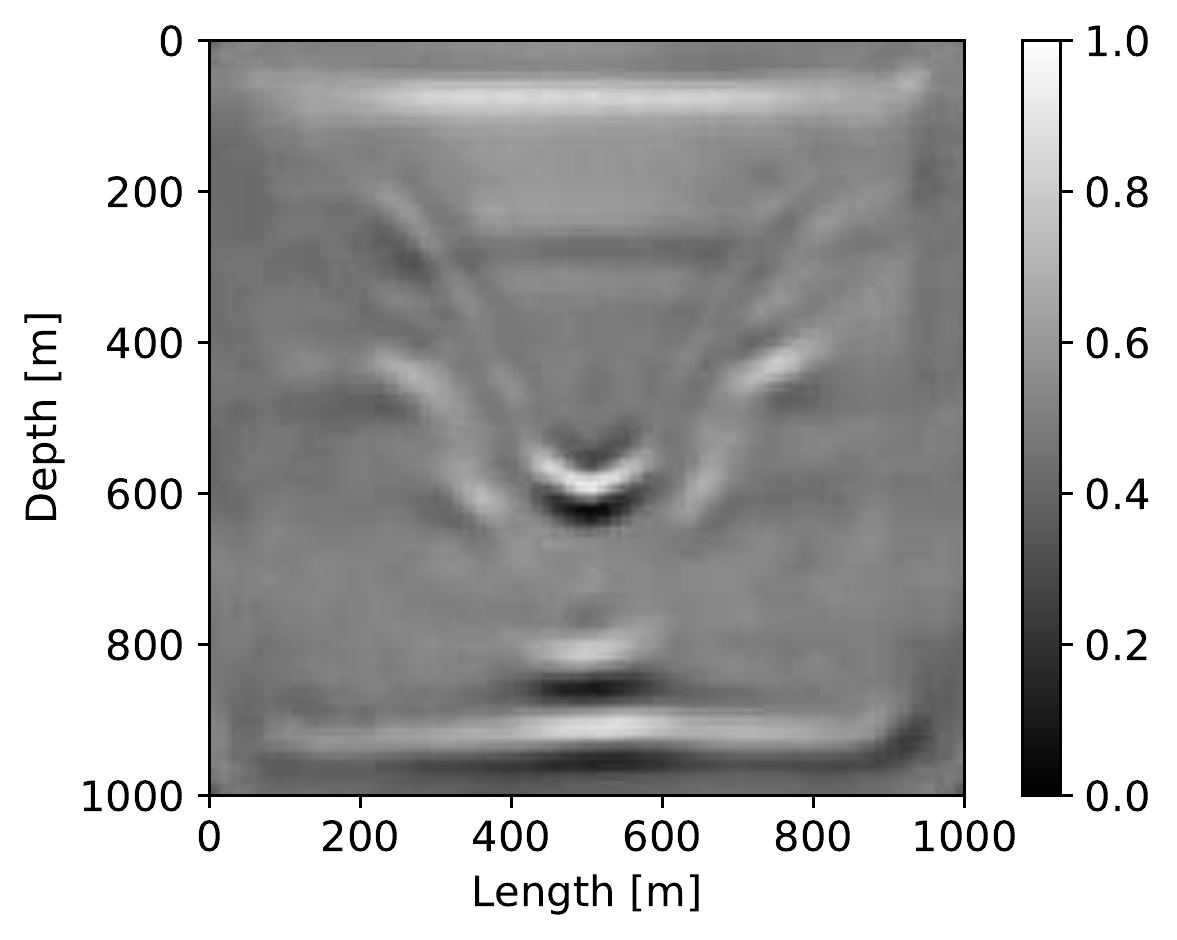}
    \end{subfigure}
    \begin{subfigure}[b]{0.4\textwidth}
        \includegraphics[width=\textwidth]{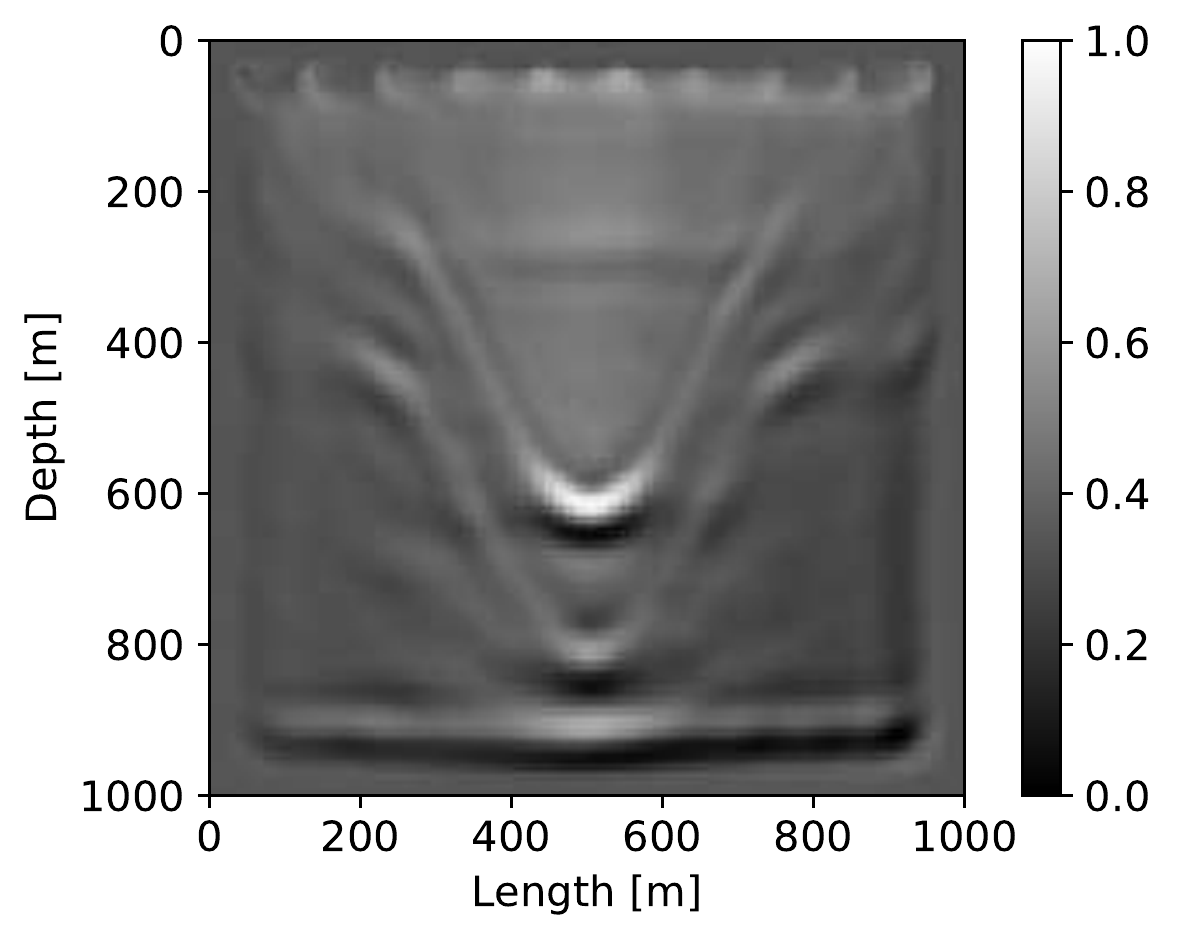}
    \end{subfigure}
    \begin{subfigure}[b]{0.4\textwidth}
        \includegraphics[width=\textwidth]{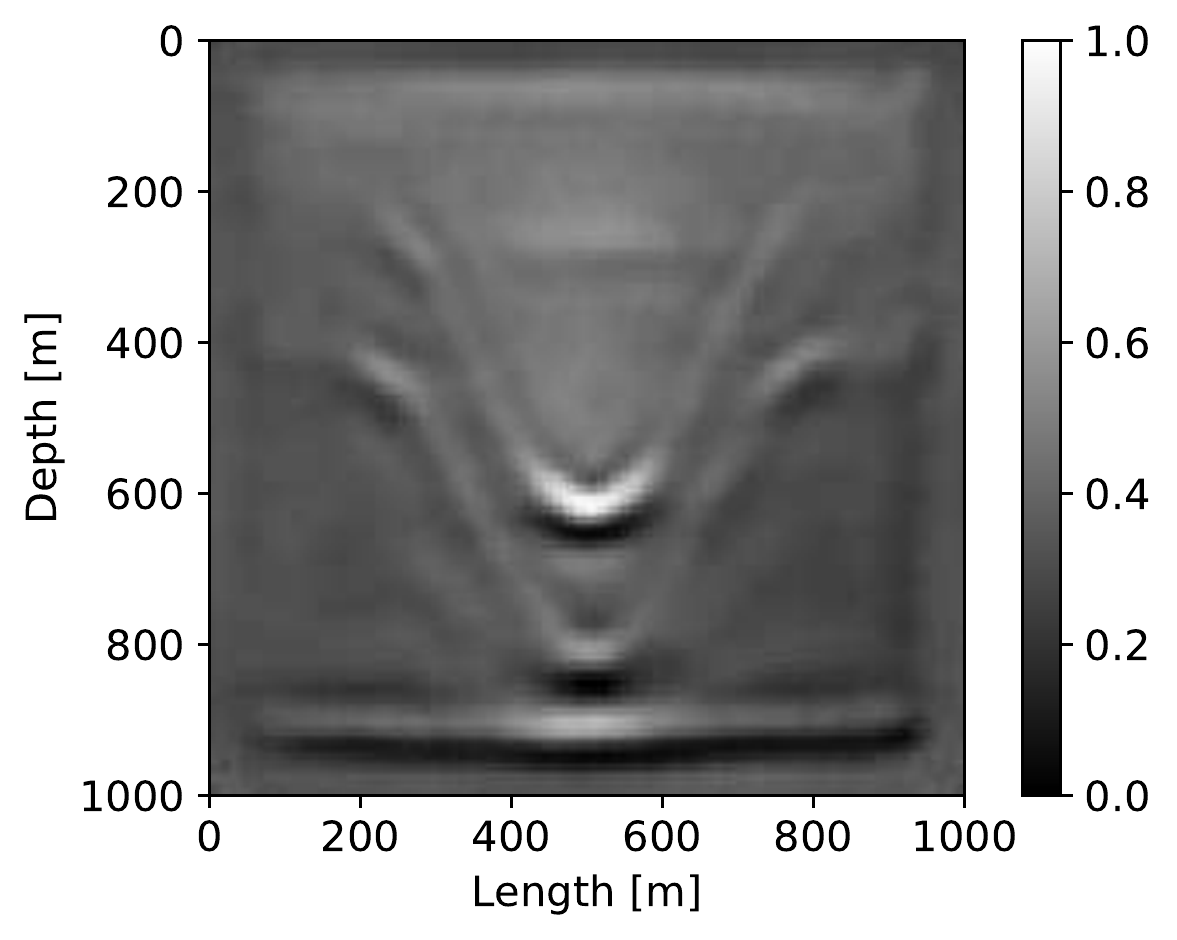}
    \end{subfigure}
    \begin{subfigure}[b]{0.4\textwidth}
        \includegraphics[width=\textwidth]{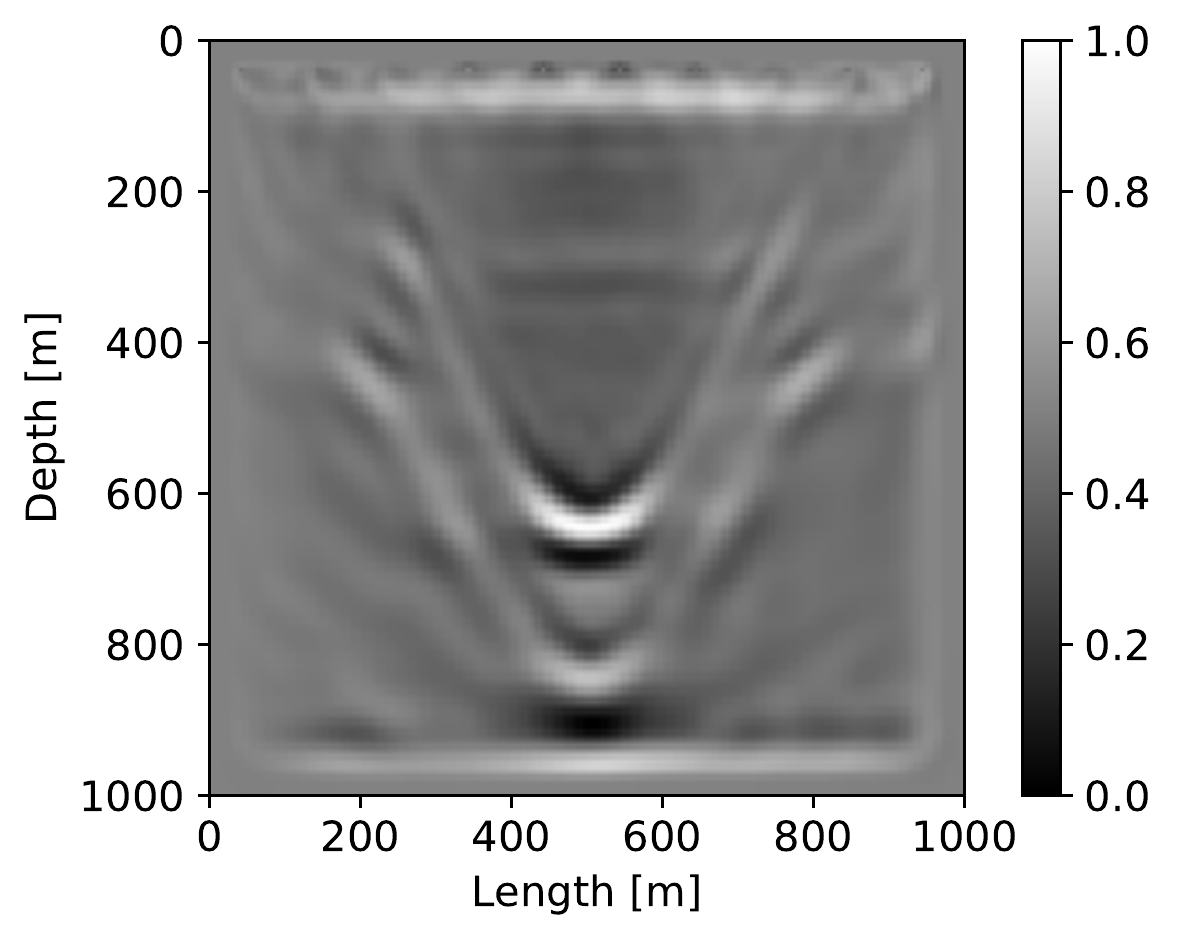}
        \caption{RTM model}
    \end{subfigure}
    \begin{subfigure}[b]{0.4\textwidth}
        \includegraphics[width=\textwidth]{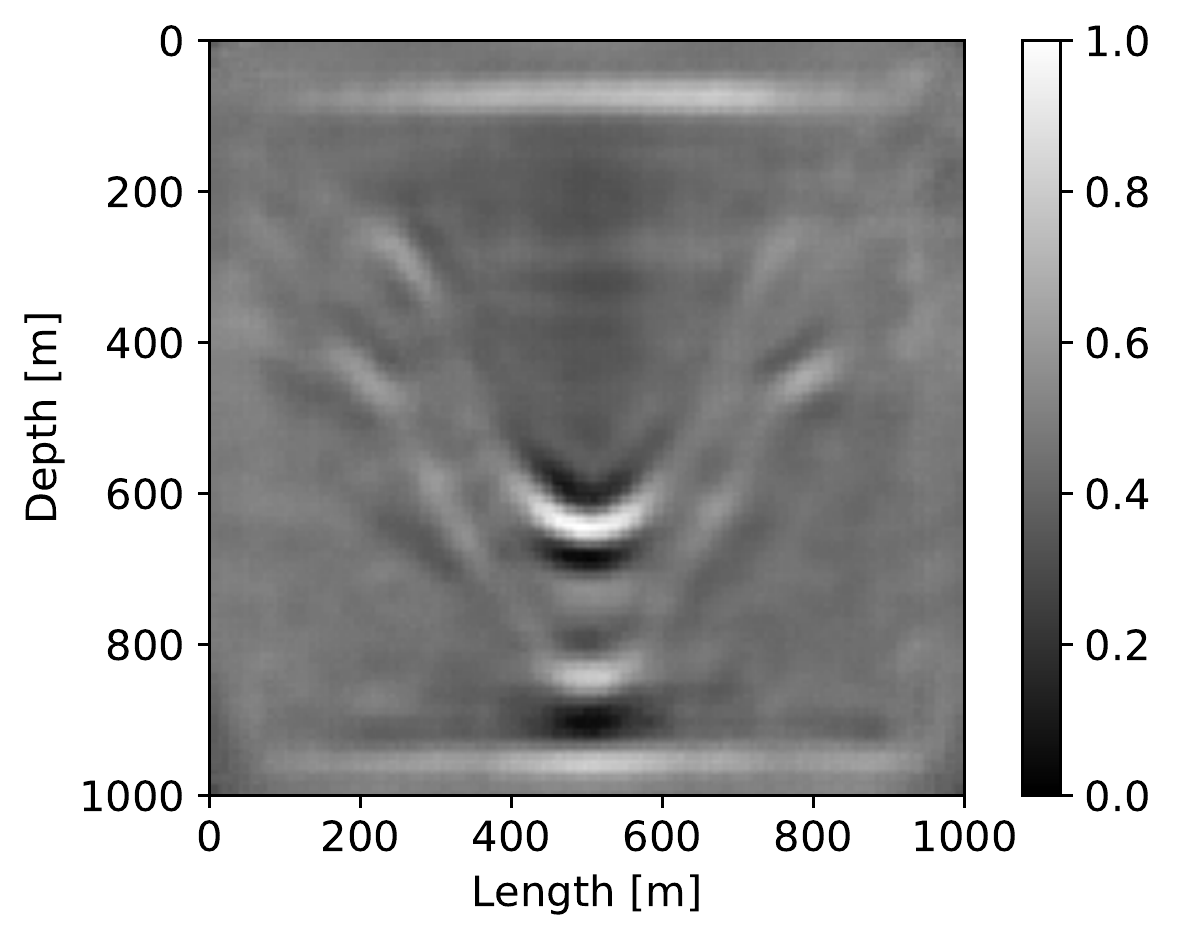}
        \caption{Surrogate model}
    \end{subfigure}
    \caption{Randomly selected images from the test data set computed by the RTM model (a) and the surrogate model (b) trained with 600 samples. The relative errors $e_I=$ are lower than 10\%.}
    \label{fig:prediction_f45}
\end{figure}

\begin{figure}
    \centering
    \begin{subfigure}[b]{0.39\textwidth}
        \includegraphics[width=\textwidth]{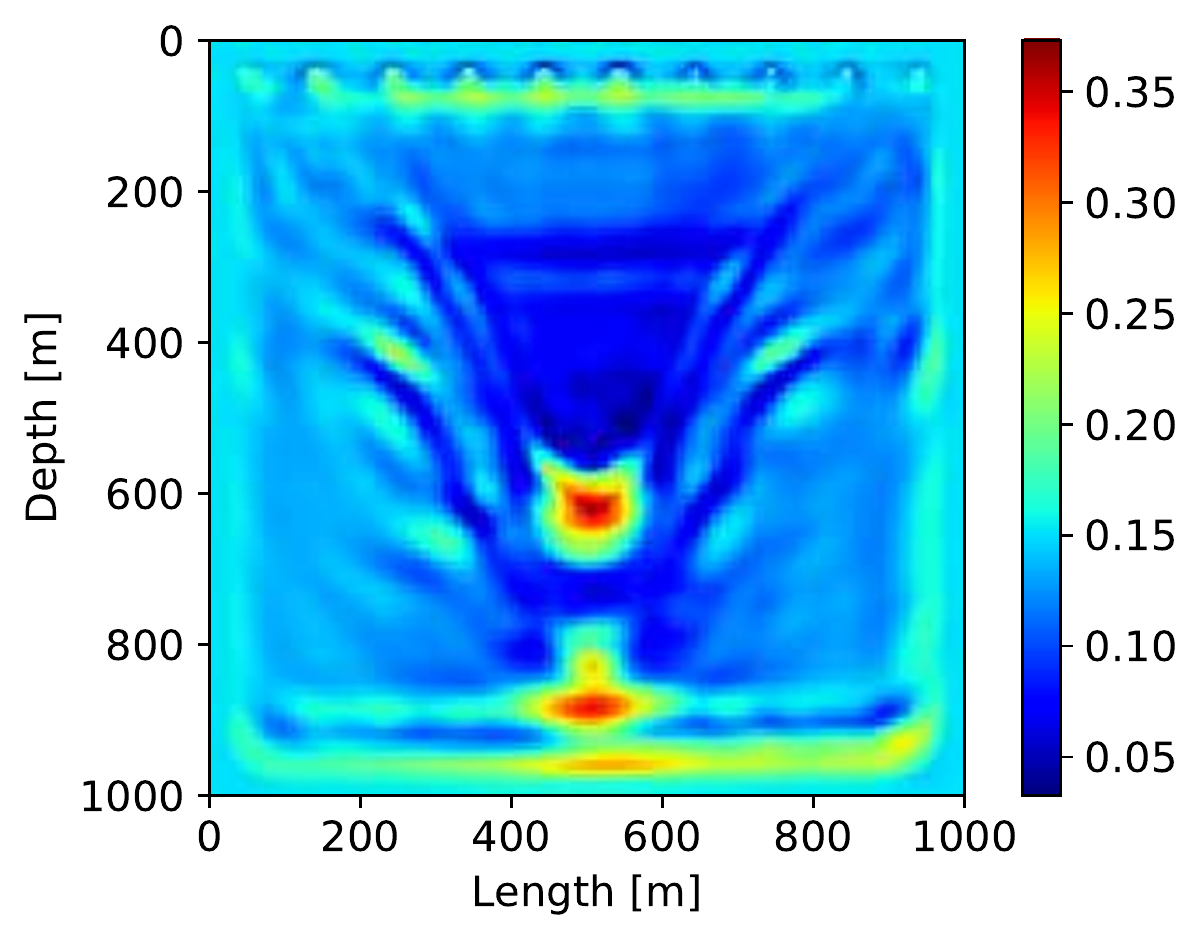}
        \caption*{Standard deviation - RTM model}
    \end{subfigure}
    \begin{subfigure}[b]{0.39\textwidth}
        \includegraphics[width=\textwidth]{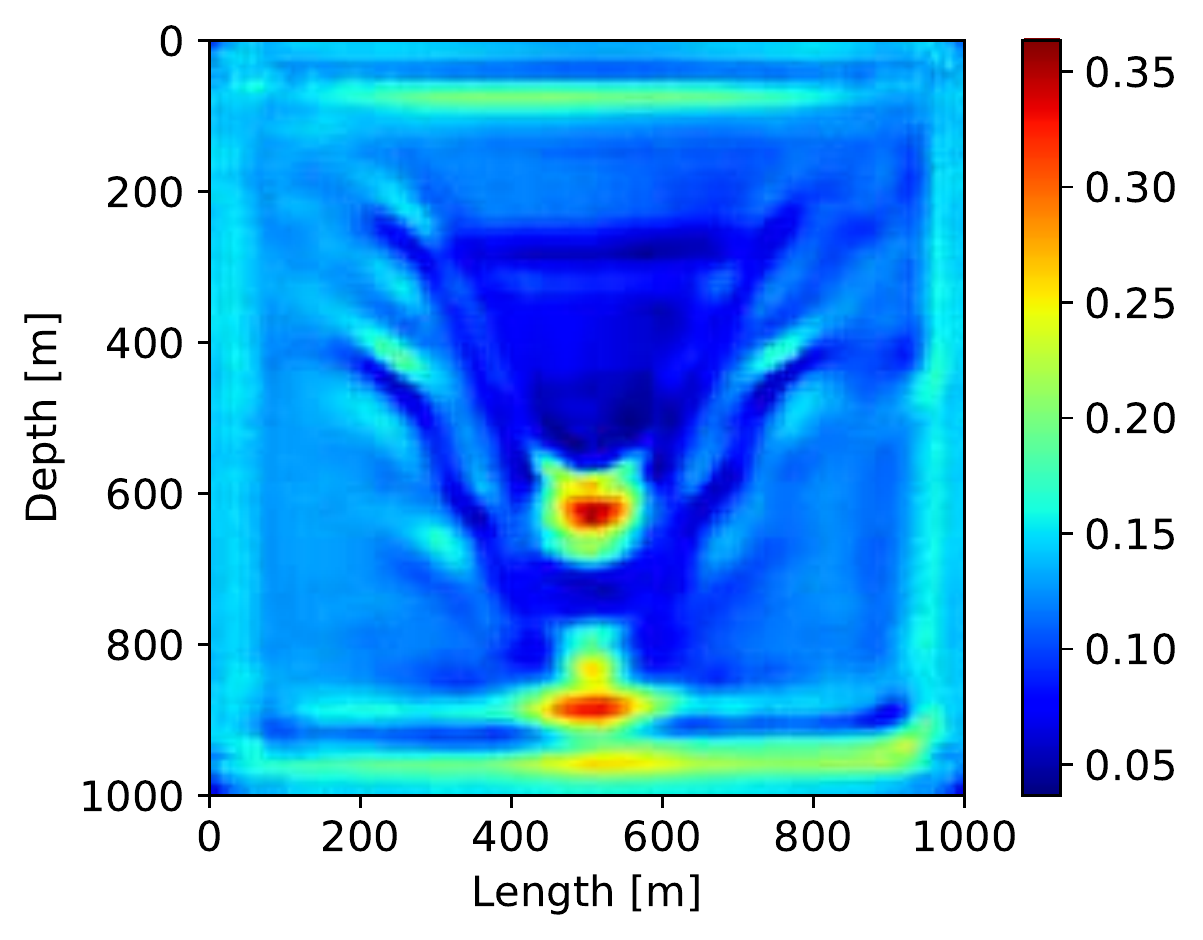}
        \caption*{Standard deviation - Surrogate model}
    \end{subfigure}
    \begin{subfigure}[b]{0.39\textwidth}
        \includegraphics[width=\textwidth]{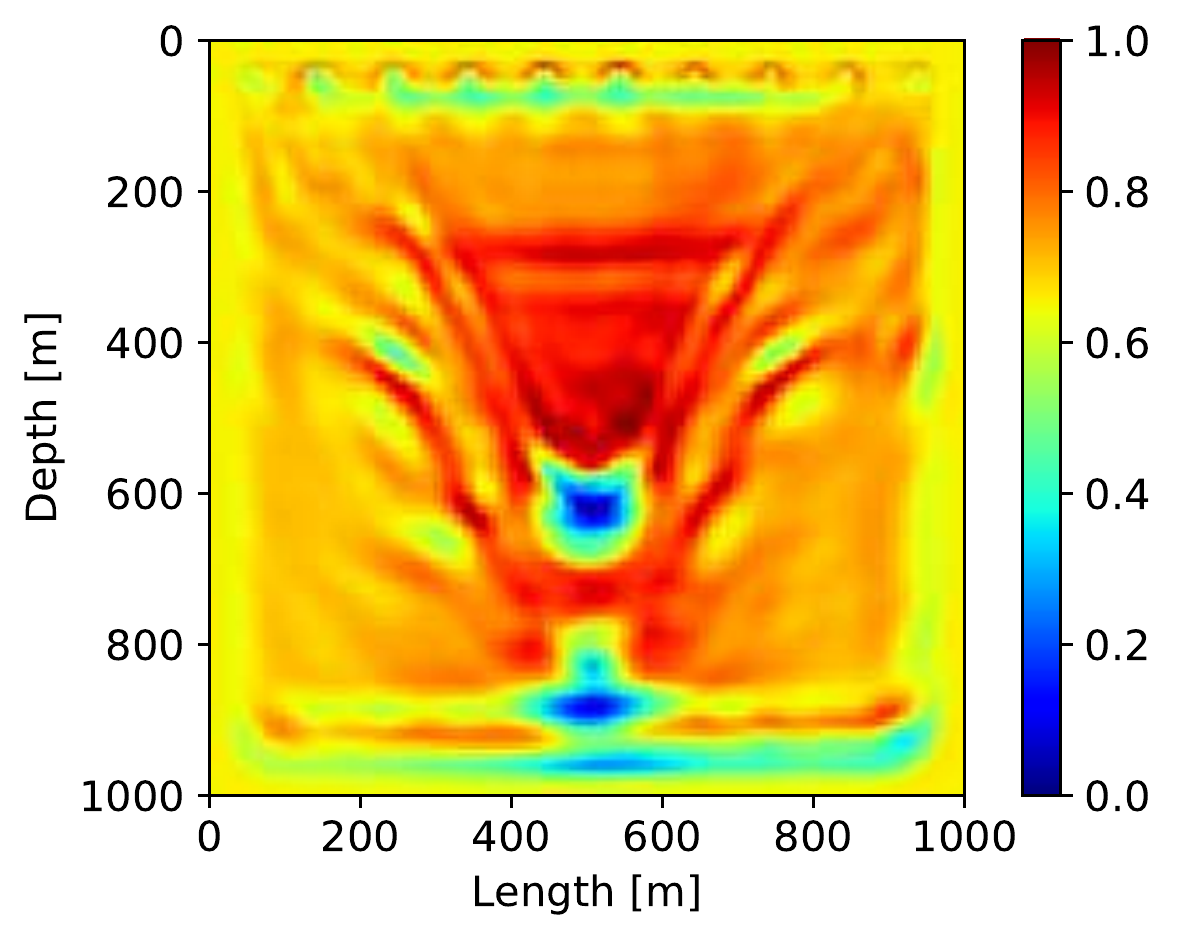}
        \caption*{Confidence index - RTM model}
    \end{subfigure}
    \begin{subfigure}[b]{0.39\textwidth}
       \includegraphics[width=\textwidth]{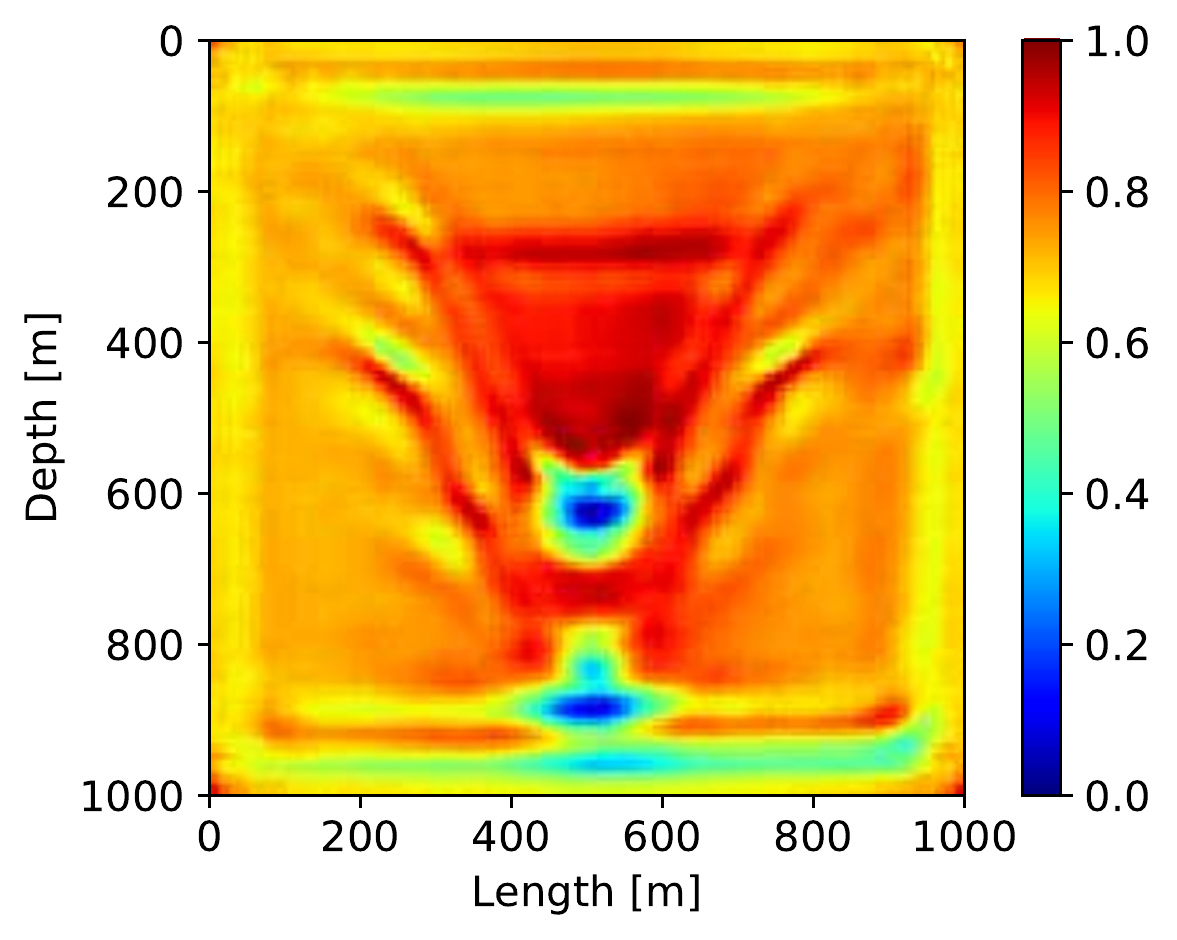}
       \caption*{Confidence index - Surrogate model}
    \end{subfigure}
    \begin{subfigure}[b]{0.39\textwidth}
        \includegraphics[width=\textwidth]{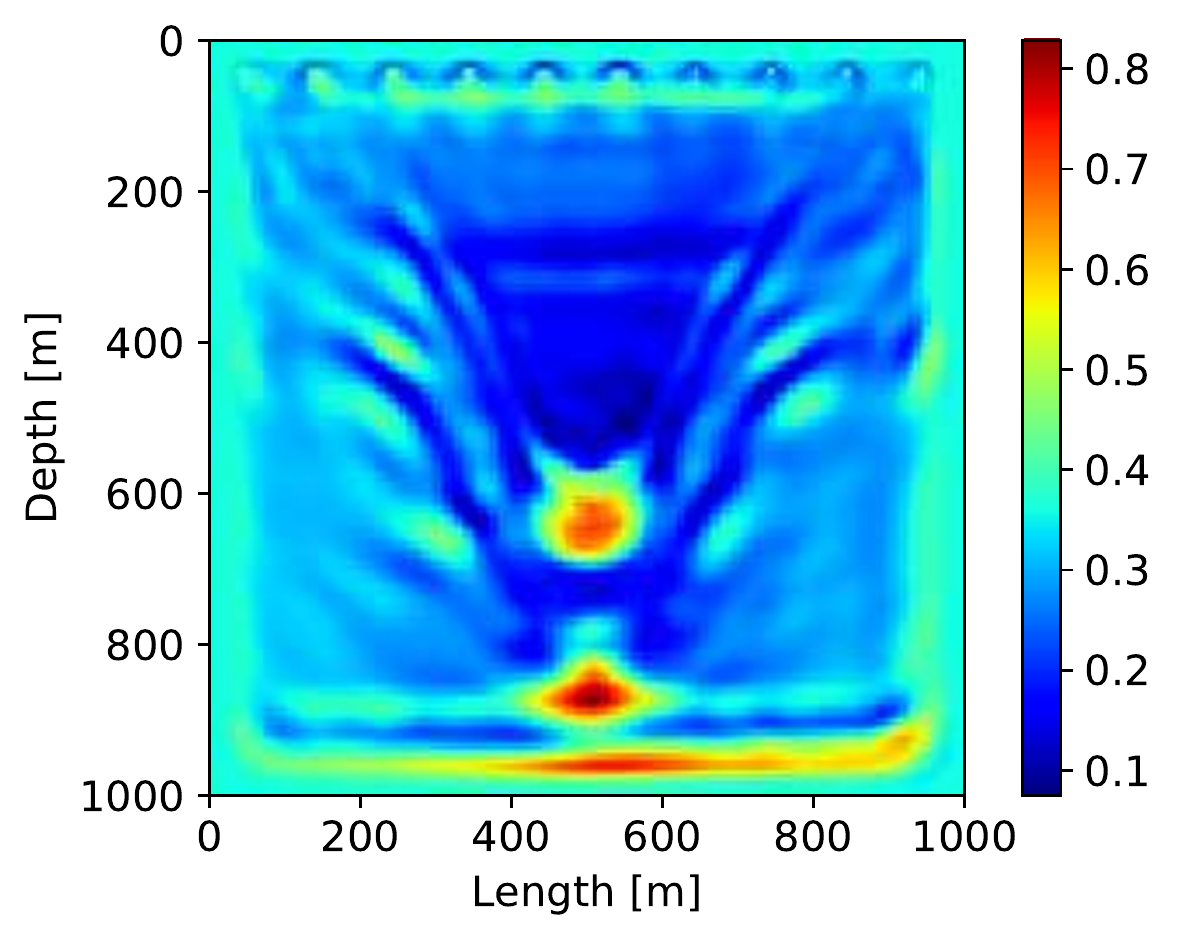}
        \caption*{Coefficient of variation - RTM model}
    \end{subfigure}
    \begin{subfigure}[b]{0.39\textwidth}
        \includegraphics[width=\textwidth]{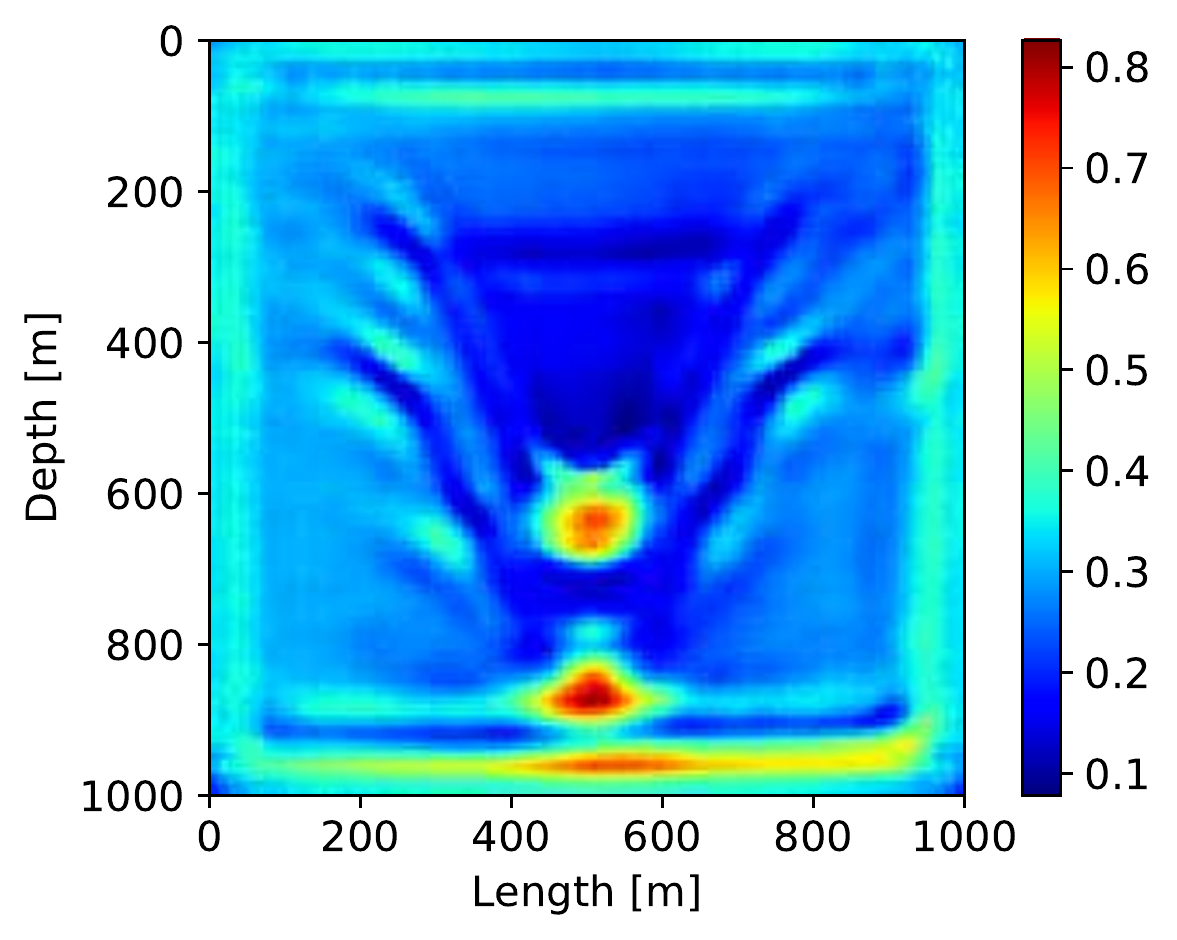}
        \caption*{Coefficient of Variation - Surrogate model}
    \end{subfigure}
    \caption{UQ indexes - standard deviation, $\sigma(\mathbf{r})$, confidence index, $c(\mathbf{r})$, and coefficient of variation, $c_v(\mathbf{r})$ - predicted by the RTM model (left) and the surrogate model (right). The relative errors between the surrogate predictions to the RTM model for the UQ indexes are lower than 6\%.}
    \label{fig:uq_maps_5layers_f45}
\end{figure}

Despite the lower accuracy presented in this scenario, the surrogate model could reach satisfactory predictions of the imaging condition, as we can see in Fig.  \ref{fig:prediction_f45}. In this Figure, we show the three randomly selected images from the test data set computed by the RTM model and the surrogate model. Note, however, that the image produced by the RTM model may not be the best image we can compute for these conditions. The grid is adjusted only to satisfy the stability and dispersion criteria for the 45~Hz cutoff frequency. We do not optimize the domain size for a proper representation of the non-reflecting boundary conditions and source/receiver arrangement. Furthermore, Figure \ref{fig:uq_maps_5layers_f45} shows a comparison between the standard deviation, $\sigma(\mathbf{r})$, confidence index, $c(\mathbf{r})$, and coefficient of variation, $c_v(\mathbf{r})$, computed by the RTM and surrogate models. We observe that the surrogate model predicts the UQ indexes with satisfactory accuracy. The relative errors between the surrogate predictions to the RTM model for the UQ indexes are lower than 6\%.

\begin{figure}
    \centering
    \begin{subfigure}[b]{0.45\textwidth}
        \includegraphics[width=\textwidth]{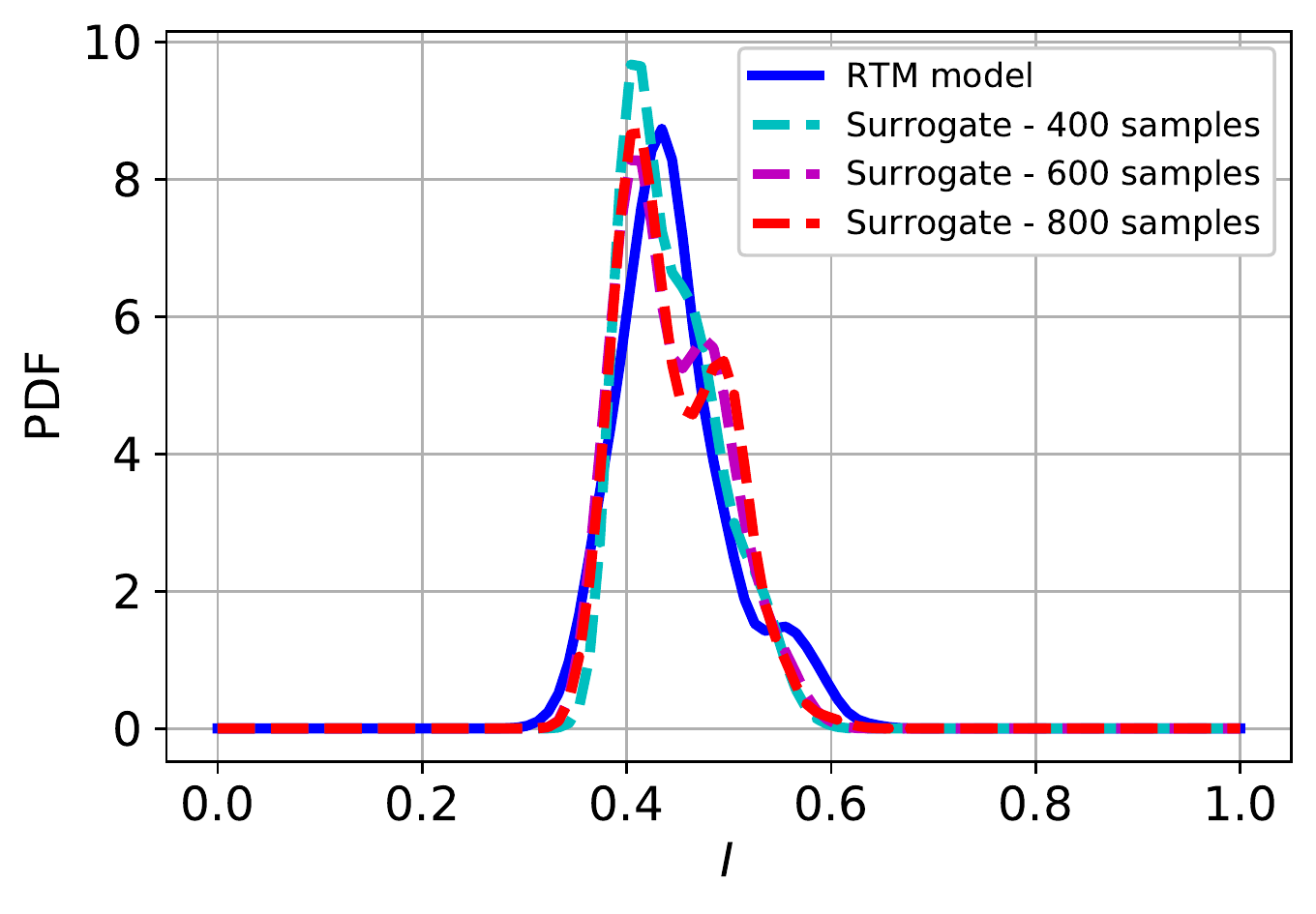}
        \caption*{$P_1$}
    \end{subfigure}
    \begin{subfigure}[b]{0.45\textwidth}
        \includegraphics[width=\textwidth]{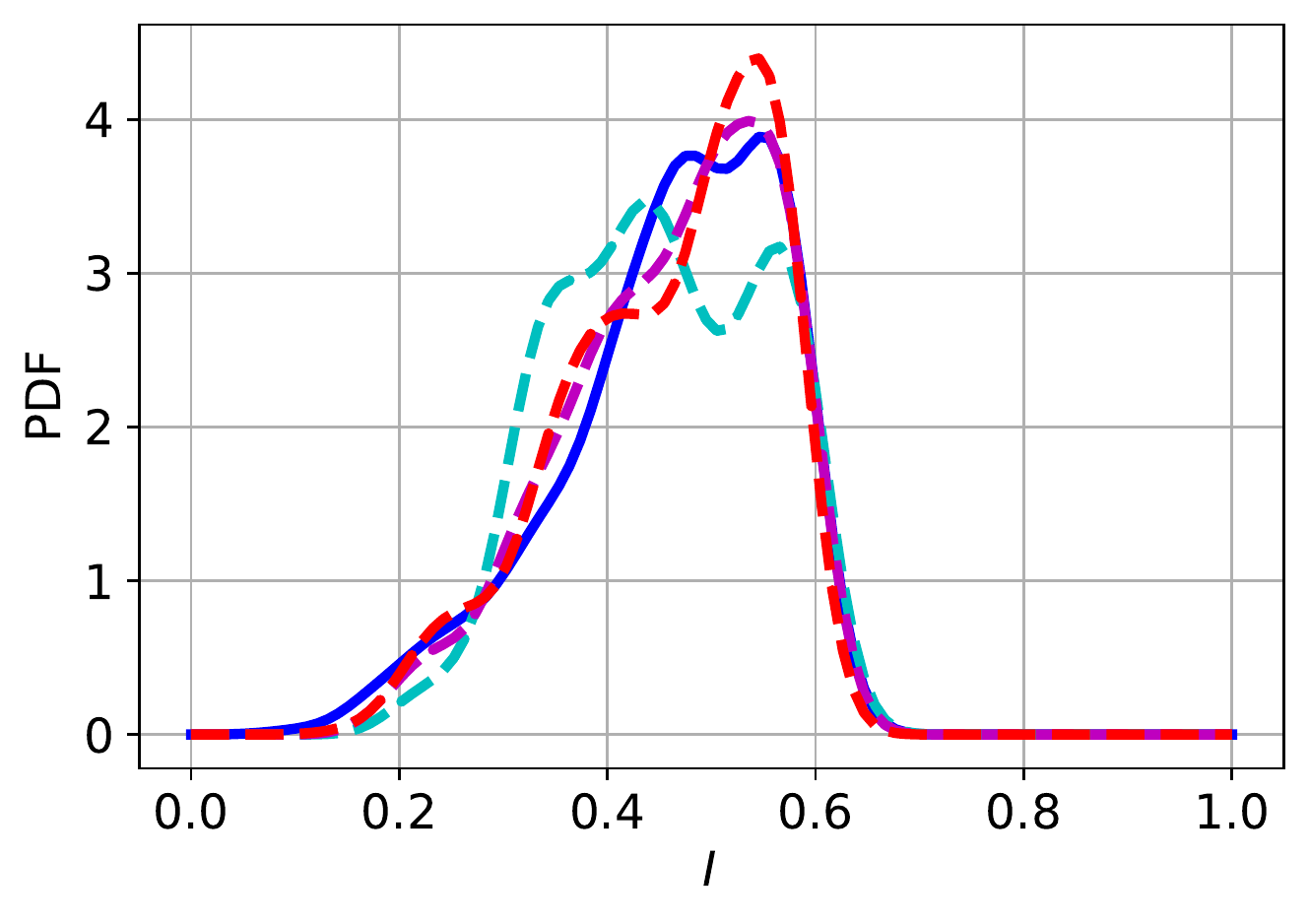}
        \caption*{$P_2$}
    \end{subfigure}
    \begin{subfigure}[b]{0.45\textwidth}
        \includegraphics[width=\textwidth]{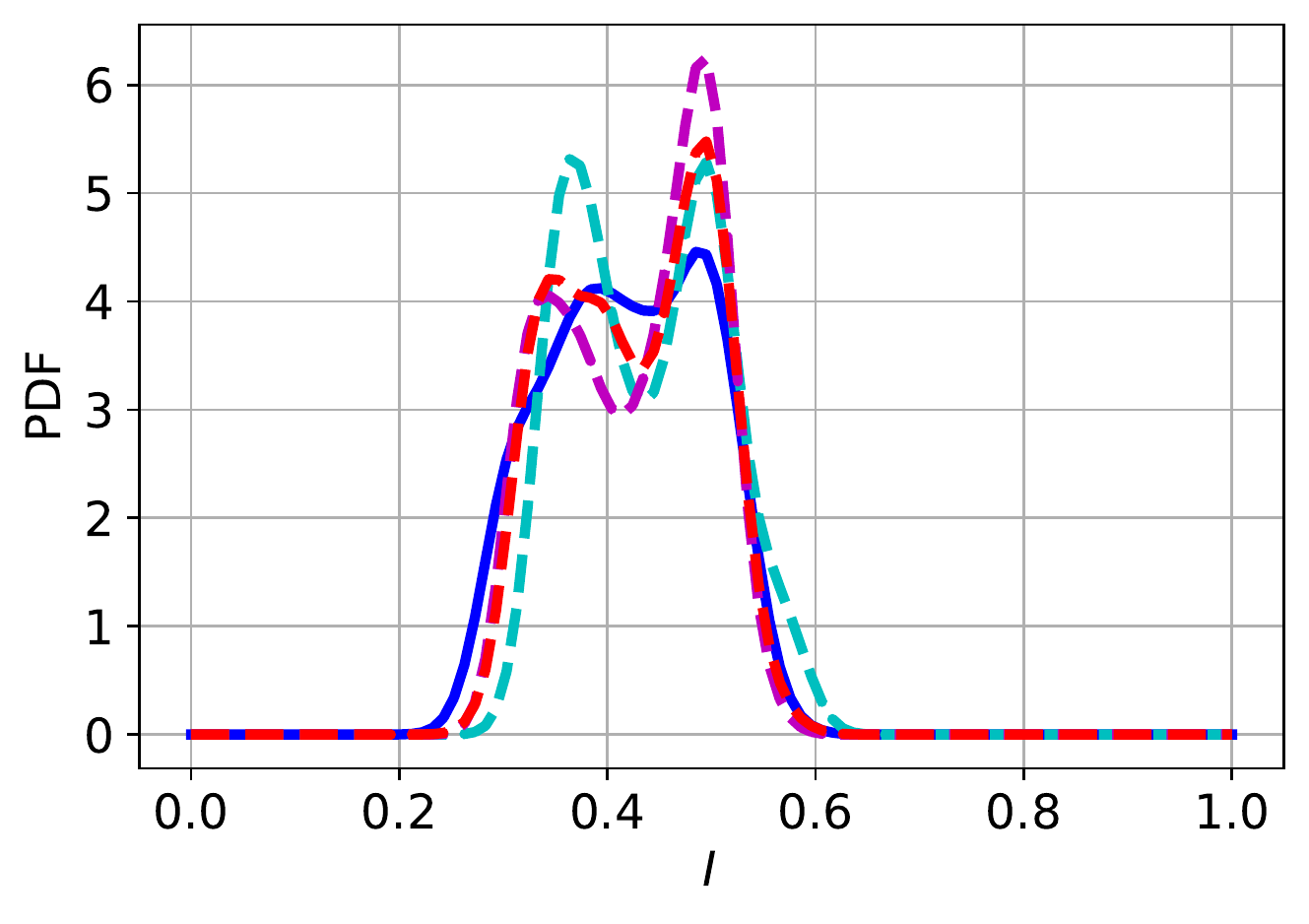}
        \caption*{$P_3$}
    \end{subfigure}
    \begin{subfigure}[b]{0.45\textwidth}
        \includegraphics[width=\textwidth]{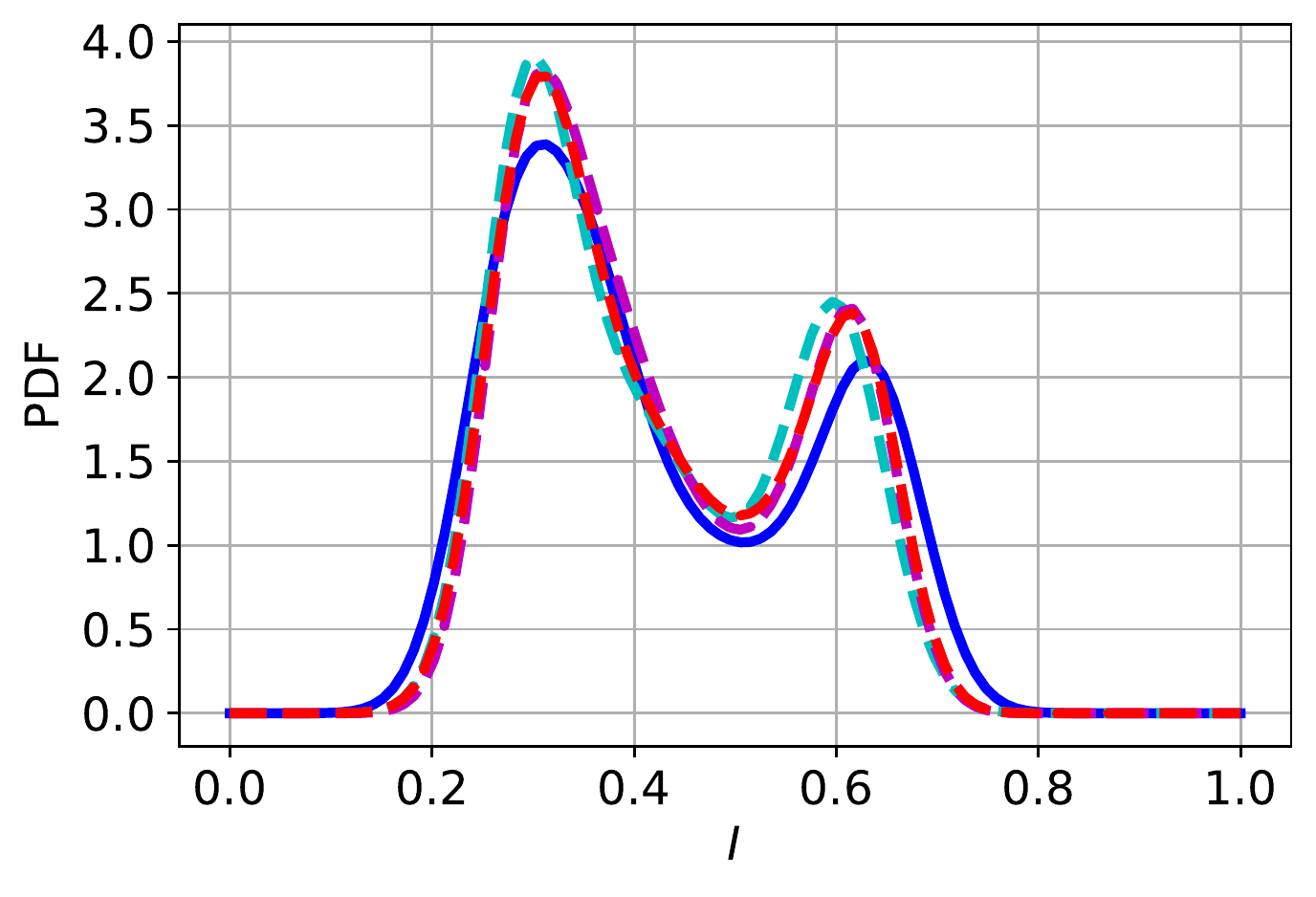}
        \caption*{$P_4$}
    \end{subfigure}
    \caption{Comparison between PDFs predicted by the RTM model and the surrogate model. }
    \label{fig:pdf_points_f45}
\end{figure}

We now investigate the probability density functions (PDFs) of the imaging condition at the control points in Fig. \ref{fig:velocity_field}. We use again as reference solution PDFs obtained by the RTM model with 500 test samples to verify the accuracy of the surrogate models trained with different datasets to estimate the PDFs at the control points. Figure \ref{fig:pdf_points_f45} depicts the imaging condition PDFs at the control points estimated by the surrogate model trained with 200, 400, 600, 800 samples, together with the reference PDFs. We observe that the PDFs obtained with the surrogate model capture well the reference PDFs in all control points, particularly for large training datasets.

\section{Conclusions}
We propose a deep learning model based on an encoder-decoder architecture to replace the costly RTM technique on producing seismic images. This approach naturally fits the framework of a computational workflow to produce seismic images with quantified uncertainty in \cite{barbosa2020workflow}. This surrogate model builds a scalable image--to--image mapping, coping with the high dimensionality of both the heterogeneous velocity fields that serve as inputs and images outputs.  Such surrogate has revealed to be very efficient in the context of UQ many-query tasks, as demonstrated by our numerical examples. Indeed, that was observed even in cases where we employ a non-optimal neural network architecture. 

We place our contribution in the emerging area of physics-informed machine learning, where the final model, in many different ways, blends two main components: often expensive computational models relying on first principles and phenomenological closure equations, and machine learning data-driven tools. Such combination not only suits perfectly to the needs required by the workflow mentioned earlier but also offers a broad spectrum of opportunities to improve performance, like employing more powerful training strategies and automatic hyperparameters optimization. 

\section*{Acknowledgements}
This study was financed in part by CAPES, Brasil Finance Code 001. This work is also partially supported by FAPERJ (grant E-26/203.018/2017), CNPq (grant 302489/2016-9), and Petrobras (grant 2017/00148-9).

\bibliographystyle{unsrt}
\bibliography{my_refs}

\end{document}